# Electron generation of leptons and hadrons
## with reciprocal α-quantized lifetimes and masses

Malcolm H. Mac Gregor

130 Handley Street, Santa Cruz, CA 95060

e-mail: mhmacgreg@aol.com

May 25, 2005

## Abstract

In elementary particle theories the fine structure constant $\alpha = e^2/\hbar c$ serves as the coupling constant for lepton interactions (QED), but is assumed to play no role in hadron interactions (QCD). However, experiments have long indicated an $\alpha$ spacing in the lifetimes of the long-lived threshold-state hadrons, and they also suggest an $\alpha$-related mass structure. The relevance of $\alpha$ to hadron interactions is an experimental question, independent of theory. In the present paper we first make a detailed analysis of the experimental lifetime data. This analysis demonstrates that out of 156 particles with well-determined lifetimes $\tau$, the 120 short *excited-state* lifetimes $\tau < 10^{-21}$ sec have a continuum of values, but the 36 long *threshold-state* lifetimes $\tau > 10^{-21}$ sec occur in $\alpha$-spaced groups that cleanly sort out the *s, c, b* quark flavors. These 36 metastable lifetimes also exhibit a factor-of-three *c*-to-*b* "flavor structure" and a pervasive factor-of-two "hyperfine structure". We then trace out an $\alpha$-defined set of mass quanta that tie together leptons and hadrons. Mass generation occurs via an initial "$\alpha$-leap" from an electron pair to a "platform state" M, and then subsequent excitations by a dominant quantum X. The low-mass "MX octet" of particles—$\mu$, p, $\tau$, $\pi$, $\eta$, $\eta'$, K, $\phi$—is reproduced to an average accuracy of 0.4%, with no adjustable parameters except a small binding energy for hadronic pairs. Without the inclusion of lepton masses, the spectrum of hadron masses is difficult to understand. These reciprocal $\alpha$-quantized results reinforce the reality of the spin 1/2 *u, d, s, c, b* quarks, and they also lead to the identification of a closely-related set of spinless mass quanta for the pseudoscalar mesons.

## I. TWIN PARTICLE MYSTERIES: MASSES AND COUPLING CONSTANTS

### A. The domain of the fine structure constant $\alpha = e^2/\hbar c$

One of the most important successes in modern physics has been the theory of quantum electrodynamics (QED). This is the theory of the interactions of electrons and muons with photons. The QED electron-photon interaction strength is expressed in terms of the dimensionless coupling constant $\alpha = e^2/\hbar c \cong 1/137$. In the QED calculation of the anomalous magnetic moment of the electron, a perturbation expansion is made in powers of $\alpha$, and the calculated value matches the experimental value to an accuracy of one part in $10^{10}$! A calculation of comparable accuracy occurs for the anomalous magnetic moment of the muon, another lepton. The theoretical framework for the much stronger hadron interactions is quantum chromodynamics (QCD), which was patterned after QED. The leptons, including the fine structure constant $\alpha$, have not been incorporated into QCD. In particular, there has been no indication from the calculations of QCD that the coupling constant $\alpha$ might play a role in the generation of hadron masses. In fact, the role of the electron in generating *lepton* masses has never been clarified. The entire problem of mass generation has remained somewhat obscure. In his classic book *QED* [1], Richard Feynman devoted the first three chapters to the interactions of electrons and photons. Then in the fourth and final chapter he took this information over and discussed how much of it applies to hadron QCD interactions. At the very end of the book, in his final observation, Feynman commented on the mass problem as follows:

*"Throughout this entire story there remains one especially unsatisfactory feature: the observed masses of the particles, m. There is no theory that adequately explains these numbers. We use the numbers in all our theories, but we don't understand them—what they are, or where they come from. I believe that from a fundamental point of view, this is a very interesting and serious problem."* [2]

Present-day theories attribute elementary particle masses to the operation of the hypothetical "Higgs particle", which seems required in order to have a renormalizable particle theory. In his book "*Facts and Mysteries in Elementary Particle Plysics*" [3], Martinus Veltman describes its salient features in the following way:

*"In short, it [the Higgs] must be coupled to **any** particle having a mass. Moreover, the coupling must always be proportional to the mass of the particle to which it is coupled.*

*To date the Higgs particle has not been observed experimentally. Unfortunately the theory has nothing to say about its mass, except that it should not be too high (less than, say, 1000 GeV) ... .*

*Because this Higgs particle seems so intimately connected to the masses of all elementary particles, it is tempting to think that somehow the Higgs particle is responsible for these masses. Up to now we have no clue as to where masses come from: they are just free parameters fixed by experiment.*

*... There is clearly so much that we do not know! ..."* [4]

In addition to the mystery of the origin of elementary particle masses, the fine structure constant $\alpha$ itself represents one of the greatest mysteries in modern physics. Feynman has summarized this situation very vividly in *QED*, where he uses the symbol *e* to stand for $\alpha$:

"*There is a most profound and beautiful question associated with the observed coupling constant, e—the amplitude for a real electron to emit or absorb a real photon. It is a simple number that has been experimentally determined to be close to –0.08542455. (My physicist friends won't recognize this number, because they like to remember it as the inverse of its square: about 137.03597 with an uncertainty of about 2 in the last decimal place. It has been a mystery ever since it was discovered more than fifty years ago, and all good theoretical physicists put this number up on their wall and worry about it.)*

*Immediately you would like to know where this number for a coupling comes from: is it related to pi, or perhaps to the base of natural logarithms? Nobody knows. It's one of the **greatest** ... mysteries of physics: a **magic number** that comes to us with no understanding by man*". [5]

In QED the dimensionless constant $\alpha$ gives the exact interaction strength for an electron to produce a photon, including all of the various pathways along which the electron can travel in generating the photon [6]. In the present studies we argue the case that the electron and its coupling constant $\alpha$ generate not only the photon, but also the spec-trum of leptons and hadrons. The domain of the fine structure constant $\alpha = e^2/\hbar c$ seems to be much larger than currently believed. This viewpoint has long been suggested by the experimental data [7].

## B. The power of reciprocal quantum mechanical variables

There is an indirect way of approaching elementary particle masses that has not been fully exploited. This involves making use of the fact that masses and lifetimes are phenomenologically-linked quantum mechanical variables. The lifetime of a particle is a measure of the stability of the mass substructure that composes the particle. Thus we can hopefully obtain global information about particle *masses* by studying the global structure of particle *lifetimes*. Specifically, if the lifetimes do or do not exhibit a dependence on $\alpha$, then the masses may be expected to do the same. In the domain of lepton physics the relevance of the coupling constant $\alpha$ has been clearly demonstrated both theoretically and experimentally, as described above in Sec. A. In the domain of hadron physics a great deal of attention has been paid to the problem of the overall mass spectrum, but far less attention has been devoted to the overall lifetime spectrum. The global lifetime spectrum is studied in detail in Part II, where we use the world-wide elementary particle data base [8] to compile a comprehensive set of elementary particle lifetimes. We then we analyze these lifetimes numerically to see what they reveal with respect to the constant $\alpha$. This analysis leads to the conclusion that $\alpha$ is a relevant scaling factor for the lifetimes of the long-lived threshold-state particles. Then, after briefly considering the reciprocal relationship between lifetimes and masses in Part III, we move on in Part IV to an analysis of the particle masses in order to ascertain if they corroborate the lifetime studies. These mass studies, in analogy to the lifetime studies, are concentrated on the long-lived particle threshold states where the mass structure is clear-cut. This study provides evidence for an $\alpha$-dependence in these particle masses that is both comprehensive and quantitatively accurate. Finally, in Part V we discuss some ramifications of these results.

The information that is contained in the elementary particle lifetime and mass data base leads to the seemingly inescapable conclusion that the domain of the fine structure constant $\alpha = e^2/\hbar c$ is not restricted to leptons, but also extends to include hadrons.

### C. Experiment, phenomenology, theory: the three steps to success

The three fundamental steps in the development of theories in physics, and in science in general, are to see *what* is there (experiment), to determine *how* it is arranged (phenomenology), and finally to explain *why* it happens that way (theory). As historians of science point out, our theories are in general forced on us step by step by the experimental data, and they are not theories that we plausibly arrive at in the absence of data.

One danger in constructing theories is that, given the experimental data base, we accidentally overlook some of the phenomenology, and thus end up developing a theory which cannot account for that phenomenology. The resulting theory may not be comprehensive enough to answer questions that we really need to know about. Historically, the need for an elementary particle theory dates back to the discovery of the electron in 1897 and the delineation of the geometry of the proton in 1911. These two particles carry precisely the same magnitude of electric charge, but with opposite signs, and they have radically different masses. The main challenge to particle theorists for the past century has been to account for this electron-to-proton mass ratio. Unfortunately, the Standard Model, the currently-prevailing elementary particle theory [8], puts electrons and protons into two apparently-unrelated categories—leptons and hadrons, and thus provides no clues as to the nature of this mass ratio.

In searching for a way out of this dilemma, we need to re-examine the elementary particle data base to see if there are any regularities that have not been recognized, and which can be used to confront the particle theories. One area that leaps to the forefront is the collection of particle lifetimes. The long lifetimes of the low-mass threshold states, right from the early days of accelerator particle physics, have exhibited a lifetime grouping, with the groups separated by fairly accurate factors of 1/137. These results were first published in 1970 [9], using an early Particle Data Group compilation [10]. These $\alpha$-spaced lifetime groups contained 13 measured lifetimes (the $\Sigma^o$, $\eta$ and $\eta'$ lifetimes, for example, had not yet been measured), and the calculations for the lifetime table were performed on a Marchand mechanical desk calculator. Over the past 34 years an additional 23 long-lived particle lifetimes have been measured and documented, and have accurately fitted into this α-spaced systematics, as described in Sec. II B below. However, the Review of Particle Physics [8], which summarizes not only the elementary particle data, but also the *physics* that presumably applies to these data, makes no mention of this phenomenology, which has not been incorporated into the Standard Model. The significance of the α-dependence of the lifetimes lies not so much in the lifetimes themselves, although that is certainly of interest, but rather in the fact that it seems logical to expect this α-dependence to carry over in some form to the masses.

The present paper is at the level of phenomenology, and it represents an unfinished task which must be carried out before a "final" theory can be obtained. After we perform this phenomenological task, as described below, we obtain a parameter-free calculation of the electron-to-proton mass ratio that is accurate to better than 1% and essentially independent of theory. This result emerges from what we can think of as a "Mendeleyev diagram" (Fig. 24) for the α-quantized "MX octet" particles (Tables 1 and 2).

We now proceed to a phenomenological global examination of the measured spectrum of elementary particle lifetimes.

## II. EXPERIMENTAL EVIDENCE FOR α-QUANTIZED PARTICLE LIFETIMES

### A. The Review of Particle Physics (RPP) lifetime data base

Research in elementary particle physics has led to the identification of roughly 200 different massive particle states that can be produced in energetic particle collisions. Most of these states are very short-lived, persisting for no more than a couple of orders of magnitude longer than the collision transit time. However, a few are long-lived metastable threshold states that signal the onset of quark excitations within the particles. In the past half century, after the advent of high-energy accelerators, the proliferation of these particle states, together with the expensive equipment required to produce and analyze them, has led to world-wide collaborations in experimental work and in the compilation of the experimental data and its systematic analysis. The Particle Data Group summarizes these results biennially in the Review of Particle Physics. The data base for the present work is RPP2004 [8].

Of the approximately 200 elementary particles and resonances listed in RPP2004, 156 have reasonably-well-established experimental lifetime values. These constitute the lifetime data base for the present studies. In this data base, a few charge multiplets with very similar and rather short lifetimes are characterized as single states. Also, some very-broad-width S-state resonances with poorly-determined widths have been excluded, as well as a few narrow-width states whose lifetimes are presently listed in RPP2004 only as upper limits. This data base is given in Appendix A.

In addition to the elementary particle data on spectroscopic properties and interactions, RPP2004 also provides summaries of the relationships between these particles, and it describes how they fit in with the general formalism of the Standard Model (SM).

### B. The historical emergence of the $\alpha$-quantized lifetime structure

The scaling of elementary particle lifetimes in factors of $\alpha$ was already apparent in the early elementary particle accelerator data, which involved only the *u*, *d* and *s* quarks. Fig. 1 shows the long-lived threshold-particle lifetimes plotted along the abscissa as logarithms $x_i$ to the base $\alpha$, with $\pi^{\pm}$ serving as the reference lifetime. The top display in Fig. 1 contains the particle data as reported in 1971 [11; also see 7, 9 and 12], when only 13 lifetimes in this range had been measured. After the emergence of the *c* quark, it was added to the lifetime $\alpha$-grid [13], and when the *b* quark was discovered, it also was added in [14]. The bottom display in Fig. 1 shows the threshold particles as reported in 2004 [8], with 36 particles included. The fact that the 1971 lifetime $\alpha$-grid, which was based on *u*, *d* and *s* quark states, also accommodates the later *c* and *b* quark states is a nontrivial result. It is in a sense a predictive success, although no one was making quantitative predictions about the *c* and *b* quarks in 1971.

As Fig. 1 illustrates, the threshold lifetimes are $\alpha$-spaced globally, but there are pronounced deviations in the spacings among the individual particles. These deviations are quite informative, as we analyze in detail in the following sections. In particular, they exhibit regularities that can be related to the mass structures of these same particles.

### C. The overall structure of the elementary particle lifetime spectrum

The elementary particle lifetimes that have been experimentally measured to date are displayed in Fig. 2, where they are plotted as exponents $x_i$ in the equation

$$\tau_i = 10^{-x_i}. \tag{1}$$

The longest-lived unstable elementary particle shown here is the neutron, with a lifetime of 886 sec. The shortest is the $Z^o$, with a lifetime of $2.64 \times 10^{-25}$ sec. This is a span of 28 orders of magnitude. A vertical dotted line is drawn at $\tau = 10^{-21}$ sec. As can be seen, a *continuum* of short lifetime values begins to the right of this line, expands rapidly at $10^{-22}$ sec, and terminates at about the transit time for an interaction event in a high energy accelerator. There are 36 long-lived particles to the left of the dotted line. These longer lifetimes occur in *widely separated groups*. As we will demonstrate, these lifetime groups correspond to thresholds where the various types of quark excitations first occur. If we didn't know about the existence of the *s*, *c*, and *b* quarks, we could deduce them from this lifetime spectrum. And this, of course, is exactly what happened in the discoveries of the *c* and *b* quarks. It was the large mass and very narrow width (long lifetime) of the J/$\psi$ vector meson that signaled the presence of a new *c*-quark type of excitation. This result was repeated a few years later with the *b* quark, in the discovery of the even-narrower-width Upsilon vector meson at three times the mass of the J/$\psi$.

An inspection of the regularities in the lifetimes $\tau$ of the 36 particles in Fig. 2 that have $\tau > 10^{-21}$ sec shows four interesting features:

*(1)* The lifetimes occur singly or in groups at spacing intervals *S*, where *S* is the lifetime ratio between groups. Visual inspection of Fig. 2 indicates that *S* is somewhat larger than a factor of 100.

*(2)* Three of the groups have partial slopes (as plotted here in an arbitrary spreading of lifetimes along the y-axis) which are roughly the same. As we demonstrate in the next section, these slopes correspond to a fairly accurate factor-of-two lifetime hyperfine (HF) structure.

*(3)* The two groups near $\tau_i = 10^{-12}$ sec in Fig. 2 form parallel lines (apart from the factor-of-two HF structure). These two groups are separated by a factor of about three in lifetimes, and also, as we will see, by a factor of about three in mass values, with the higher-mass states having the longer lifetimes.

*(4)* The Λ—Ω hyperon pair and the $\Lambda_c$—$\Omega_c$ charmed hyperon pair each contain 1 and 3 flavored quarks, respectively, and both pairs have lifetime ratios of about 3 to 1, thus suggesting that the flavored quarks in the Λ and Ω independently trigger the decays.

Our main interest here is to pin down feature *(1)*, the overall group spacing interval *S*, both to ascertain its numerical value and also to see how far it extends in fitting these lifetimes. But in order to do this, we should first deal with feature *(2)*, the factor-of-two HF structure, which logically represents a quite different lifetime effect than the overall spacing. In Sec. D we plot the lifetime HF ratios and phenomenologically correct for them, which enhances the sharpness and clarity of the overall group spacing. Then in Sec. E we vary the spacing interval *S* and statistically select the best fit to the lifetime data. In considering these results, it should be kept in mind that Fig. 2 contains *all* of the experimentally well-determined lifetimes. The *comprehensiveness* of this lifetime systematics is as important as its accuracy and span of values.

### D. The lifetime factor-of-two (and three) hyperfine (HF) structure

The three lifetime groups in Fig. 2 that showed indications of a regular "fine structure" are displayed in detail in Fig. 3, where the particles are identified and their lifetime ratios are indicated numerically. These lifetime ratios are all approximately factors of two (apart from the matching Λ—Ω hyperon pairs, which we discuss below), and are denoted here as a lifetime "hyperfine" (HF) structure to distinguish them from the factor-of-three c-to-b quark "flavor structure" discussed in Sec. G and the much larger "factor-of-α" structure deduced in Sec. E. As can be seen in Fig. 3, the particles in each of these three HF groups are closely-related with respect to masses and spins, and also with respect to their *s* and *c* quark content. The HF spacings are both slightly larger and slightly smaller that the value 2, and an average over the factor-of-2 pairs shown in Fig. 3 gives a value of almost exactly 2, thus suggesting that the small deviations from 2 are due to smaller random effects. One nearby particle that is shown in Fig. 2 but not in Fig. 3 is the tau lepton, which is indicated by an arrow in Figs. 2 and 5. Both the mass and the lifetime of the tau closely match those of the *c*-mesons shown in the right-hand group in Fig. 3.

Phenomenologically, it seems clear that the HF factors of two displayed in Fig. 3 correspond to some kind of mass and/or charge substructure in these particles. In the case of the $\Sigma$ hyperons, for example, the $\Sigma^+$ decays equally into $p + \pi^o$ and $n + \pi^+$ final states, whereas the $\Sigma^-$ decays only into $n + \pi^-$. Thus it seems logical for the $\Sigma^-$ lifetime to be longer than the $\Sigma^+$ lifetime by a factor of two, as shown in Fig. 3. However, in the case of the $\Xi$ hyperons, the $\Xi^o$ and $\Xi^-$ each decay into a single $\Lambda\pi$ mode, and yet they also have lifetimes that differ by a factor of two. In the absence of a comprehensive theory of these decays, we can empirically "correct" for this HF structure by simply applying factors of two in order to make all of the lifetimes in a group move into the "central" value of the group. The effect of these HF corrections on the lifetime group structure is illustrated in Fig. 4. If the overall structural features which lead to the large separation interval *S* between groups are independent of the HF features that dictate the factors of two, then removing the effect of the factors of two should sharpen the determination of the spacing interval *S*.

The $\Lambda$ and $\Omega$ lifetimes represent a special situation. We have the lifetime ratios

$$(\Lambda^o = uds)/(\Omega^- = sss) = 3.19,$$

$$(\Lambda_c^+ = udc)/(\Omega_c^o = ssc) = 2.90,$$

where $\Lambda$ is the lowest-mass *single-flavored* hyperon and $\Omega$ is the lowest-mass *triple-flavored* hyperon. Their lifetime ratios of about three reflect their flavor structure, and they represent a unique "factor of three HF structure". In the case of the *charmed* $\Lambda_c$—$\Omega_c$ pair, this indicates that the proper HF correction for moving the $\Omega_c$ into the "central" lifetime group is a factor of 3 x 2 = 6, which we apply here. The same correction should logically be applied to the *strange* $\Lambda$—$\Omega$ pair, but it isn't clear whether the $\Lambda$ or the $\Omega$ represents the more accurate reference lifetime, so we just treat them with the averaged factor-of-two corrections of Fig. 3 as an approximate adjustment factor.

We should summarize what we are attempting to accomplish with these HF hyperfine lifetime corrections. The *ansatz* we make here is that the global lifetime spacing parameter *S* that is observed in the 36 long-lived particles of Fig. 2 actually represents a spacing in powers of $\alpha^{-1} \cong 137$. The deviations from these α-spacings include characteristic factor-of-two HF perturbations which are caused by the mass and/or charge substates in the constituent quarks, and which are logically independent of the global α-spacings. We can conceptually remove this HF structure, as shown in Figs. 3 and 4, in order to clarify the α-structure. This is a purely phenomenological correction, but we can test its consequences, as follows: we form two data sets, one with and one without HF corrections, and we then analyze both of them numerically to see how closely they match an α-spaced grid. This procedure answers two questions: *(1)* does the empirical spacing depend strongly on the HF corrections? *(2)* do the HF-corrected data give a better fit to the α-grid than the uncorrected data? As we show in Sec. E, the answers to these two questions are *no* and *yes*, respectively. This indicates that even approximate corrections for the small-scale HF effects represent an improvement in the overall determination of the large-scale lifetime grouping interval. Fig. 5 shows how the lifetimes of Fig. 2 are modified by the incorporation of the HF corrections of Figs. 3 and 4.

In the next section we carry out a statistical analysis which demonstrates that the proper scaling factor for these threshold particle lifetimes is $S \cong \alpha^{-1} \equiv \hbar c/e^2 \cong 137$.

### E. The α-quantization of the elementary particle threshold lifetimes

In this section we examine Figs. 2 and 5 to see what values they yield for the lifetime periodicity *S* that is evident in the 36 long-lived ($\tau > 10^{-21}$ sec) threshold elementary particles, particularly with respect to the reciprocal of the fine-structure constant, $\alpha^{-1} \cong 137$. Our first task is to select a basic reference lifetime, which we take to be the $\pi^{\pm}$ lifetime. The pion is the lowest-energy unstable hadron, and $\pi^{\pm}$ is its longest-lived and most accurately measured charge state. Using the $\pi^{\pm}$ as a reference, we write the lifetimes in the form

$$\tau_i/\tau_{\pi^\pm} = S^{-x_i}, \tag{2}$$

where $S$ is the scaling factor that characterizes the periodicity of the lifetime groupings. The "correct" value for $S$ is the value that *minimizes* the deviations of the exponents $x_i$ in Eq. (2) from integer values. To determine $S$, we define the *average absolute deviation* (AAD) of the $x_i$ exponents as

$$\text{AAD} = \frac{1}{N}\sum_{i=1}^{N}|\Delta x_i|, \quad \Delta x_i \equiv x_i - I_n, \quad I_n = \text{nearest integer}. \tag{3}$$

The value of $S$ we seek is the one that minimizes AAD. Hence we must do calculations over a range of $S$ values and plot out the function AAD($S$). Since the maximum absolute deviation $|\Delta x_i|$ of an individual lifetime exponent $x_i$ from an integer value is 0.5, a *random* distribution of $x_i$'s will yield AAD $\cong$ 0.25. But if the lifetimes being tested correspond to a definite $S$ value, then AAD should show a dip at this value if an appropriate reference lifetime is used.

We start with the 156-member lifetime sets of Fig. 2 (uncorrected) and Fig. 5 (HF-corrected). Fig. 6 displays calculations of AAD for $S$ values $S = 10 - 200$ in these two cases. The solid curve and dashed curve are for the uncorrected and corrected lifetimes, respectively. As can be seen, both curves are essentially identical, and they each show an oscillatory pattern centered on the value AAD = 0.25, which is what we expect for a random distribution of $x_i$ values. Thus both the uncorrected and corrected $x_i$ distributions are dominated by the continuum of 120 short-lived (uncorrected) states in Figs. 2 and 5, which exhibit an aperiodic distribution of lifetime values.

Now we divide the RPP2004 lifetimes into the 120 short-lived excited states with lifetimes $\tau < 10^{-21}$ sec and the 36 long-lived threshold states with lifetimes $\tau > 10^{-21}$ sec, and we repeat the above $S = 10 - 200$ calculations of AAD($S$). The results of this periodicity analysis are displayed in Fig. 7. The top figure (7a) shows the AAD values for the 120 aperiodic states, which form an oscillating curve that is centered at AAD = 0.25 and extends from about AAD = 0.2 to AAD = 0.3. This represents a random distribution of elementary particle lifetime values. The bottom figure (7b) is for the 36 grouped thresh-old states of Figs. 2 and 5, and it give a quite different result. The solid curve in (7b) is for the uncorrected lifetimes of Fig. 2, and the dashed curve is for the HF-corrected lifetimes of Fig. 5. At small values of $S$, both curves fluctuate around the random value AAD = 0.25. But then they dip down at larger values of $S$, with broad minima around the value $S \cong 135$. As expected, the HF-corrected lifetimes show a deeper minimum, reflecting the greater compactness of the lifetime groups in Fig. 5 as compared to Fig. 2. The factor-of-two HF corrections have a small but significant effect on the precise location of the minimum in the AAD($S$) curve, as we now demonstrate.

Fig. 8 gives expanded representations of the minima in the uncorrected and corrected AAD($S$) lifetime curves of Fig. 7b. As can be seen from these expanded representations, a computer calculation with small $S$ increments does not yield an AAD($S$) curve that has smooth local derivatives. This is a consequence of the granularity of the lifetime data set. The uncorrected lifetimes in Fig. 8a give $S = 133.5$ at AAD $\cong 0.15$, and the HF-corrected lifetimes in Fig. 8b give $S = 136.1$ at AAD $\cong 0.12$. If we extend the asymptotes bordering the plateau minimum in Fig. 8b to more sharply define its location, we obtain an inferred minimum for $S$ at the value 136.3. Thus if the true theoretical value for the spacing $S$ of these lifetime groups is $S = \alpha^{-1} \cong 137.0$, as seems logical, then the factor-of-two HF corrections represent a significant improvement in the numerical analysis, as well as a clarification in the visual presentation of the lifetime groups. These results will be further illustrated in the next section, where we identify the Standard Model quark content of these lifetime clusters.

The goal we have been working toward in these AAD($S$) minimization studies is to establish the fact that the lifetimes of the long-lived threshold-region elementary particles are in fact functions of the fine structure constant $\alpha$. The best way to demonstrate the relevance of this AAD analysis is to replot the uncorrected lifetimes of Fig 2 and the factor-of-two corrected lifetimes of Fig. 5 on an $\alpha$-spaced lifetime grid, using the equation

$$\tau_i / \tau_{\pi^\pm} = \alpha^{x_i}, \tag{4}$$

with the $\pi^\pm$ meson as the reference lifetime. These results are displayed in Figs. 9a and 9b, respectively, where the accuracy of the $\alpha$-spacing for the $\tau > 10^{-21}$ sec threshold states is apparent. Two features of Fig. 9 should be noted with respect to the 36 threshold states:

*(1)* the wide range of lifetimes encompassed—ten powers of α; *(2)* the comprehensiveness of the results—all measured lifetimes in this range are included, and there are no "rogue" (randomly-oriented) lifetime values. These 36 lifetimes furnish direct experimental verification of the relevance of the fine structure constant α to elementary particle decays, irrespective of any theoretical considerations.

The α-quantization of elementary particle lifetimes is manifested in its most accurate and comprehensive manner in the low-mass spin 0 pesudoscalar mesons. Fig. 10 shows both the uncorrected (10a) and HF-corrected (10b) pseudoscalar meson lifetime plots on an α-spaced grid, using Eq. (4). The numerical values of the lifetime logarithms $x_i$ are shown below the data points. With the $\pi^\pm$ as the reference lifetime, the $\pi^o$ is displaced from it by almost precisely a factor of $\alpha^4$, and the η and η' are in turn displaced by additional factors of α. This a span of 12 orders of magnitude, and the α-scaling is accurately maintained for these unflavored mesons over the entire range. The strange $K_L^o$ and $K^\pm$ mesons are displaced from the $\pi^\pm$ by factors of two (see Fig. 3). The $K_S^o$, which shares a common ππ decay channel with the $K^\pm$, is displaced from the $K^\pm$ by a factor of 1/138.3, which is very close to the factor α = 1/137.0. The precision of this α-quantization is visually apparent, especially in Fig. 10b. The universality of the α-quantization is also apparent, with all members of the pseudoscalar octet falling into this pattern. These low-mass spin 0 meson states are logically the simplest meson excitations, which gives added significance to the α-dependence. If particle lifetimes and masses are related quantities, we might expect the masses of these particles to exhibit an equally-accurate α-dependence, and, as we will see in Sec. IV N, this is indeed the case.

### F.  The SM quark structure of the α-quantized lifetimes

One of the most interesting features of the α-quantized lifetime clusters in Fig. 9, and one of the most theoretically important, is the fact that they are sorted into dominant-quark decay groups. This is illustrated in Fig. 11, where the particles are identified in terms of their Standard Model quark flavors (*u,d*), *s*, *c*, *b*. As can be seen, the very-long-lived particles with lifetime values $x_i$ < 3 cluster in lifetime groups that have all members of the group containing the same type of dominant SM quark (*i.e.*, dominant from the standpoint of decay systematics). This grouping is straightforward in the cases where the particle contains a single *s* or *c* or *b* quark, multiple *s* quarks, or matching $c\bar{c}$ or $b\bar{b}$ quark-antiquark pairs. In the cases where a particle contains a mixed $c\bar{s}$ or $c\bar{b}$ quark-antiquark pair, the lifetime is dictated by the *c* quark, and in the case of a mixed $b\bar{s}$ quark-antiquark pair, the lifetime is dictated by the *b* quark, as we show in Sec. G. These homogeneous quark flavor groups serve to confirm the reality of these SM quarks, quite apart from their quark significance in uniquely leading to the correct isotopic spin rules.

Another feature which can be noted in Fig. 11 is that the quark groupings in the lifetime region $x_i \cong 0-2$ are repeated in the lifetime region $x_i \cong 4-6$. To bring out this feature more clearly, we select the 36 $\tau > 10^{-21}$ sec threshold particles, use the uncorrected lifetimes of Fig. 9a, and separate the particles along the *y*-axis into their various family types. This plot is given in Fig. 12. In each quark family there is a group of long-lived particles that have single-quark decays, and another group of corresponding short-lived particles that have matching quark-antiquark or radiative decays. The lifetime separation between the long-lived and the short-lived group in each family is approximately a factor of $\alpha^4$, which is more than eight orders of magnitude. Furthermore, as the overall mass values of these quark families increase, their lifetimes are successively shifted from family to family towards shorter lifetimes in quantized steps of $\alpha$. Thus the fine structure constant $\alpha \cong 1/137$ dominates the lifetime behavior of these quark families, and hence the stability of their quark substructures.

The family relationship between unpaired-quark decays and paired-quark decays is delineated in detail in Fig. 13, which uses selected lifetimes from Figs. 9 and 12 and identifies the individual particles. It features sets of related particles that have characteristic lifetime separation intervals of approximately $\alpha^4$. The $\pi^o/\pi^\pm$ lifetime ratio is accurately given as $\alpha^4$. The $\Sigma^o$ and ($\Sigma^+,\Sigma^-$) lifetime separation interval is slightly longer than $\alpha^4$. The B and D unpaired *b* and *c* quark states are separated from their paired $\Upsilon$ and J/$\psi$ counterparts by slightly less than $\alpha^4$ in lifetimes. And the excited states $\Sigma_c$ and $\Lambda^*_c$ at the top right (which were not included in Fig. 12) are factors of $\alpha^4$ shorter in lifetimes than the corresponding ground states. Somewhat surprisingly, the neutron (a weak decay) and the muon (a lepton) also fit on this α-spaced lifetime grid, and they have an accurate lifetime separation of $\alpha^4$. They also have a very accurate mass ratio of nine to one. The universality of these $\alpha^4$ lifetime intervals between related sets of particles means that they must play an essential role in any comprehensive theory of long-lifetime ($\tau > 10^{-21}$ sec) hadronic decays.

We now move on to more detailed lifetime features of these threshold elementary particle decays.

### G. Quark dominance and the factor-of-three *b*-*c* quark lifetime ratio

Figures 11 - 13 show two closely-spaced lifetime groups located at and just above $x_i = 2$ on the α-spaced grid of Eq. (4). These correspond to *b*-dominant and *c*-dominant decays, respectively. The *b*-dominant group at the left contains five meson and baryon excitations that have virtually the same lifetimes, as shown in Figs. 12 and 13. The *c*-dominant group at the right contains four meson and baryon excitations that have essentially the same lifetime, plus four additional excitations that exhibit factor-of-two hyperfine deviations (see Fig. 3). For the purposes of the present discussion, we select all five of the *b*-quark excitations but just the four *c*-quark excitations that do not require HF corrections. These nine long-lived elementary particles are plotted together on an expanded $x_i$ scale in Fig. 14.

One interesting feature of Fig. 14 is that it clearly illustrates the dominance order of the flavored *s*, *c*, and *b* quarks [8] in triggering hadronic decays. If the hadron contains just one flavored *s* or *c* or *b* quark in addition to unflavored *u* and *d* quarks, then that flavored quark dictates the lifetime of the particle. As Fig. 11 shows, the *s*-triggered decays yield a lifetime group that, after factor-of-two HF corrections have been applied, all live about one power of $\alpha^{-1}$ longer that the *b*-triggered decays. The *b*-triggered decays in turn live (as we will show) a factor of three longer than the *c*-triggered decays. But what happens if a particle contains a two different flavored quarks, such as $s\bar{c}$, $s\bar{b}$, or $c\bar{b}$ ? Does one flavored quark still dominate, or does the lifetime fall somewhere in between the two single-flavor groups? Fig. 14 provides the answer. In the *b*-quark lifetime group at the left, the $B_s^o = b\bar{s}$ meson has almost the same lifetime as the $B^o = b\bar{d}$ and $B^\pm = b\bar{u}$ mesons, and the $\Xi_b^o = bsu$ and $\Lambda_b^o = bdu$ hyperons fall in this same *b*-quark lifetime grouping. Since the *b*-quark lifetime group is removed from the *s*-quark lifetime group by a factor of 137 in lifetime ratios, it is clear that the *b* quark, and not the *s* quark, is the trigger for these decays. When we move to the *c*-quark lifetime group at the right in Fig. 14, we find the $D^o = c\bar{u}$, $D_s^\pm = c\bar{s}$ and $B_c^\pm = c\bar{b}$ mesons with very similar lifetimes. Thus whenever a *c* quark is present, it serves as the decay trigger, and dominates over the *s* and *b* quarks. Hence we have a *flavored quark decay hierarchy*: *c* > *b* > *s*.

Another interesting feature of Fig. 14 is the lifetime ratio that is exhibited between the *c*-quark and *b*-quark decays. The dotted lines in Fig. 14 represent the averages of the $x_i$ exponents in the two groups. They correspond to an average lifetime ratio of 3.23. The mass ratio of the $\Upsilon(1S) \equiv b\bar{b}$ and $J/\Psi(1S) \equiv c\bar{c}$ resonances is 3.05, which serves to characterize the mass ratio of the *b* and *c* quarks. Thus the factor-of-three flavor spacing in the lifetime ratios of the *b*-quark and *c*-quark groups logically relates to the mass ratio of these quarks. We could envision that the *b* quark is formed (except for its charge) as a combination of three *c* quarks, and that the decay of the *b* quark, if mediated by a decay of any one of its three *c*-quark constituents, would therefore be three times as fast as the decay of an individual *c*-quark. But this would mean that the *b*-quark lifetimes would be a factor of three shorter than the *c*-quark lifetimes, whereas it is actually the other way around—the *b* quarks have the longer lifetimes. As a general rule in the hadron excitations, higher-mass states are more unstable and decay faster than lower-mass states. Thus the longer lifetimes for these threshold *b*-states as compared to the shorter lifetimes of the threshold *c*-states represents an anomaly, which should provide us with some information about the stability of the *b*-quark structure as compared to the *c*-quark structure.

The final feature of interest in Fig. 14 is the manner in which the *b*-quark and *c*-quark lifetime groups fit into the overall α-quantization grid displayed in Figs. 9 and 11. The factor-of-three flavor difference in these lifetime groups is a fine-structure effect that resembles the factor-of-two HF structure shown in Fig. 3. Thus it is an effect that is independent of the overall α-quantization. Should the *b*-quark group accurately match the α-spaced grid of Fig. 9 and the *c*-quark group be displaced by a factor of three, or should it be the other way around, or should it be halfway between? As Fig. 9 shows, the *b*-quark group is right on the α-grid and the *c*-quark group is displaced. This is not what we might logically expect. The non-flavored quarks start at $x_i = 0$ in Fig. 9, and the *s*-flavored baryons appear right at $x_i = 1$. So the *c*-flavored quarks ought to appear right at $x_i = 2$. But it is the *b*-flavored quarks which instead show up there. The *s* and *b* quarks both carry the same −1/3 *e* fractional charge, whereas the c-quark has a charge of +2/3 *e*. Thus the 1/3 *e* charge state may in some sense be more "fundamental" or stable than the 2/3 *e* charge state, at least as applied to hadronic decays. A lot of interesting "quark" physics is embodied in the particle states shown in Fig. 14.

In connection with the question of quark dominance in hadronic decays, it is of interest to examine the effect on particle lifetimes of replacing one flavored quark by another. Fig. 15 shows a set of baryon octet particles at the left, and a corresponding set of charmed baryons at the right which have an *s* quark replaced by a *c* quark. This quark substitution increases the mass of the particle, and also increases its charge by one unit. As can be seen, the *s* to *c* quark replacement characteristically decreases the lifetime of the particle state by a factor of α/9, with an additional fine structure factor-of-two correction being required in some cases. Fig. 16 shows two sets of baryons that illustrate the effect of replacing an *s* quark with a *b* quark. This quark substitution does not change the charge of the state, and it leads to a characteristic decrease in lifetime by a factor of 2α/3. Since the *b* and *c* particles have an intrinsic factor of three flavor difference in lifetimes, as shown in Fig. 14, the intrinsic scaling factor for the *s* to *c* quark replacements shown in Fig. 15 should probably be 2α/9. These individual quark replacements collectively add up to form the overall quark groupings shown in Figs. 9 and 11.

### H. Chi-squared determinations of the lifetime α-quantization

In Sec. E we used average absolute deviation (AAD) minimization of the exponents $x_i$ in Eq. (2) to determine the scaling factor *S* that best applies to the lifetime groups of the long-lived $\tau > 10^{-21}$ sec particle states shown in Fig. 2. After the application of factor-of-two HF corrections (Fig. 2) to these data, the AAD(*S*) curve exhibited a minimum at the value *S* = 136.1, as shown in Fig. 8b. This AAD minimum is close to the QED scal-ing value $\alpha^{-1} \cong 137.0$, and it indicates that the coupling constant $\alpha$ is pertinent to these hadronic decays. The AAD minimization is based on the measured lifetime values themselves, and not on the experimental uncertainties in the values.

An alternative way to determine $S$ is to write down mathematical expressions for the lifetimes $\tau_i(S)$, and then find the value of $S$ that minimizes the chi-square sum [15]

$$\chi^2(S) = \sum_i \left( \frac{\tau_i(S) - \tau_i(\exp)}{\Delta \tau_i(\exp)} \right)^2 . \tag{5}$$

This method assigns the most importance to the most accurately measured data points. It only works well in a statistical sense if all of the data have roughly comparable error limits. Otherwise, the $\chi^2$ sum is dominated by the fits to the data that have very small error bars, which may not be intrinsically of more importance than the other data points. The neutron and muon have very accurately determined lifetimes, and do not fit in naturally with the other threshold particles in this type of analysis. But the chi-squared analysis is of interest here as a comparison to the AAD analysis in trying to determine if the lifetime scaling factor $S$ is in fact $\alpha^{-1}$ for these hadronic particles.

The $\chi^2$ analysis was carried out in the following manner. The pseudoscalar $\pi^\pm$ meson was selected as the reference lifetime, as in the AAD analysis. Three lifetime data sets were used, consisting of 3, 10, and 23 particles, respectively, where the 3-particle set contains only non-strange pseudoscalar mesons, and the 10-particle and 23-particle sets also contain mesons and baryons that have lifetime logarithms $x_i \cong 1$ and $x_i \cong 2$ in Fig. 9. These three sets are defined as follows:

*Set A*: $\pi^o$, $\eta$, $\eta'$ (no HF corrections required) (3 particles);

*Set B*: $\pi^o$, $\eta$, $\eta'$, $\Sigma^-$, $\Xi^-$, $B^\pm$, $B^o$, $B_s^o$, $\Xi_b$, $\Lambda_b$ (no HF corrections required) (10 particles);

*Set C*: Set B plus $K_S^o$, $\Lambda^o$, $\Xi^o$, $\Sigma^+$, $\Omega^-$ (HF corrections required) plus eight charmed
    mesons and baryons near $x_i \cong 2$ (HF and flavor corrections required) (23 particles).

The set A and set B lifetimes have no HF corrections, and thus involve only the *macroscopic* scaling factor $S$. Some of the set C lifetimes require the *microscopic* HF and flavor corrections displayed in Figs. 3 and 14, respectively. The three spin 0 pseudoscalar mesons in set A exhibit the most accurate $\alpha$-quantization of any of the hadron families. The twenty three mesons and baryons in set C include all of the hadrons that have lifetime exponents $x_i \cong 1$ and 2 in Fig. 9. The theoretical values for $\tau_i(S)$ in Eq. (5) are calculated as follows. The reference lifetime $\tau_{\pi^\pm}$ divided by $S$ macroscopically represents the $x_i \cong 1$ lifetimes, and $\tau_{\pi^\pm}$ divided by $S^2$ represents the $x_i \cong 2$ lifetimes. Then microscopic factor of 2 HF corrections are applied to some calculated $s$ and $c$ particle lifetimes to compensate for the experimental HF deviations shown in Fig. 3. Finally, all of the calculated $c$ particle lifetimes are shifted to longer values by factors of 3 in order to account for the systematic factor of 3 $b$-quark to $c$-quark flavor displacements shown in Fig. 14. These calculated values for $\tau_i(S)$ are inserted into Eq. (5), together with the experimental cross sections $\tau_i(exp)$ and errors $\Delta\tau_i(exp)$ from RPP2004 [8], and the $\chi^2(S)$ sum is minimized by varying the macroscopic scaling factor $S$.

Fig. 17 shows two calculations of $S$ for the three-particle ($\pi^o$, $\eta$, $\eta'$) Set A data. An AAD($S$) Set A calculation is shown in Fig. 17a, and it has a minimum at $S = 136.1$. (This minimum is at the same value as the one shown in Fig. 8b for the 36-particle HF-corrected data set.) A matching $\chi^2(S)$ Set A calculation is shown in Fig. 17b, and it has a $\chi^2$ minimum at $S = 138.8$. Since no microscopic corrections are required for these data, the calculations reflect just the macroscopic scaling factor $S$. These AAD and $\chi^2$ minima at the values $S = 136.1$ and $S = 138.8$ neatly bracket the QED scaling factor $\alpha^{-1} \cong 137.0$, and they indicate very strongly that the numerical coincidence between $S_{min}$ and $\alpha^{-1}$ is not accidental. Furthermore, it should be noted that the scaling of these three pseudoscalar meson lifetimes, together with that of the $\pi^\pm$ reference lifetime, accurately extends over a lifetime range of six powers of $\alpha$ (Fig. 10).

Fig. 18 shows the $\chi^2(S)$ calculations for the 10-particle Set B and 23-particle Set C data. Set B has the $\chi^2$ minimum at $S = 142.9$, which shows that corrections to the scaling factor $S$ are small (much smaller than a factor of two) for all of these ten particles. Set C, which includes empirical factor-of-2 and factor-of-3 corrections, has the $\chi^2$ minimum at $S = 145.9$, which is in reasonable agreement with the analyses based solely on uncorrected lifetimes. This suggests that the microscopic fine-structure factors of 2 and 3 in the particle lifetimes actually occur roughly as presented here, and are essentially independent of the overall macroscopic $\alpha$-quantization of these quark threshold particle states.

# I. Mass values of the α-quantized quark lifetime groups

We have been studying the overall elementary particle lifetime systematics, and in particular the α-quantized quark threshold region. As Fig. 11 shows, the particle lifetimes in this region occur in discrete α-spaced groups which are dominated by their *s*, *c*, *b* quark flavors. Although Fig. 11 clearly reveals the quark flavor structure in these lifetimes, it gives no information about the quark masses. The ordinate in this figure is used to conveniently spread out the lifetimes so that the data points within a group are not overlaid. We can illustrate the correlation between particle lifetimes and particle masses by keeping the same lifetime abscissa as in Fig. 11, but now using the ordinate to display the corresponding masses of these particles. This procedure gives the data plot shown in Fig. 19.

As can be seen in Fig. 19, there is a correlation between particle masses and quark flavors. The unflavored hadrons start at 135 MeV, the *s* hadrons start at 494 MeV, and both of them extend up to roughly 2.5 GeV. The *c* hadrons start at 1865 MeV and extend up to about 4.5 GeV, except for the B$_c$ meson, which has a mass of 6.4 GeV. And the *b* hadrons start at 5280 MeV and extend up to about 11 GeV. The particle masses in the quark threshold region $x_i \leq 6$ occur in groups, but these mass groups do not exhibit the sharp α–quantizations that are evident in the lifetime groups in this region. In the $x_i \geq 6$ region, both the masses and lifetimes have continuum-like distributions that reflect the quite different nature of the short-lived excitation processes for these states.

We can extract a little more information from these mass and lifetime correlations by selecting just the meson excitations that involve *c* and *b* quarks. These are shown in Fig. 20. The unpaired *c*-quark states start at 1865 MeV, and the paired $c\bar{c}$ states appear a factor of two higher. Similarly, the unpaired b-quark states start at 5280 MeV, and the paired $b\bar{b}$ states also appear a factor of two higher. This is the behavior we expect for constituent quarks, where the mass of the composite meson is essentially the sum of the masses of its quarks. We can also see that the masses of the *b*-quark states are about a factor of three larger than the masses of the *c*-quark states, which echoes the factor of three difference in their lifetimes shown in Figs. 11 and 14.

## J. The significance of α-quantized lifetimes

Experimentally, elementary particle lifetimes are straightforward quantities to measure. Their inherent α-dependence is also straightforward to ascertain: if these lifetimes, or a selected subset of lifetimes, scale in powers of 1/137, then they are in some manner related to α. As we have demonstrated here, this is the case for the long-lived threshold-state particles with lifetimes $\tau > 10^{-21}$ sec. There are two significant aspects of this result: *(1)* if particle lifetimes depend on α, then the associated particle masses may also depend on α; *(2)* the threshold-state particles considered here are mostly *hadrons*, which are widely believed to be independent of α. Thus it is important that we now examine the masses of these α-dependent threshold-state hadrons to see if they also display an α-dependence. If they do not, then the lifetime results may be misleading. But if they do, then the expected reciprocal relationship between these variables is fulfilled, and the evidence that α applies to the hadronic domain would seem to be compelling. As we will see in Sec. IV, the lifetime α-systematics are in fact supported by the corresponding mass systematics. The lifetime α-dependence is in agreement with the main QCD features of the spin 1/2 Standard Model quarks, as we can see for example in Figs. 11 - 13. But it signals some further changes in the still-evolving history of quark masses. Where the admittance of α into the hadronic domain has a more decisive effect is in the development of grand unified theories, which would have to be modified. The lifetime studies of the present section represent the starting point for this sequence of events, which is the reason we have gone into so much detail in Part II to examine their α-dependence from as many viewpoints as possible.

The lifetime studies of Part II have consisted of displays of the experimental data, together with various analyses that bring out the α-quantization of the quark threshold states, both qualitatively and quantitatively. In Part III we briefly discuss how the concept of linked lifetime and mass variables is manifested in the α-quantization process. Then in Part IV we present evidence of a mass α-quantization, using the same general approach as in the lifetime studies—letting the experimental data tell their own story, and giving both general qualitative results and precise quantitative values.

# III. THE RECIPROCAL α-QUANTIZATION OF LIFETIMES AND MASSES

## A. Conjugate variables in atomic physics and particle physics

One of the major challenges in physics during the *first* half of the twentieth century was to understand the structure of the atom. A key to this understanding was the discovery of the conjugate relationship between *coordinates* and *momenta*, a discovery which has become a cornerstone of quantum mechanics, and is embodied in Heisenberg's equations. A major challenge during the *last* half of the twentieth century was to understand the structure of the elementary particle. A key to this understanding, which has yet to be fully implemented, is the conjugate relationship between *mass widths* and *lifetimes*, in the sense that it may lead to a better understanding of the particle *masses* involved.

Quantum mechanics is central to both of these endeavors, and its use of conjugate coordinates is important in each. Historically, the effort to understand atomic structure was a crucial factor in the development of quantum mechanics, and it led to the discovery of the role played by non-commuting observables. The first decisive step in unraveling atomic structure was the demonstration by Rutherford that the positive charge of the atom is concentrated at the center. The next step was the development of the Bohr orbitals, with electrons placed in planetary orbits around the positively-charged nucleus. These orbits were quantized in units of angular momentum, so that velocities combined with spatial positions (angular momenta) became the variables of interest. The subsequent discovery of the de Broglie wavelength of the electron explained the orbital angular momentum quantization in terms of the coordinate wavelength quantization, which led directly to the creation of wave mechanics. Heisenberg clarified the non-commuting nature of conjugate coordinate ($x$) and momentum ($p$) variables (operators) in studies that led to the formulation of the Uncertainty Principle, $\Delta x \cdot \Delta p \geq \hbar$. It is interesting to note that Planck's constant $h$, which was initially introduced in the quantization of black body radiation, subsequently appeared in the quantization of the atomic orbitals, and also in the quantification of the uncertainty that arises from the use of conjugate variables. The knowledge obtained from the *angular momentum* spectra in the atom gave us important information about the conjugate *position coordinates* that describe the geometry of the atom.

In atomic physics the conjugate nature of *coordinate* and *momentum* variables is exploited. Relevant conjugate quantities in elementary particle physics are *lifetimes* and *mass widths*, which are the uncertainties in the non-commuting variables $t$ and $m$. Specifically, the uncertainty in the lifetime of a particle ($\Delta t$), which is expressed in the form of its *mean life* $\tau \equiv \Delta t$ [8], is related to the uncertainly in the mass of the particle ($\Delta m$), which is expressed as the width of the measured resonant state, and these are quantitatively tied together by the Heisenberg uncertainty principle

$$\Delta t \cdot \Delta m \geq \hbar \quad (c = 1), \tag{6}$$

where the numerical factor in front of $\hbar$ depends on the choice of variables (*e.g.*, full width [8] or half width at half maximum for Gaussian distributions), and where the equality that represents the lower limit of the uncertainty in Eq. (6) is used to define the transformation of variables. Experimentally, the lifetimes $\tau \equiv \Delta t$ of the long-lived particles are measured directly, but the lifetimes of the short-lived particles, which do not have measurable flight paths, have to be deduced from the resonance (full) widths $\Delta m$, using the equality in Eq. (6). Hence Eq. (6) is used to transform from $\Delta m$ coordinates (mass widths) to $\Delta t$ coordinates (mean lives). It seems plausible that we can invert this conjugate $\Delta m$-$\Delta t$ relationship and use the widths $\Delta m$ to infer properties of particle masses $m$. Specifically, if there is a discernable periodicity $f(\tau)$ in the particle $\Delta t$ lifetime structure, with a (reciprocal) periodicity $g(m)$ in the corresponding particle $\Delta m$ width structure, then the periodicity $g(m)$ may carry over and also apply to the particle mass $m$.

Lifetime structures are in some respects easier to measure than mass structures, since elementary particles appear to have a (quark) mass substructure that is not amenable to direct observation (single quarks are not observed in collision or decay events). Thus to investigate the "periodicity" $g$ in the quark mass structure, we must determine the observable periodicity $f$ in the global lifetime structure, and then use $f(\Delta t) \cdot g(\Delta m) = \hbar$ to infer a mass quantization $g(m)$. In particular, if $f = f(\alpha)$, then we look for $g = g(\alpha^{-1})$.

When we investigated the global elementary particle lifetime spectrum in Sec. II, we discovered a flavored-quark structure (Fig. 11) that emerges very clearly. This flavored-quark lifetime structure may correspond to a flavored-quark mass substructure.

## B. The "α-leap" α-quantization of elementary particle lifetimes

*A priori*, we have no obvious reason to expect elementary particle hadron lifetimes and masses to be functions of the fine structure constant $\alpha \cong 1/137$. And we have no direct link between *hadron* lifetimes and masses and *lepton* lifetimes and masses. However, as we demonstrated in Part II, the long-lived elementary particles, with lifetimes $\tau > 10^{-21}$ sec, occur in well-separated lifetime groups which have lifetime ratios that are in powers of α. These long-lived particles represent the threshold states where the various types of quarks first appear. As the mass values within a particle "family" increase, we move from one lifetime group to another by taking single or multiple lifetime "α-leaps" that are factors of 1/137. Several of these lifetimes groups exhibit an additional factor-of-two HF hyperfine structure (Fig. 3), and the *b*-dominated and *c*-dominated unpaired-quark groups exhibit a factor-of-three flavor-structure (Fig. 14). These small lifetime corrections are superimposed on the overall α-spaced grid structure (Fig. 9), which includes both hadrons and leptons.

The search for an α-quantization in these lifetimes is straightforward. We calculate lifetime *ratios* with respect to a standard reference lifetime, which here is the $\pi^\pm$ lifetime, and we see if they are multiples of 1/137. This is a global mapping, and the lifetime intervals as powers of α are not exact, since various small final-state effects can enter in. Also, factor-of-two HF perturbation corrections have to be handled empirically (Fig. 4), since we have no comprehensive theory for them as yet. And the experimental lifetimes are not always precisely measured. Nevertheless, the linear AAD studies (Figs. 6-8) and quadratic $\chi^2$ studies (Figs. 17 and 18) displayed in Part II indicate statistically that α is the underlying factor which characterizes the scaling of these lifetimes, and it extends over a broad range of values. This is an empirical conclusion that is based directly on the experimental lifetime data, and it applies independently of any theoretical considerations.

**C. The "α-leap" and "α-mass" α-quantizations of elementary particle masses**

Elementary particle lifetimes and mass widths are conjugate variables, so the observed lifetime α-dependence must necessarily carry over to the widths. Hence the theoretical explanation for this α-spaced lifetime structure may be related to the mass structure. Thus we should examine elementary particle masses in the same manner as we did with lifetimes in order to ascertain their dependence on α. Since lifetimes and mass widths, as defined for example by the Particle Data Group [8], are tied together by the lower-limit equality in Eq. (6), their α-dependences are reciprocal in nature, and this may extend to included the masses themselves. Thus, since we observed $\alpha^{-1} \cong 1/137$ lifetime "α-leaps" with increasing mass, we should look for a mass "α-leap" of 137 in the ratio of two related masses. However, there is an obvious difference in these two cases. Lifetimes are spread out over a wide range of values, so they can exhibit a scaling over many powers of α—a scaling that contains many α-leaps. But the masses of these same particles, which range from 105 MeV for the muon to 11 GeV for the Upsilon (excluding the W and Z), encompass less than one factor of 137 in total mass ratios. In order to find a mass α-leap as large as 137, we have to introduce the electron, the other massive lepton, and its antiparticle, the positron. (We will frequently denote a symmetric particle-antiparticle "electron-positron pair" as simply an "electron pair".) The electron did not appear in the lifetime systematics because it is stable and has an infinite lifetime. When we do include the electron, we obtain two minimum-mass α-leaps, as we describe in Sec. IV C: *(1)* from a spin 0 electron-positron pair to a spin 0 pion, with a mass increase of 137; *(2)* from a spin 0 or 1 electron-positron pair to a spin 0 or spin 1 muon pair, with a mass increase that is a factor of 3/2 larger. (We will often describe the *muon* α-leap as being in the electron-pair "spin 1" channel, to distinguish it from the electron-pair "spin 0" *pion* channel.) Since the lifetimes exhibit an α-dependence for all types of particles, and not just for pions and muons, there must be an α-dependence which pervades all of the threshold particles that have α-spaced lifetimes. Empirically, the way this occurs is that the "spin 0" and "spin 1" mass "α-leaps" define the spin 0 and spin 1 "α-masses" $m_\pi$ and $m_{\mu\mu}$. The α-dependent mass quanta $m_\pi$ and $m_{\mu\mu}$ are composite excitations that have sym-metric particle and antiparticle components. These components can be later separated into asymmetric particle and antiparticle "α-mass" subunits, which occur in *boson* and *fermion* forms as the "α-masses" $m_b$ (spin 0) and $m_f$ (spin 1/2), with mass values of 70 MeV and 105 MeV, respectively. These $m_b$ and $m_f$ "α-masses" serve as the building blocks for quarks. Since the threshold particles are formed as combinations of quarks, the α-dependence of the masses is spread through all of them, thus satisfying the quantum mechanical lifetime and mass reciprocity requirement. This does not lead to a transparent relationship between the mass α-structure and the lifetime α-structure, but it indicates the direction in which an explanation of the α-dependent lifetime behavior must be sought.

The interest in particle *lifetimes* as a probe for studying particle *masses* comes from the peculiar nature of the elementary particle masses. The two hundred odd elementary particle states that have been identified to date appear to have a clear quark substructure, so that the number of constituent quarks is much smaller than the number of observed particles. And yet it has proved to be impossible to separate the particles into their individual quarks, so that we could measure the individual quarks masses. This result was completely unanticipated theoretically, and it seems to be unique to this branch of physics. Also, measurements of the sizes of quarks, and of leptons, show only a point-like spatial extent, whereas the composite particles have the expected Compton-like sizes we are familiar with. Thus the fact that elementary particles exhibit a well-defined quark lifetime structure (Fig. 11) assumes special significance. It raises the question as to what this lifetime structure can tell us about the quark masses. Also, this lifetime systematics serves as an important test for any comprehensive theory of elementary particle structure. These lifetimes are experimental regularities, which exist independently of any particular elementary particle theory. Until a particle theory can account for these first-order lifetime relationships, it cannot be regarded as complete.

We now move on to a discussion of the elementary particle mass spectrum.

## IV. EXPERIMENTAL EVIDENCE FOR α-QUANTIZED PARTICLE MASSES

### A. The elementary particle mass problem

Before the age of high-energy accelerators, only a handful of elementary particles were known to exist. However, we now have about 200 measured particle states, which suggests that some of these must be more "elementary" than others. A significant step towards simplifying this plethora of particles was the recognition that they can be constructed from a much smaller set of "quark" substates. The Standard Model [8] currently features six spin 1/2 quarks (*d, u, s, c, b, t*), together with their corresponding antiquarks. The fractional charge states for these quarks are (−1/3, +2/3, −1/3, +2/3, −1/3, +2/3), respectively, and they accurately reproduce the measured isotopic spins of the various particles. This has been one of the major successes of the Standard Model. As shown in the lifetime studies of Part II, the long-lived elementary particle lifetimes clearly reflect this SM quark structure. In particular, Fig. 11 demonstrates the manner in which unpaired *s*, *c*, and *b* quarks dominate the lifetimes of the particles that contain them. The mass of the *t* quark is too large for *t* quarks to play a role in this lifetime systematics, and the low-mass *u* and *d* quarks are spread throughout all of the particle groups, so their effects are not immediately apparent. Elementary particle lifetimes *per se* furnish an independent confirmation of the reality and importance of the *s*, *c*, and *b* quarks.

Measurements of both the lifetimes and the overall masses of the elementary particles are straightforward. Since the particle lifetimes show evidence of a quark substructure, and since lifetimes and masses are related quantities, we logically expect the particle masses to show experimental evidence of this same quark substructure. That is, we expect a particle in a high energy collision to break apart into separated *u, d, s, c, b* quarks or antiquarks, so that we can measure the masses of individual quarks. But experimentally this does not occur. Specifically, it has been established that particle break-ups or decays into *fractionally-charged* particles do not occur. The *quark charges*, which seem to exist as 1/3 and 2/3 entities on the quarks inside of a particle, cannot be extracted as fractional charges outside of the particle: only integrally-charged particles are found in the decay products. Thus if the electric charges do actually separate into 1/3 and 2/3 frac-tions on the quark states, as is indicated by the isotopic spin rules, they are evidently still tied together by unbreakable confining forces that seem to increase with separation distance. This type of binding energy is denoted as "asymptotic freedom". The question then remains as to whether the *quark masses*, apart from the charges, can be separated and dislodged in collision or decay processes. Empirically, the masses of the final-state particles do not seem to be those of the initial-state *u, d, s, c, b* quarks. Thus Standard Model quarks are not observable as isolated entities, so that it is difficult to ascertain their intrinsic masses experimentally.

There are two different reasons we can advance to explain why single-quark masses are not observed in particle annihilations. Either *(1)* the quark *masses* are bound so strongly that quarks cannot be pulled apart, or *(2)* the quarks have lightly-bound mass *substructures*, so that when a particle is shattered or decays, the quarks also shatter into their mass substates (but not with the fractional quark charges). The first reason (which applies with respect to the quark *charges*) requires a rubber-band type of mass binding energy that increases with separation distance. However, the second reason is the one that seems to emerge from the present studies, which show the SM quarks as having weakly-bound (a few percent) $\alpha$-generated mass substructures. The $\alpha$-dependent lifetime structures were described in detail in Part. II. We now proceed in Part. IV with a study of the reciprocal $\alpha$-dependent mass structures of these same particles.

## B. The "$\alpha$-leap" $(m_b, \bar{m}_b)$ and $(m_f, \bar{m}_f)$ boson and fermion excitation quanta

The observed $\alpha$-quantization of elementary particle *lifetimes* is logically accounted-for by a corresponding $\alpha$-quantization (or, more accurately, $\alpha^{-1}$ quantization) of the related particle *masses*. In the case of the lifetimes, this $\alpha$-dependence is only observed in the 36 long-lived threshold particle states with lifetimes $\tau > 10^{-21}$, and it takes the form of *lifetime decreases* by factors of 1/137 as the masses of the particles increase. Since conjugate lifetimes and mass widths are tied together by Planck's constant $\hbar$ (Eq. 6), their reciprocal $\alpha$-dependence suggests mass $\alpha$-quantizations that involve *mass increases* by factors of 137, and they should occur in the same 36 threshold particles. If this is actually the case, it makes this mass α-dependence easy to identify, because there are very few massive elementary particle states whose mass ratios are 137 or larger. In order to search for an α-quantization in particle masses, we plot the mass values of the 36 particles that have α-quantized lifetimes, and we add in the stable electron and proton. This plot is displayed in Fig. 21.

Fig. 21 shows the long-lived massive elementary particles arranged in *excitation towers* according to spin states. The lowest-mass ground states are spin 0 and spin 1 electron-positron pairs for integer-spin bosons, and spin 1/2 electrons for half-integer-spin fermions. The excited states are the 36 long-lived particles of Fig. 2, together with the stable proton. The $J = 0$ and 1 boson towers represent production channels for the direct transformation of energy into mass. The $J = 1/2$ fermion tower contains unpaired fermion *particle* states that are not produced singly, so it by itself is not a production channel: it is the particle half of matching particle-antiparticle towers. However it is useful to treat the $J = 1/2$ towers as separate entities in studying their excitation systematics. An important feature of these spin 1/2 towers is that binding energies are not involved in the separate channels, so their fermion excitation masses can be more accurately evaluated than those of the hadronically bound bosons in the spin 0 and spin 1 towers.

The mass values displayed in Fig. 21 demonstrate that the only mass ratios among these particles which are 137 or larger are those involving electrons. The first particle state above the 1 MeV $e^-e^+$ level is the spin 0 pion level, which is occupied by the $\pi^o$ and $\pi^\pm$ mesons, and which has an average pion mass of 137 MeV (see Sec. J). This 137 MeV pion level represents an "α-leap" of one power of $\alpha^{-1}$ above the 1.02 MeV $e^-e^+$ level. In detail there is an "α-leap" of $m_e/\alpha = 70$ MeV, for each of the $e^+$ and $e^-$ electrons, where $m_e$ is the electron mass, together with small hadronic binding energy (~ 3%) and charge-splitting effects (Sec. J). Since the pion is a spin 0 particle, the electron-positron pair that generates it must be in a $J = 0$ spin state. Also, we can deduce from the spin 0 kaon (which contains an odd number of 70 MeV quanta) that the substates of the pion are also spin 0 states. Hence this 70 MeV α-leap represents a *boson excitation quantum* $m_b$.

The second particle level above the 1 MeV electron-positron pair is the leptonic muon pair at 211 MeV, which is a factor of 3/2 higher in energy, and which is produced with no binding energy. The α-leap here is $(3/2)(m_e/\alpha) = 105$ MeV for each electron as it generates a spin 1/2 muon. Thus this 105 MeV α-leap represents *a fermion excitation quantum $m_f$*. The muon pair in the 211 MeV level is produced with zero binding energy, and it can be in either a J = 0 or J = 1 spin configuration. In Fig. 21 it is displayed as a member of the J = 1 excitation tower. Extending these results, we can conceptually separate the *muon-pair* α-leap excitations into two paired J = 1/2 *single-muon* α-leap excitation towers, particle and antiparticle, and then display the spin 1/2 fermions in an individual tower, as shown in Fig. 21.

As Fig. 21 indicates, there are two dominant α-quantized mass intervals in the mass spectrum of the 36 threshold-state particles—a 70 MeV boson α-leap excitation quantum $m_b$ and a 105 MeV fermion α-leap excitation quantum $m_f$. We will demonstrate that these two α-leap mass intervals define the basis states that accurately reproduce the threshold elementary particles. In particle and antiparticle forms, they represent the particle *mass* α-quantization that is reciprocal to the *lifetime* α-quantization. As building blocks for quarks they distribute this mass α-quantization through all of the quark states, just as the reciprocal lifetime α-quantization is distributed through all of the quark threshold-particle states. Furthermore, we will demonstrate in Part V that $m_b$ and $m_f$ can be mathematically related to one another as spinless (spin 0) and spinning (spin 1/2) forms of the same basic mass quantum.

We now formally define the boson and fermion "α-leap mass intervals" that are displayed in Fig. 21. Guided by the mass values of Fig. 21, we define two basic elementary particle α-mass excitation quanta:

$$(m_b, \bar{m}_b) \equiv (m_e, \bar{m}_e)/\alpha \quad (70.025 \text{ MeV "boson α-mass", spin 0 channel}), \quad (7)$$

$$(m_f, \bar{m}_f) \equiv (3/2)(m_b, \bar{m}_b) \quad (105.038 \text{ MeV "fermion α-mass", spin 1/2 channel}), \quad (8)$$

where $m_e$ is the electron mass. These are mass excitations produced by the action of the operator α (or $α^{-1}$) on the electron. This raises the question as to whether these $m_b$ and $m_f$ excitations correspond to absolute energy *levels* or to energy *intervals* above the electron mass. Are the pion and muon masses proportional to $m_b$ and $m_f$, respectively, or are they proportional to $(m_e+m_b)$ and $(m_e+m_f)$? Since the electron mass $m_e$ is much smaller than $m_b$, its effect in the pion excitation channel is obscured by much larger binding-energy and charge-splitting effects (Sec. J). However, in the muon excitation channel neither of these effects occurs, and we can use the observed mass of the muon to determine the answer to this question. Assuming that the fermion α-mass $m_f$ is an excitation interval energy which adds to the $m_e$ electron energy of 0.511 MeV, we obtain the following calculated and experimental muon masses

$$m_\mu(\text{calc}) = m_e + m_f = m_e(1 + 3/2\alpha) = 105.549 \text{ MeV}, \quad m_\mu(\text{exp}) = 105.658 \text{ MeV}. \quad (9)$$

The high accuracy (0.1%) of the agreement between the calculated and experimental values demonstrates that $m_f$ is in fact an additive *interval* excitation mass above the ground state mass $m_e$. By extension, we assume that $m_b$ is also an interval excitation.

The excitation quanta $m_b$ and $m_f$ are mass (energy) quanta. They carry the particle and antiparticle quantum numbers of the electron or positron from which they were created. The particle-antiparticle symmetry of compound states constructed from these quanta can be ascertained from their masses: that is, *unpaired* particle or antiparticle states occur right on the "mass shell" (with zero binding energy BE), whereas *paired* particle-antiparticle-symmetric hadron bound states have a "hadron binding energy" (HBE) of 2-4% (Sec. J). Leptons do not form compound states. Phenomenologically, the ($m_e+m_b$) and ($m_e+m_f$) energy levels generated by these $m_b$ and $m_f$ electron α-leaps serve as $M_b$ and $M_f$ "platform masses" upon which "excitation towers" are erected, as we now describe.

### C. The α-leap "platform masses" $M_\pi$, $M_\phi$, $M_K$ and $M_\mu$

The search for an α-quantization of particle masses turned up two candidates—the 137 MeV α-leap from an electron pair to the pion, and the 211 MeV α-leap from an electron pair to a muon pair. In terms of *single-electron* excitations with a precise α-spacing, these correspond to 70 MeV and 105 Mev α-leaps respectively (Eqs. 7 and 8). We might think that further electron-pair α-quantization excitations would consist of additional ~140 or ~210 MeV α-leaps, but this is not what is observed experimentally. Instead, the combination of the initial electron mass plus the mass excitation that corresponds to the α-leap (see Eq. 9) creates what we denote as a "platform mass" M, which establishes the basic quantum numbers for the production channel, and which leads to the construction of an "excitation tower" on top of it that is formed from a different α-quantized mass unit—the "supersymmetric" excitation quantum X = 420 MeV defined in Sec. D. This combination of an α-leap to a platform M plus additional X-quantized excitations constitutes the electron-positron two-step "MX" production process described in Sec. E, which uniquely and accurately reproduces the masses of the basic threshold states.

In this section we define the various platform masses M that are created by electron α-leaps. We later use these platforms and the MX production process to create the excitation towers that generate threshold-state particles. These are purely phenomenological results. We generate the platforms M and then select the required number of X quanta to reproduce the particle masses. However, the results we obtain are in the form of a pattern that clearly is not random (Sec. I). And the accuracy of the results attests to their relevance. The most remarkable aspect of these results is that they combine leptons and hadrons together in one unified mass formalism. In fact, without combining leptons with hadrons we could not ascertain the details of this formalism.

There are two spin-dependent *asymmetric generic platform masses*:

$$M_b \equiv m_e + m_b, \quad \overline{M}_b \equiv \overline{m}_e + \overline{m}_b, \quad 70.54 \text{ MeV } (J=0) \text{ "boson platform masses"}; \quad (10)$$

$$M_f \equiv m_e + m_f, \quad \overline{M}_f \equiv \overline{m}_e + \overline{m}_f, \quad 105.55 \text{ MeV } (J=1/2) \text{ "fermion platform masses"}. \quad (11)$$

Since the direct conversion of energy into particle mass is a particle-antiparticle-symmetric process, these asymmetric $M_b$ and $M_f$ platform masses cannot be produced singly. An electron-based α-leap excitation process must start with a symmetric electron-positron pair, and the platform states it produces must be particle-antiparticle symmetric. The two *symmetric generic platforms* are

$$M_b + \overline{M}_b \equiv (m_e + \overline{m}_e)(1 + 1/\alpha) = 141.07 \text{ MeV}, \quad \text{"symmetric boson platform"}, \quad (12)$$

$$M_f + \overline{M}_f \equiv (m_e + \overline{m}_e)(1 + 3/2\alpha) = 211.10 \text{ MeV}, \quad \text{"symmetric fermion platform"}. \quad (13)$$

We will denote these as the platforms $M_b\overline{M}_b \equiv M_b + \overline{M}_b$ and $M_f\overline{M}_f \equiv M_f + \overline{M}_f$. These symmetric platforms occur experimentally in production modes that we label by the particles produced in these modes. The $M_\pi = M_b\overline{M}_b$ platform is produced by matching $m_b$ and $\overline{m}_b$ α-leaps from an electron-positron pair, and it corresponds to the pion. Its $M_b$ sub-states are hadronically bound together. The pion is a spin 0 particle, so the initial electron pair must be in a J = 0 spin state. The $M_\phi = M_f \overline{M}_f$ platform is produced by matching $m_f$ and $\overline{m}_f$ α-leaps from an electron-positron pair. It does not represent an observed particle, but when an X quantum is added to each of its $M_f$ substates, the hadronic $\phi = s\overline{s}$ vector meson is formed, where $s$ is the strange flavored quark. The $\phi$ is a spin 1 meson, so the electron-positron pair in the $M_\phi$ production mode must be in a J = 1 spin state. These two symmetric platforms are formally defined as follows:

$$M_\pi \equiv (m_e + \overline{m}_e)(1 + 1/\alpha) = 141.07 \times (1 - \text{HBE}) \text{ MeV}, \quad J = 0 \text{ "pion platform"}; \quad (14)$$

$$M_\phi \equiv (m_e + \overline{m}_e)(1 + 3/2\alpha) = 211.10 \times (1 - \text{HBE}) \text{ MeV}, \quad J = 1 \text{ "phi platform"}; \quad (15)$$

where HBE is the hadronic binding energy. There is a third symmetric platform,

$$M_{\mu\mu} \equiv (m_e + \overline{m}_e)(1 + 3/2\alpha) = 211.10 \text{ MeV}, \quad J = 0 \text{ or } 1 \text{ "mu-mu platform"}, \quad (15a)$$

which is also produced by matching $m_f$ and $\overline{m}_f$ α-leaps from an electron-positron pair, but the muon and antimuon leptons that occupy it (Eq. 9) are unbound. It can be in either a J = 0 or J = 1 spin state. Since the J = 1/2 $M_f$ and $\overline{M}_f$ elements of the $M_{\mu\mu}$ platform act independently, we can work with them separately, as described below, instead of using the combined form shown in Eq. (15a).

The asymmetric single-electron $M_b$ and $M_f$ platforms of Eqs. (10) and (11) are not produced directly, but they appear in reaction products and play an important role in the MX systematics. The $M_b$ platform mass does not correspond to an observed particle, but when an X quantum is added it becomes the strange flavored K meson, or kaon. Thus we label it as the $M_K$ platform. Since the kaon is spinless and the X quantum is also spinless, the $M_k$ mass quantum is also spinless, even though it is generated from the spin 1/2 electron. The reason this can happen is that the $M_K$ or $M_b$ mass quanta are always produced in pairs from electron-positron ground states, and if the electron pair is in a J = 0 spin configuration, angular momentum is conserved in the α-induced $M_K$ excitation process.

The $M_f$ platform mass is the muon (Eq. 9), and hence is labeled as $M_\mu$. It is the basic mass quantum for all of the spin 1/2 quarks. The $M_\mu$ and $\overline{M}_\mu$ substates in the $M_{\mu\mu}$ platform of Eq. (15a) are unbound, and it is more significant phenomenologically to dia-gram the $M_{\mu\mu}$ excitation tower in terms of separate $M_\mu$ and $\bar{M}_\mu$ particle and antiparticle excitation columns than to use it directly. In these separated particle and antiparticle columns the particles have zero binding energy, which makes their masses easy to calculate.

The asymmetric $M_K$, $\bar{M}_K$, $M_\mu$ and $\bar{M}_\mu$ platforms are defined as follows:

$$M_K \equiv m_e(1+1/\alpha), \bar{M}_K \equiv \bar{m}_e(1+1/\alpha), 70.54 \text{ MeV } J = 0 \text{ "kaon platforms"}, \quad (16)$$

$$M_\mu \equiv m_e(1+3/2\alpha), \bar{M}_\mu \equiv \bar{m}_e(1+3/2\alpha), 105.55 \text{ MeV } J = 1/2 \text{ "muon platforms"}, \quad (17)$$

Having defined the $M_\pi$, $M_\phi$, $M_K$ and $M_\mu$ platforms, we now define the excitation quantum X that appears in the excitation towers erected on these platforms.

### D. The α-quantized "supersymmetric" excitation quantum X

One of the interesting results to emerge from the search for α-quantized masses is the discovery of the dominant excitation quantum X that serves as the basic mass element for the "excitation towers" erected on the "platform states" M of Sec. C. The quantum X is "supersymmetric" in the sense that it is the lowest-mass excitation unit which can be particle-antiparticle symmetric when composed of either all boson α-mass excitation quanta $m_b$ and $\bar{m}_b$ of Eq. (7) or all fermion α-mass quanta $m_f$ and $\bar{m}_f$ of Eq. (8). X must contain even numbers of $m_b$ or $m_f$ subquanta to be particle-antiparticle symmetric, and it must contain $m_b$ or $m_f$ quanta in the ratio of three to two to be isoergic in the two representations. Thus X is formally defined as follows:

$$X = 3(m_b + \bar{m}_b) = 2(m_f + \bar{m}_f) = 420.15 \text{ MeV} \quad (18)$$

Although X is shown here in these *symmetric* representations, it appears in different excitations with different forms. In hadronic channels X has a small (2-3%) binding energy and appears with an experimental mass of about 410 MeV. In leptonic or unpaired hadron channels X has essentially zero binding energy and appears with a mass of about 420 MeV. In the kaon, which is the excitation $K = M_K + X$, the quantum X seems to have different and possibly mixed forms for $K^\pm$, $K^o_L$ and $K^o_S$, as indicated by their $K^o_L \to \pi\pi\pi$, $K^o_S \to \pi\pi$ and $K^\pm \to \pi\pi$ decay modes, and their $K^o_L/K^\pm = 4$ and $K^o_S/K^\pm = \alpha$ lifetime ratios (Fig. 3). In its role as an excitation quantum, X denotes the increase in mass in a production channel, and hence does not represent a basic alteration of the channel quantum numbers. Thus X shares the same characteristics as the $m_b$ and $m_f$ excitation quanta out of which it is formed (Eqs. 7 and 8), which are particle and antiparticle mass elements. We demonstrate in Part V that the mass quanta $m_b$ and $m_f$ can be reproduced as spinless and relativistically spinning (J = 0 and 1/2) modes of the same basic mass quantum, so the quantum X may facilitate transformations betweens these two types of particles, as for example in the decay modes $\pi \rightarrow \mu + \nu$ and $\tau \rightarrow \pi + \nu$, and thus serve as a "cross-over" excitation. The creation of α-quantized 420 MeV mass quanta X above the α-quantized platform masses of Eqs. (14 - 17) constitutes the MX particle generation process described in the next section.

### E. Two-step "MX" excitation towers and the "MX octet"

When a symmetric $m_b \bar{m}_b$ or $m_f \bar{m}_f$ two-electron α-leap is generated from an electron-positron pair, the resulting energy level acts as an $M_\pi$ or $M_\phi$ "platform state" upon which further excitations can be made (Eqs. 14 and 15). Asymmetric one-electron α-leaps can also be identified as components of the production process, and they lead to the asymmetric platforms $M_K$ and $M_\mu$ (Eqs. 16 and 17). Subsequent excitations are not in single α-leaps of $m_b$ = 70 MeV or $m_f$ = 105 MeV, but are instead in units of the supersymmetric "multiple-α-leap" quantum X = 420 MeV (Eq. 18). This is the two-step "MX" production process. The observed *one-electron* MX excitations are:

$$(1)\ m_{electron}/\alpha \rightarrow M_K,\ (2)\ M_K + nX \rightarrow \text{spin 0 particle},\ n = 1; \quad (19)$$

$$(1)\ m_{electron}/(2\alpha/3) \rightarrow M_\mu,\ (2)\ M_\mu + nX \rightarrow \text{spin 1/2 particle},\ n = 0, 1, 2, 4. \quad (20)$$

The observed *electron-pair* MX excitations are:

$$(1)\ m_{electron\text{-}positron}/\alpha \rightarrow M_\pi,\ (2)\ M_\pi + nX \rightarrow \text{spin 0 particle},\ n = 0, 1, 2; \quad (21)$$

$$(1)\ m_{electron\text{-}positron}/(2\alpha/3) \rightarrow M_\phi,\ (2)\ M_\phi + nXX \rightarrow \text{spin 1 particle},\ n = 1. \quad (22)$$

Eq. (19) gives the kaon and antikaon; (20) gives the muon and antimuon, $s$ and $\bar{s}$ quarks, proton and antiproton, and tau and antitau; (21) gives the pi, eta and eta' mesons; and (22) gives the phi(1020) meson. These MX excitations generate the "MX octet"

$$\pi, \eta, \eta', K, \mu, p, \tau, \phi \tag{23}$$

and the related antiparticles $\overline{K}, \overline{\mu}, \overline{p}, \overline{\tau}$. These are the basic low-mass threshold states. The $u, \overline{u}, d$ and $\overline{d}$ quarks correspond to $n = 1/2$ levels in Eqs. (20) and (22), and are not allowed as direct excitations in this X-quantized formalism.

As a notation for describing MX production in detail, we attach a superscript $n$ to the platform M and denote the hadronic binding energy (HBE) that is used. This gives

$$M_\pi^n \equiv (M_\pi + nX)(1 - HBE); \quad M_\phi^n \equiv (M_\phi + nXX)(1 - HBE); \tag{24}$$

$$M_K^n \equiv (M_K + nX); \quad M_\mu^n \equiv (M_\mu + nX); \tag{25}$$

This notation is used for the precision mass calculations of Sec. K.

### F. The "TR" mass-tripling process for $s\overline{s} \to c\overline{c} \to b\overline{b}$ flavor generation

The three Standard Model flavored quarks are $s$, $c$ and $b$. The threshold state for $s$ quark pair production is the $\phi(1020) = s\overline{s}$ vector meson, with a binding energy of about 3% (Sec. J). Since the $s$ quark and K meson both carry the strangeness quantum number $s$, and since the strangeness-preserving $K\overline{K}$ final state, which is the dominant $\phi$ decay mode, lies below the $\phi$ mass, the $\phi$ has a more rapid decay than generally occurs for a threshold state. Thus its lifetime is slightly shorter than those of the 36 long-lived threshold particles shown in Fig. 21. The $M_\phi$ platform that produces the $\phi(1020) = s\overline{s}$ meson also produces the other two flavor threshold states, the vector mesons J/$\psi$(3097) = $c\overline{c}$ and $\Upsilon(9460) = b\overline{b}$. But the production of the higher-mass $c\overline{c}$ and $b\overline{b}$ thresholds does not follow the MX excitation scheme that generated the $s\overline{s}$ threshold. Instead, a process of successive mass triplings (TR) occurs in which the $s\overline{s}$ flavored-quark pair is first created in an MXX process, and then this MXX excitation (both the platform mass M and the excitation XX) is tripled twice to produce first the $c\overline{c}$ pair and then the $b\overline{b}$ pair. The resulting mass ratio of three between $c$ and $b$ quark states (Fig. 20) echoes the lifetime ratio of three between states with unpaired $c$ and $b$ quarks (Fig. 14).

It is interesting to characterize this TR mass tripling process in terms of mass X-units. The $M_\phi$ platform (Eq. 15) has an energy, apart from HBE binding energy effects, of 0.5 X. The $s\bar{s}$ excitation (Eq. 22) requires another 2X of energy, so the $\phi = s\bar{s}$ threshold state represents 2.5 X-units of energy. If we triple this amount, we get 7.5 X-units for the $c\bar{c}$ threshold, which is an increase of 5 X-units. Tripling this again gives 22.5 X-units for the $b\bar{b}$ threshold, which is a further increase of 15 X-units. Since the $M_\phi^n$ production channel of Eq. (24) features particle excitations that increase in *even* numbers of X-units above the $M_\phi$ platform, the TR tripling process is not just a different way of describing MX excitation processes, but represents a different type of particle excitation. In the absence of a theory for this α-quantized systematics, both the MX and TR excitations represent empirical results within the framework of α-generated mass units.

The MX production mechanism and the TR flavored-quark tripling mechanism combine to accurately reproduce the main elementary particle threshold states, including both leptons and hadrons, in one comprehensive α-quantized formalism.

### G. MX generation of the unflavored threshold particles

The mass values of all 36 threshold-state elementary particles were displayed in Fig. 21, and the *unflavored* members of this group were labeled. These are low-mass nonstrange states whose mass systematics should be particularly transparent. In Fig. 22 we plot the excitation towers for these states, which feature the $M_\pi$, $M_\mu$ and $\bar{M}_\mu$ platforms of Eqs. (14) and (17). This mass plot displays the X-quantization of the excitation towers in a very clear manner. The excitations in the J = 0 boson tower, where particle and antiparticle substates are hadronically bound together, occur via single X steps (Eq. 24), and the experimental X intervals of 410 MeV reflect a small hadronic binding energy of about 2.4% (Eq. 18). The excitations in the matching J = 1/2 fermion and antifermion towers result in the production of particle-antiparticle fermion pairs. As can be seen, the experimental X intervals of 417 and 419 MeV (Eq. 18) in these towers indicate that the individual particles in each channel are formed with essentially zero binding energy.

The excitation formalism displayed in Fig. 22 accurately ties together the masses of the three leptons, the three nonstrange pseudoscalar mesons, and the proton and neutron. This primitive-looking result represents the only viable parameter-free explanation known to the author for the proton-to-electron mass ratio, which has been a mystery for the past century. It also uniquely reproduces the muon-to-electron and tau-to-electron mass ratios.

We can find more examples of X excitations, and also extend this formalism to include the *flavored* threshold particles, by displaying the platform states $M_K$ and $M_\phi$ and their excitation towers, which we do in the next section.

### H. MX and TR generation of the flavored threshold particles

The *flavored* threshold-state particles generated in the MX process are shown in Fig. 23, together with the TR tripling process. The strange spin 1/2 quark $s$ and the strange spin 0 K meson both carry the strangeness number *s*, and are created in particle-antiparticle "strangeness pairs" in associated production, as for example $\pi^- + p \to \Lambda + K^o$. The $\alpha$-quantized structures of these strange-flavored states are similar: $K = M_K+X$ (J = 0) and $s = M_\mu+X$ (J = 1/2), with masses of 496 and 526 Mev, respectively. The $\phi(1020)$ spin 1 meson is an $s\bar{s}$ bound state with a binding energy of 3.1%, and the $\eta'$ spin 0 meson is a $K\bar{K}$ bound state with a binding energy of 3.4%. After the $s\bar{s}$ threshold state has been created, the $c\bar{c}$ = J/$\psi$ "charm" threshold state is created by adding an $s\bar{s}$ excitation unit to each column of the excitation tower, which triples the $c\bar{c}$ mass with respect to the $s\bar{s}$ mass. This process is repeated to create the $b\bar{b}$ = $\Upsilon$ "bottom" threshold state.

The excitation formalism displayed in Fig. 23 ties together the flavored quarks K, *s*, *c* and *b*. The experimental X masses of 425 MeV in the $M_K$ production channels come from the fact that they are unpaired hadron channels which do not bind together in the production process. The X masses of 407 MeV in the $M_\phi$ channel reflect the $\phi = s\bar{s}$ binding energy of 3.1%.

### I. The $\alpha$-quantized MX octet mass matrix

The unflavored and flavored MX excitations of Figs. 22 and 23 exhibit complementary particle groupings. To illustrate this more clearly, we combine them together in

Fig. 24. The goal here is to demonstrate that although we have arbitrarily selected the number of X quanta for each particle state, the pattern that emerges is not random. It must be kept in mind that these are not just carefully-selected particle states which happen to fit into this pattern, but constitute *all* of the long-lived threshold states that have been observed in this mass region. There are no "rogue" states that lie outside of this α-quantized mass formalism, just as no rogue states are observed in the α-quantized lifetime plot of Fig. 11.

We can array the MX octet particles of Eq. (23) in an excitation matrix, as shown in Table 1. This shows how the different platforms sort out the types of particles that are produced. This is a short list of particles, but it provides key relationships between particles that we need in order to understand the structure of the elementary particle mass spectrum. Crucial aspects of this structure are its α-quantization, its interleaving of leptons and hadrons, and its use of the electron and positron, the lowest-mass particles, as the foundation for these masses. It is a general rule in physics that the lowest-mass state—the ground state—is the one that dictates the properties of the system, and elementary particle physics is no exception.

In addition to its symmetry characteristics, the other thing we need to know about the MX octet mass matrix is how accurately it reproduces the masses of these particles, which we show in Sec. K. But first we need to discuss the handling of mass charge splittings and mass binding energies, which we do in Sec. J.

## J. Charge splittings and hadronic binding energies (HBE)

Some elementary particles exist in single charge states, whereas others have two or more charge states (isotopic spins) and different mass values in those states. The MX octet mass matrix that is summarized in Table 1 and Eqs. (24) and (25) assigns the same mass value to all isotopic spin states of a particle. This limits the accuracy of the mass calculations. If we are going to construct a particle from a set of substates, then the masses of these substates, or at least the mass of the overall structure, should depend on the charge of the particle. However, when we examine the experimental variations in the charge mass splittings, it becomes apparent that this is not an easy task to acccomplish.

The lowest-mass—and thus in some sense the simplest—isotopic spin multiplets occur in the pseudoscalar mesons. The pions are the lowest-mass multiplet, and they have an intriguing charge splitting of the mass. The measured $\pi^\pm - \pi^o$ mass difference is 4.5936 ± 0.0005 MeV, or 3.3%. This is the largest percentage charge splitting of any of the elementary particles except the $W^\pm$ and $Z^o$. In the present α-quantized mass formalism, the mass of the pion is derived from the mass of the electron by making an α-leap from a spin 0 electron-positron pair to an $m_b \bar{m}_b$ bosonic α-mass pair, thus relating the mass of the pion to the mass of the electron. The charge splitting in the pion also seems to be related to the mass of the electron. Fig. 25 is a plot of the $\pi^\pm - \pi^o$ mass difference, shown in comparison to a mass interval of nine electron masses—4.5990 MeV. These two mass intervals agree to an accuracy of 0.1%. In view of the factors of three and nine that are observed in the particle masses, and particularly in the particle lifetimes (Figs. 14 - 16), the agreement shown in Fig. 25 is probably not accidental. Since we represent the pion here as $\pi = m_b \bar{m}_b$, we can reproduce the $\pi^\pm - \pi^o$ mass difference by assigning charge-dependent masses to $m_b$ and $\bar{m}_b$. But if we now move on to the kaon isotopic spin multiplet we find a $K^o - K^\pm$ mass difference of 3.995 ± 0.034 MeV. This mass interval is different, and the sign of the mass difference has been reversed! Thus we can't readily reproduce the kaon charge splitting by using the same $m_b$ and $\bar{m}_b$ charge-dependent masses that we would use for the pion. And when we move on to the mass differences of the proton and neutron and other isotopic spin multiplets, the charge-splitting becomes even more obscure. In lieu of a comprehensive theory for handling charge splittings, we use *average isotopic spin masses* as the best representation of experimental mass values.

Another factor affecting mass calculations is binding energies. The α-quantized masses that are deduced from the lifetime and mass relationships of the present paper are constituent-quark masses, where the mass of the particle comes mainly from the masses of the quarks that make up the particle. If the quarks have enormous binding energies, as was originally assumed in QCD [16], then the quark masses can have a wide range of energies. But if the quarks have small binding energies (a few percent), then their masses are delimited by the particle masses. The calculated MX octet masses are close to the observed masses, which indicates that the quark hadronic binding energies (HBE) are small.

We can cite three experimental results that indicate the magnitude of these binding energies. *(1)* The annihilation of the $\bar{p}n$ bound state into pions shows a binding energy of 4.4% [17]. *(2)* The assumption that $\eta' = K\bar{K}$ (see Table 1), where K is the average kaon mass, gives a binding energy of 3.4%. *(3)* The MX excitation diagram of Fig. 24 shows five independent values of the quantum X in an unbound configuration with an average mass of 419.4 MeV, and four values of X in a bound configuration with an average mass of 408.5 MeV, giving a binding energy of 2.6%. These three examples set the scale for the range of hadronic binding energies.

### K.  Precision mass values for the MX octet

In Part II we dealt with two questions about the $\alpha$-quantization of elementary particle *lifetimes*: *(1)* its comprehensiveness, and *(2)* its accuracy. This $\alpha$-dependence only applies to the long-lived ($\tau > 10^{-21}$ sec) threshold state particles, but in that domain it is comprehensive. All 36 threshold particles fit onto the $\alpha$-spaced lifetime grid, as shown in Figs. 9 and 11. And when "corrections" are applied to remove the factor-of-two "hyperfine" structure (Figs. 3 and 4) and allowances are made for the factor-of-three "flavor" structure (Fig. 14), the global accuracy of the lifetime $\alpha$-quantization is impressive: the $\alpha^4$ lifetime spacing between unpaired-quark and paired-quark decays, which is a spacing of more than eight orders of magnitude, applies to essentially all of the threshold particles, as shown in Figs. 12 and 13; and the $\pi^{\pm}$, $\pi^o$, $\eta$ and $\eta'$ mesons have highly accurate $\alpha$-spacings that span almost thirteen orders of magnitude (Figs. 10 and 17).

Since elementary particle lifetimes and masses are phenomenologically-linked quantities, we expect the $\alpha$-quantization of elementary particle *masses* to also be comprehensive and accurate. In Part IV we have already addressed the problem of comprehensiveness, although our focus here is a little narrower, centering on the pseudoscalar mesons, the leptons, the proton, and the $\phi = s\bar{s}$ vector meson. These can all be accounted-for by the MX production mechanism of Sec. E. When we add in the flavored TR tripling mechanism of Sec. F, we reproduce the $J/\psi = c\bar{c}$ and $\Upsilon = b\bar{b}$ vector mesons. And when we extend these results to account for the rest of the 36 threshold particles (Sec. O and Fig. 33), the α-quantized basis states qualitatively reproduce the observed masses. Thus the mass α-quantization of the threshold particles is quite comprehensive. The remaining question is that of mass accuracies. The most accurate and parameter-free mass calculations occur for the MX octet states of Table 1, as we demonstrate in Table 2. The flavor TR tripling excitations (Sec. L) are also quite accurate. But complexities set in for the higher-mass hyperon and meson threshold particles, which involve a mixture of α-mass basis states.

The MX octet states represent the simplest particle excitations, and should be the most straightforward to calculate. In making these calculations, we use average values for the experimental pion, kaon and nucleon masses, as discussed in Sec. J. We also use a single value to represent the hadronic binding energies HBE of the π, η, η' and φ mesons. We obtain this HBE value by first selecting a range of binding energies and calculating the particle masses, then evaluating the *absolute percent deviations* (APD) of the individual calculated masses from the experimental values, and finally determining the average APD value for the data set. The results of this calculation, which are displayed in Fig. 26, show a flat minimum from HBE = 2.5% to 2.7%. Thus in Table 2 we use HBE = 2.6%.

The mass calculations of Table 2 are parameter-free except for the HBE. Starting with electrons, an initial α-leap is made to a platform M, using $m_b$ or $m_f$ excitation quanta. This generates the muon and the pions. Then an excitation tower in units of X is constructed on M, and the levels are filled with particles. This is the MX production process, and it generates the η, η' and K mesons, the proton, the τ lepton, and the φ vector meson. As Table 2 demonstrates, the average absolute mass accuracy of this procedure is 0.4%, or 3.6 MeV in mass. Since the $m_b$ and $m_f$ quanta are 70 and 105 MeV, respectively, and the X quantum is 420 MeV, the experimental α-quantization of this MX octet is apparent. Equally important is the manner in which the leptons and baryons are seamlessly interleaved, thus demonstrating the universality of the mass generation operator α. The mass calculations shown here are simple, but they are not simplistic. They yield the electron-proton mass ratio, which is perhaps the oldest unattained goal in elementary particle physics. And they answer I. I. Rabi's rhetorical question about the muon [18]: "Who ordered that?" The muon mass is the basic building block for all of the fermions.

**L. Excitation doubling and tripling and extrapolation to the W and Z**

Does the α-quantized low-mass excitation systematics of the present paper also accommodate the super-massive W and Z vector mesons? That is the question we address in the present section.

The *pion* excitation tower in Fig. 24 has two X levels above the $M_\pi$ platform, the matching *kaon* excitation towers each have one X level above the unoccupied $M_K$ platforms, the *phi* excitation tower has one (paired) X level and two TR levels above the unoccupied $M_\phi$ platform, and the matching *muon* excitation towers each have two XX levels above the $M_\mu$ platforms. In order to see what this implies for the *fermion* excitation systematics, we plot the basic MX quark and particle states on a single J = 1/2 excitation curve, which gives the results displayed in Fig. 27. As can be seen, the excitations above the muon are quantized in units of X (the *u* and *d* quarks, which correspond to 1/2 X units, are not generated directly), and the observed excitation intervals are 1X, 2X and 4X. This suggests an *excitation-doubling* process.

If we extend this result to the J = 1 *vector meson* tower, which features mass-tripling (TR) excitations, and make a plot similar to Fig. 27, we obtain Fig. 28, where the observed excitation intervals are 1TR and 2TR. An extrapolation to the 4TR excitation gives a calculated mass of about 85 GeV. The only two vector mesons anywhere in this mass domain are the W at 80 GeV and the Z at 91 GeV. Thus it is of interest to examine their possible relevance to a mass-tripled excitation-doubling systematics.

The experimental J/ψ to ϕ mass ratio is 3.038, and the ϒ to J/ψ mass ratio is 3.055. If we assume that the TR mass-tripling process is exact, then these two experimental ratios, which are each a little greater than three, may indicate that the binding energies of these states decrease with increasing mass. If we take the theoretical MX mass quantum for the ϕ in Table 2 and use it to generate mass-triplings, then the theoretical excitations for these three vector meson states are $\phi = M_\phi^1$, $J/\psi = 3\,M_\phi^1$ and $\Upsilon = 9\,M_\phi^1$. When the calculated mass values for these excitations are compared to the experimental values, the binding energies turn out to be 3.04%, 1.82% and 0.02%, respectively. Thus the binding energy does appear to get smaller as the excitation mass gets larger, and the high-mass ϒ is essentially a zero-binding-energy state. Hence we can extrapolate to higher energies by using the experimental ϒ mass and applying successive TR excitations. Fig. 28 shows a TR excitation curve that is anchored on the ϒ mass and extends from the ϕ to the W and Z, and also a plot of the experimental mass values for these states. As can be seen, the excitation curve passes right between the W and Z. The W–Z mass difference is more than 10 GeV, which is enormous on the mass scale of the threshold particles, but not so large in this high-mass regime. The W–Z average mass is quite close to the TR excitation mass. In detail, the theoretical TR mass, which is nine times the experimental ϒ mass, is 85.14 GeV, and the W–Z average mass is 85.81 GeV. The agreement here is better than 0.8%. Also, the fact that the W–Z mass in Fig. 28 is at the location of the 4TR excitation, just as the τ mass in Fig. 27 is at the 4X excitation, and both the 3TR and 3X excitation levels are unoccupied, suggests that the mass agreement in Fig. 28 is not accidental. Hence the α-quantization of elementary particle masses, which embraces both the lepton and hadron low-mass states may extend up to the ultra-high mass region.

## M.  Charge exchange (CX) reactions and proton stability

In addition to providing accurate mass values, the MX excitation towers have another interesting property. They delineate the charge-exchange (CX) transfers that take place during tower excitation. These are illustrated in Fig. 29, which shows the $M_\phi$ platform excitations that produce the $\phi = s\bar{s}$, $J/\psi = c\bar{c}$ and $\Upsilon = b\bar{b}$ flavor threshold states, plotted together with the $M_{\mu\mu}$ platform excitations (divided here into separate $M_\mu$ and $\bar{M}_\mu$ platform channels) that produce the $\mu\bar{\mu}$, $p\bar{p}$ and $\tau\bar{\tau}$ fermion threshold states. In the $M_\phi$ tower, the $s$ quark is created (in a pairwise manner) with a fractional charge $e = -1/3$. Then the mass is tripled and the $c$ quark is created with a charge $e = +2/3$. Finally the mass is tripled again and the $b$ quark is created with a charge e = – 1/3. These flavored-quark excitations involve the exchange of a unit $e$ of electric charge during each mass-tripling (TR) step. In the $M_{\mu\mu} \equiv M_\mu + \bar{M}_\mu$ tower, the $\mu\bar{\mu}$ pair is created by the initial α-leap from an electron-positron pair. Then two X excitations occur in each $M_\mu$ and $\bar{M}_\mu$ channel and are accompanied by a 2$e$ double-charge exchange, which creates the p$\bar{\text{p}}$ pair with reversed charges. Finally, two more X excitations occur in each channel with another double-charge exchange, so that the $\tau\bar{\tau}$ pair is created with unreversed charges.

The CX transfer is interesting in its own right as a way of accounting-for the charges on the flavored quarks. However, its real interest is in its implications for the proton. When the proton and antiproton are formed, with masses equal to nine muon masses, they decompose into $u$ and $d$ quark substates, which have masses equal to three muon masses. In this decomposition their integer charges are also decomposed into fractional quark charges. But now the positively charged proton is trapped in the negatively charged muon and tau channel, and it has no way to decay back down to the negatively charged electron ground state. So the proton is stable. The tau lepton has a mass equal to seventeen muon masses, and it remains a lepton because seventeen is a prime number which cannot be trisected into equal substates. The tau cannot decompose, and its charge cannot fragment.

### N. The low-mass spin 0 unflavored particle states: an accurate $\alpha$-quantized system

The simplest and therefore probably most informative hadronic elementary particle states are the low-mass spin 0 pseudoscalar mesons—$\pi^\pm$, $\pi^0$, $\eta$, $\eta'$, $K^\pm$, $K_L$, $K_S$. In the lifetime studies these states show the most accurate $\alpha$-quantization, as is displayed in Fig. 10. Two important aspects of Fig. 10 are that *(a)* all of the particles in this category are included, and *(b)* the accuracy of the $\alpha$-scaling is maintained over six powers of $\alpha$, or almost 13 orders of magnitude. In addition to the $\alpha$-scaling, the flavored pseudoscalar kaons exhibit a factor-of-two fine structure with respect to the $\pi^\pm$ reference lifetime (Fig. 3). For clarity, we remove these from our analysis, leaving just the unflavored $\pi^\pm$, $\pi^0$, $\eta$ and $\eta'$ particles. With this data base, the AAD($S$) analysis of Sec. II E gives the scaling interval $S = 136.1$, and the $\chi^2(S)$ analysis of Sec. II H gives $S = 138.8$, as displayed in Fig. 17. The average value of these two independent analyses is $S = 137.45$, as compared to the fine-structure-constant scaling value $\alpha^{-1} = 137.04$. This close matching of values is a strong argument in favor of $\alpha$ as the relevant scaling factor for particle lifetimes.

The lifetime results are interesting in themselves, but perhaps of greater interest are the implications they carry with respect to the masses of these particles. Since lifetimes and mass widths are conjugate quantum mechanical variables that are related by the Heisenberg uncertainly principle of Eq. (6), an α-dependence for the lifetimes mandates a reciprocal α-dependence for the mass widths, and thus plausibly also for the masses. And since the lifetime α-dependence for the nonflavored pseudoscalar meson states of Fig. 17 is very accurate, we might expect the mass α-dependence for these particles to be equally accurate. Specifically, the lifetimes scale in powers of 1/137, so we look for some kind of mass dependence that is in factors of 137. And, thanks to the system of units commonly employed in particle physics, this reciprocal mass α-dependence emerges in an accurate and spectacular manner, as depicted in Fig. 30. Fig. 30a shows the accurately α-spaced lifetimes of the unflavored PS mesons from Fig. 10. Fig. 30b is a display of the corresponding masses of these particles, shown plotted on a linear mass grid that is in units of 137 MeV. As can be seen, the $\pi^\pm$, $\pi^0$, η and η' masses fall squarely on this grid. In order to account for the relevance of using a 137 MeV mass grid, we have also included in Fig. 30b another spin 0 unflavored low-mass particle system—the J = 0 electron-positron pair. The mass of this $e^-e^+$ pair is 1.022 MeV, and it sets the mass scale for the other particles. To see why this is so, we note that the $e^-e^+/\pi$ mass ratio is almost precisely equal to α. Thus the $e^-e^+$ to π mass excitation is the mass "α-leap" (Fig. 21) that is reciprocal to the lifetime "α-leaps" which occur experimentally for these particles. The ~137 MeV mass quantum $m_\pi$ that is generated by this α-leap serves as a mass unit for the η and η' higher excitations, as displayed in Fig. 30. The α-spaced lifetimes and masses of Fig. 30 stand as a demonstration of the concept of reciprocal lifetime and mass α-quantization in a transparent, complete, and accurate manner.

The interest in Fig. 30 is due not only to its display of an α-dependence in the unflavored PS meson lifetimes and masses, but also to the *patterns* of this α-dependence. In the lifetimes we see a basic reference lifetime followed by a lifetime that is a factor of $α^4$ shorter, and then lifetimes that are additional factors of α shorter. In the masses we see a basic spin 0 "ground state" mass followed by a mass that is one $m_\pi$ α-mass larger, and then masses that are additional X ≡ $3m_\pi$ mass units larger. These α-dependences are indi-cated in Fig. 30. The lifetime factor of $\alpha^4$ is a characteristic lifetime spacing that is seen in many particle families, as displayed in Figs. 12 and 13. The mass unit X ~ 411 MeV appears in Fig. 24.

The reciprocal α-quantized lifetimes and masses shown in Fig. 30 are based on the spin J = 0 $e^-e^+$ ground state. A similar pattern of reciprocal α-quantized lifetimes and masses is observed for the spin 0 D meson and its excited states. Fig. 31 displays both of these systems, which are placed together for comparison purposes. The lifetimes are shown in Fig. 31a, with all lifetimes expressed as ratios to the $\pi^\pm$ reference lifetime. The D-meson lifetimes use the logarithmic $x_i$ scale at the top of the figure, which is shifted by two powers of α with respect to the pion lifetime scale at the bottom of the figure. The "basic" $D^\pm$ and $D^o$ lifetimes are separated from one another by about a factor of two (see Fig. 3), and the $D^o$ exhibits the characteristic factor of three $c$-quark displacement from the $\pi^\pm$-based $\alpha^{x_i}$ lifetime grid (Figs. 11 and 14). The $D^{*\pm}$ lifetime is about a factor of $\alpha^4$ shorter (the $D^{*o}$ lifetime has not been accurately measured), and the $D^o_1$ lifetime is another factor of α shorter. Thus these D-meson lifetime intervals closely approximate those of the pion lifetime intervals, although they differ somewhat in detail. The fact that similar lifetime patterns occur in two systems whose actual lifetimes are shifted by two powers of α (four orders of magnitude) with respect to one another indicates that the observed α-quantization of lifetimes is a general result, and not specific to one hadron family. The masses are displayed in Fig. 31b, where the pion and D-meson spin J = 0 "ground states" are placed at the origin of the relative-mass excitation scale shown along the ordinate. In each case there are first-excited-states at $m_\pi$, followed by a second-excited-state at X ≡ $3m_\pi$. The interest in Fig. 31 is that it displays two matching reciprocal lifetime and mass systems that feature the same "α-leap" lifetime intervals and the same "α-mass" energy intervals. The PS pion system contains the α-quantization $m_\pi$ mass generator in the form of the 137 MeV "α-leap" from the spin 0 leptonic $e^-e^+$ pair to the spin 0 hadronic pion, and the hadronic D meson system uses this $m_\pi$ ~ 137 MeV "α-mass" and the X ≡ $3m_\pi$ excitation quantum in its own excitation processes. The lifetime α-quantizations of these states logically follow from the mass α-quantizations, although the details remain to be worked out.

As the final result in this section, we show yet another spin 0 system with this same mass quantization—the $D_s$ meson, whose excited-state lifetimes have not been measured. Fig. 32 combines the mass diagrams for all three of these systems—the unflavored PS pions, the singly-flavored D mesons, and the doubly-flavored $D_s$ mesons, all of which have spin 0 "ground states". As can be seen these three flavor-types each have a first excitation level about $m_\pi \sim 140$ MeV above the ground state, and then a second excitation level about X = 420 MeV above the first level. The first excitation level for the PS mesons is an "α-leap" of $m_\pi = e^-e^+/\alpha$, and the first excitation levels for the D and $D_s$ mesons are $m_\pi$ "α-masses". The second levels for all of these systems are X excitations.

The main conclusion we would like to draw from Figs. 30 - 32 is their experimental validation of the concept of the reciprocal scaling of particle masses and lifetimes. This concept is of course embodied in Eq. (6).

### O.  Higher mass and "mixed" threshold particles

The final step in this examination of threshold-particle α-quantization is to see how the low-mass systematics applies to the other long-lived particles. When this step is carried out, the same overall mass quantization appears, but as the excitations get more complex other factors enter in. To illustrate this in a general way, we take the Standard Model spin 1/2 quarks, assign them the α-mass values

$$(u, d), s, c, b \cong 3, 5, 15, 45\ m_\mu \cong 315, 525, 1575, 4725 \text{ MeV}, \qquad (26)$$

and then combine them with *zero binding energy* to create the long-lived *s, c* and *b* hyperons and mesons. The results of this calculation are displayed in Fig. 33. As shown there, the calculated masses of the Λ and Ξ hyperons are too high by about 3%. We could phenomenologically account for this as a binding energy effect, although we did not apply any binding energies in calculating the *p* and *n* baryon masses of Table 1. However, the calculated Σ hyperon masses are *lower* than the experimental masses, which is certainly not a binding energy effect: the Σ − Λ mass difference of about 70 MeV does not follow from the intrinsic *u, d* and *s* quark masses that make up these strangeness −1 hyperons. Additional excitation effects are clearly entering in. We would expected approximately equal Ξ to Λ and Ω to Ξ mass intervals, but the experimental Ω to Ξ mass interval is larger by about 140 MeV, so extra excitations are occurring here also. The $\Xi_c$ to $\Lambda_c$ and $\Omega_c$ to $\Xi_c$ mass intervals are approximately equal. When $c$ or $b$ quarks are added to the hyperon states, the calculated masses are all somewhat below the experimental values, but qualitatively correct. Moving to the charmed mesons, the calculated mass of the D is about 1.2% too high, which is in rough agreement with the binding energies required for the Table 2 states, but the calculated mass of the $D_s$ is 6.7% too high, so the $D_s$ – D mass difference does not correspond to the intrinsic quark mass difference. The calculated masses of the B mesons are all qualitatively correct, but are below the experimental masses. We can see from these results that using the SM quarks with the α-mass values assigned in Eq. (26) gives reasonable estimates for the experimental masses of these states, but not to the 1% precision of the threshold states of Table 2. The low-mass MX octet particles are the ones that exhibit a simple α-mass structure.

The thresholds for producing $s\bar{s}$, $c\bar{c}$ and $b\bar{b}$ quark pairs are the $\phi$, J/ψ and ϒ vector mesons (Fig. 28), which are produced in the X-quantized $M_\phi$ platform tower displayed in Figs. 23 and 24. The generation of $u\bar{u}$ and $d\bar{d}$ quark pairs logically occurs in this same tower, but their required 1/2 X quantization (Fig. 28) does not conserve particle-antiparticle symmetry in each channel of the tower. Thus no narrow 630 MeV $u\bar{u}$ or $d\bar{d}$ peak is observed. However, a "mixed" $u\bar{u}\pi$ or $d\bar{d}\pi$ excitation can conserve this symmetry, and the narrow-width ω(782)→πππ meson appears at this excitation energy. The D and $D_s$ charmed quark excitations shown in Figs. 31 and 32 represent other examples where α-quantized fermion and boson basis states are combined together.

A conclusion we draw from these analyses is that the threshold elementary particles do in fact have an α-quantized mass structure, but these particles have to be properly arrayed in order to bring out the aspects of this structure. In the same way, the α-quantized structure of particle lifetimes becomes apparent only when these lifetimes are properly arrayed. The α-spaced lifetimes and masses are experimental regularities, whose interpretation requires only quantum mechanics combined with the role played by the coupling constant α.

P. The Palazzi 35-MeV meson mass grid

The production of α-quantized elementary particles from an electron-positron ground state yields two basic mass quanta—the boson mass $m_b = m_e/\alpha \cong 70$ MeV (Eq. 7) and the fermion mass $m_f = (3/2)\, m_e/\alpha \cong 105$ MeV (Eq. 8). These can appear in pure boson (spin 0) or pure fermion (spin 1/2) substate combinations, unpaired or paired, and also in mixed boson-fermion combinations. The largest common factor for these excitations is the mass quantum $M_c^\alpha = 35$ MeV, so if a universal quantized mass plot is made which can accommodate assorted particles, it should be quantized in units of $M_c^\alpha$.

As a completely independent approach to elementary particle masses, Paolo Palazzi [19] started with a computerized version of the RPP meson data base [8], divided the mesons into the PDG spin-parity-flavor families, and analyzed each family separately, using integer multiples P of the mass quantum $u \cong 35$ MeV to reproduce the masses in each family, where P = even for mesons (bosons) and P = odd for baryons and leptons (fermions). The values for $u$ were varied, and detailed statistical fits were made to obtain an optimized mass $u$ for each family. These analyses demonstrated that all of the meson families can be fit with values of the mass unit $u$ which are close to 35 MeV. Palazzi's results for the low-mass threshold states are similar to those of the present paper, but he extends them beyond the present work by demonstrating that this 35 MeV mass quantization also applies to the higher-mass and shorter-lived particles. This result suggests that the higher-mass states are not rotational excitations, which would have a different mass quantization.

Palazzi's mass analysis also gives (from the viewpoint of the present paper) some information about meson binding energies. The various spin-parity-flavor families were statistically analyzed in order to obtain the best average mass fits as functions of $u$ and P (where P is an even integer). The most strongly-bound family has a $u$-value of 33.86 MeV, which in terms of the mass values of Eqs. (7) and (8) in the present paper corresponds to a (negative) binding energy of 3.3%. This is in close agreement with the empirical binding energies cited in Sec. IV J above. Other spin-parity-flavor families have smaller binding energies, which seem to be characteristic for each of the various members within the family, and which plausibly reflect (again from the present viewpoint) a blend of particle-antiparticle symmetries, isotopic spin effects, and lifetime constraints within the families.

## V. A MODEL FOR THE FERMION-TO-BOSON 3/2 MASS RATIO

In demonstrating the validity of a theory in physics, it helps to establish *cross links* between the different elements of the theory. These cross links serve to strengthen the entire theoretical framework. The main theoretical idea we are attempting to establish in the present work is the relevance of the fine structure constant $\alpha = e^2/\hbar c$ as a coupling constant in *hadron* interactions. Its relevance as a coupling constant in *lepton* interactions has already been clearly established in quantum electrodynamics (QED) [1], and we now empirically investigate its role in quantum chromodynamics (QCD). One line of evidence in favor of this idea is the observed $\alpha$-quantization of the long-lived threshold-state elementary particle lifetimes (Part II), which has been apparent in the experimental data for a long time [9, 11-14]. Another line of evidence comes from the pattern of $\alpha$-quantization that is observed in the threshold-state elementary particle masses (Part IV), which is an area of physics that has been pursued essentially independently of the lifetime results [20, 21, 11, 12, 14]. Cross-linkage between these two lines of evidence is suggested by the implementation in the present studies of the fact that lifetimes and mass widths are *conjugate* Heisenberg variables (Eq. 6), and the systematics of mass widths can carry over and apply to the masses themselves. The long-lived threshold elementary particle *lifetimes* exhibit an accurate and comprehensive quantization in powers of $\alpha$, so the *masses* of these same particles may exhibit a reciprocal quantization in powers of $\alpha^{-1}$, and they do! When this cross-linkage is invoked, the combined lifetime and mass $\alpha$-quantized data provide a more compelling argument for the relevance of the fine structure constant $\alpha = e^2/\hbar c$ in *hadron* interactions than either data set does separately.

The threshold particle mass $\alpha$-quantization that is displayed in Part IV offers yet another example of cross-linkage. The mass systematics of these threshold states reveals the existence of two basic *$\alpha$-masses*—$\alpha$-quantized mass units: the boson mass quantum $m_b = m_e/\alpha \cong 70$ MeV of Eq. (7), which occurs in the J = 0 spin channel; and the fermion mass quantum $m_f = (3/2)\, m_e/\alpha \cong 105$ MeV of Eq. (8), which occurs in the J = 1/2 spin channel. Calculational cross-linkage between these two mass quanta is provided by a simple mathematical model that ties them together as spinless and spinning configurations of the same fundamental mass quantum. This model is the relativistically spinning sphere (RSS) [22]. The details of this model have been presented elsewhere, and here we give just the pertinent results in the form of a series of lemmas. The physical relevance of this model to elementary particle physics, apart from its spectroscopic properties, is that particles which are composed of $m_b$ quanta (the pseudoscalar mesons) decay into particles which are composed of $m_f$ quanta (*e.g.* the muon), and vice versa. These decays are easier to account for if the same mass quantum is involved on both sides of the decay process.

Experimentally, a spin 0 pion (which is its own antiparticle, and hence possesses particle-antiparticle symmetry) appears at about 137 MeV, and a spin 1/2 muon-antimuon pair (which also possesses particle-antiparticle symmetry) appears at about 211 MeV—a factor of about 3/2 larger in energy. These are the lowest-mass symmetric particle states above the 1 MeV mass of a symmetric electron-positron pair. The muons are essentially unbound particles, and they each have only a single charge state, so their threshold energy is easy to interpret (Eq. 9). The pion, on the other hand, is a composite state which has matching bound particle and antiparticle substates, and also mass-dependent charge states (Fig. 25). Thus the "intrinsic" mass of the pion that should be compared to the "intrinsic" mass of a muon pair in order to determine their precise experimental mass ratio is not easy to ascertain.

The theoretical $m_b$ and $m_f$ mass quanta of Eqs. (7) and (8) have a mass ratio of precisely 3/2. The spin 1/2 $m_f$ quantum is more massive than the spin 0 $m_b$ quantum, which suggests that these could be spinning and spinless configurations of the same basic mass unit. To investigate this supposition mathematically in the simplest manner, we represent this basic mass quantum as a uniform sphere of matter. When the sphere is set into rotation at an angular velocity ω, its mass increases relativistically in accordance with the equation

$$m(r) = m_o(r)/\sqrt{1-\omega^2 r^2/c^2} \,, \tag{27}$$

where $m(r)$ is a mass element at a distance $r$ from the axis of rotation. In the context of special relativity, this mass increase arises from the instantaneous velocity v = ω$r$ of the mass element. It can also be viewed in the context of general relativity as arising from the gravitational potential of the rotational motion [23]. To get the total mass increase, we integrate over the volume of the sphere. The case of interest here is when the sphere is rotating at the *relativistic limit*, which we define as follows:

*A relativistically spinning sphere (RSS) is fully relativistic when its equator*

*is moving at, or infinitesimally below, the velocity of light, c.*

Assuming this limiting rotation, we have the following lemmas:

Lemma 1: A relativistically spinning sphere is half again as massive as its non-spinning counterpart: $M_s = 3/2\, M_o$.

Lemma 2: The volume of an RSS in the rotating frame is half again as large as that of its nonspinning counterpart: $V_s = 3/2\, V_o$.

Corollary A: The density distribution $\rho(r) = m(r)/v(r)$ of an RSS in the rotating frame is an invariant, independent of its rotational velocity.

Corollary B: Since the density distribution $\rho(r)$ of a massive object is a measure of its applied stresses, the invariance of $\rho(r)$ in the rotating frame can be taken as an indication of the absence of relativistic stresses.

Corollary C: The relativistic radius of curvature of an RSS vanishes at the equator of the sphere as measured in the rotating frame, so that a mass element placed there is effectively in linear motion, and an electric charge placed there will not radiate.

Lemma 3: The calculated moment of inertia of an RSS is $I = 1/2\, M_s R^2$, where R is the radius of the sphere.

Lemma 4: If the RSS radius is the Compton radius $R = \hbar/M_s c$, then its calculated spin angular momentum is $J_s = 1/2\, \hbar$.

Lemma 5: An equatorial massless point charge *e* on the RSS corresponds to a current loop with a calculated magnetic moment $\mu = e\hbar/2mc$.

Corollary D: The classical RSS model correctly reproduces the gyromagnetic ratio of the spin 1/2 electron.

Lemma 6. An equatorial point charge on an RSS gives rise to an electric quadrupole moment, but if the axis of rotation of the RSS is at the quantum-mechanically prescribed spin 1/2 angle of 54.7° to the *z* quantization axis, the electric quadrupole moment vanishes identically in the z direction, and it also vanishes in the x and y directions when averaged over a cycle of precessional motion.

Lemma 7: A spinning sphere that is given a translational velocity $w$ specified by the relativistic parameter $\gamma = \sqrt{1 - w^2/c^2}$ has the correct $M_\gamma = \gamma M_s$ (mass), $J_\gamma = J_s$ (spin) and $\mu_\gamma = \mu/\gamma$ (magnetic moment) relativistic transformation properties if, and only if, it is spinning at the full RSS relativistic limit $\omega = c/R$.

Lemma 8: The invariance of the spin angular momentum $J_s$ of an RSS under translational motion is due to offsetting relativistic length contraction and mass increase effects, whereas the relativistic variation of the magnetic moment $\mu$ is due to effects of the same sign. The relativistic mass increase $M_\gamma$ is independent of the length contraction.

The electromagnetic properties of this RSS model are of interest spectroscopically, but they do not have direct relevance to the mass systematics of the present paper. The proofs for Lemmas 1 - 6 are contained in the literature [24, 25] and follow analytically from the equations of classical mechanics. The mass equation in cylindrical polar coordinates $(r, \theta, z)$ is

$$M_s = \frac{3M_o}{R_s^3} \int_0^{R_s} \sqrt{\frac{R_s^2 - r^2}{1 - \omega^2 r^2/c^2}} r\, dr, \qquad (28)$$

where $M_s$ and $M_o$ denote spinning and spinless masses, and $\omega$ is the angular velocity. The corresponding moment of inertia equation is

$$I_s = \frac{3M_o}{R_s^3} \int_0^{R_s} \sqrt{\frac{R_s^2 - r^2}{1 - \omega^2 r^2/c^2}} r^3\, dr, \qquad (29)$$

These integrals are convergent. If the actual equatorial velocity of the spinning sphere is infinitesimally less than $c$, then the corrections to the relativistic equations are also infinitesimal. The proofs for Lemmas 7 and 8 cannot be done analytically, but require the separate relativistic transformation of each mass element. These calculations were carried out [26] by using approximately 37,000 mass elements, and were applied to a number of rotational and translational velocities.

The main relevance of this RSS model to the present results is that it ties together the $J = 0$ and $J = 1/2$ $\alpha$-mass excitation quanta of Eqs. (7) and (8) as rotationless and rotational states of the same basic spherical mass quantum, $m_b = m_e/\alpha = 70.025$ MeV. The supersymmetric "cross-over" excitation quantum $X = 6m_b = 4m_f$ = 420.15 MeV is the lowest-mass configuration that can thus accommodate $m_b \leftrightarrow m_f$ transitions via isoergic and particle-antiparticle-symmetric $m_b[m_b]\overline{m}_b + m_b[\overline{m}_b]\overline{m}_b \Leftrightarrow 2m_f\overline{m}_f$ mass transformations, where the brackets denote mass annihilations. Since spinless pions decay into spinning muons, and spinning tau leptons decay into spinless pions, these two-way transformations between spinless and spinning mass quanta are required experimentally.

The RSS model accounts for the 3/2 mass ratio between the J = 1/2 and J = 0 electron-generated α-mass units $m_f$ and $m_b$ that appear in the α-quantized mass systematics of the threshold-state elementary particles, and it thus provides another cross-linking element in this lifetime and mass formalism. This cross-linkage serves to confirm the significance of the experimentally-indicated 3/2 mass ratio, and the experimental mass results serve in turn to verify the relevance of the mathematical RSS model.

## VI. RAMIFICATIONS

### A. Cosmological masses and the vacuum state

In the present paper we have been studying elementary particles, which represent the observable world. But we now know from recent astrophysical discoveries that this *baryonic matter* seems to be only about a twentieth of the actual matter in the universe. Studies on the rotation of large spiral galaxies, which are held together by gravity, reveal that their observable matter is too small by an order of magnitude to supply the needed gravitational attraction. Hence these galaxies must also possess clouds of invisible *dark matter* that contain the requisite gravitational mass. Current estimates are that this dark matter constitutes about a quarter of the total mass of the universe, and is gravitationally lumped in the galaxies. Its composition is a matter of conjecture. Even more mysterious is the concept of *dark energy*. The universe is expanding outward in all directions, propelled by the force of the initial "big bang" and moving against the pull of its own gravity. This raises the question as to whether this expansion is constant or even increasing, in which case the universe will simply fly apart, or whether the expansion is slowing down, in which case the universe will ultimately stabilize or suffer a gravitational collapse. Re-cent astrophysical evidence suggests that the expansion rate is actually increasing, and cosmologists account for this fact by introducing dark energy, which has the property that it is gravitationally repulsive instead of attractive, and thus is spread out uniformly throughout the cosmos. This dark energy constitutes the remaining three quarters of the mass in the universe, and its composition and repulsive mechanism are unknown.

A somewhat related cosmological problem is the nature of space itself. When the wave-like nature of light and electromagnetism was first discovered, physicists assumed that space must be full of *ether*, since a wave must be a wave in something. However, Michelson and Morley demonstrated the surprising result that no detectable effects of the motion of this ether relative to the motion of the earth could be observed. This led to the disappearance of the ether. Waves could travel without any propagating medium. Space was simply a void. But developments in twentieth physics have now resurrected a more material concept of space. In quantum electrodynamics, space contains vacuum fluctuations in which virtual electron-positron pairs are continually being created and annihilated. The charge on an electron in space polarizes these pair fluctuations and thereby changes the effective value of the electron charge. This "vacuum polarization" produces measurable effects which are handled by "renormalization" theory [27]. The QED calculations that have emerged from renormalization theory, and the matching experimental values, are among the most accurate results ever obtained in physics. These vacuum fluctuations also produce the Casimir effect, the attractive force that is measured between two closely-spaced metal plates in a vacuum. A more indirect indication of a material space is suggested by the general theory of relativity, which attributes gravity to a curvature of space—the curvature ought to be a curvature of something. Today the vacuum state is regarded as a dynamic entity, a restless sea of appearing and disappearing electron pairs, the zero-point fluctuations of stochastic electrodynamics (SED).

There is one feature of the present α-quantized mass formalism—the relativistically spinning sphere RSS—that might have relevance to these cosmological masses. In addition to its role in accounting for the 3/2 mass ratio between spin 1/2 and spin 0 states, the RSS has another intriguing property: as it loses energy it *expands in size* in order to conserve angular momentum. That is, it retains its Compton radius R = $\hbar/M_s c$, which varies inversely with its mass $M_s$. The elementary particle excitation process we have dis-displayed in Figs. 21-24 and 29 starts with the creation of an electron-positron pair out of the vacuum state, which suggests that a Dirac sea of "proto-electron pairs" (PEP) may be spread throughout the cosmos, since this electron-pair creation process can take place anywhere. Suppose that the "fabric of space" is composed of this PEP sea. Then if electron pairs are excited out of the vacuum, create various particles depending on the available energy, lose some of this energy through the creation of stable protons (baryonic matter), low-energy photons (the microwave background), and neutrinos or other neutral particles (dark matter), and then cascade back down into the PEP sea at a somewhat lower energy, they will expand beyond their original size and thereby expand the fabric of space (dark energy).

## B. The enigmatic pseudoscalar mesons: the strongest example of reciprocal $\alpha$ quantization; the weakest example of Standard Model systematics

In the lifetime and mass studies of the present paper, the PS pseudoscalar mesons—the "PS octet" of quantum chromodynamics—exhibit the clearest and most accurate examples of reciprocal $\alpha$-quantization. They are the "crown jewels"—the best results we can offer. Fig. 10 illustrates the stunning accuracy of the unflavored $\pi^{\pm}$, $\pi^o$, $\eta$ and $\eta'$ PS lifetime ratios when expressed in powers of $\alpha \cong 1/137$, and Fig. 30 displays the equally stunning reciprocal accuracy of the masses of these particles when plotted on a 137 MeV mass grid. The accuracy of these results should not be surprising, in the sense that these PS mesons are the lowest-mass spinless states in the elementary particle zoo, and they should have simple and relatively stable structures, which they do. The surprising result about these particles is the difficulties they have historically created for the Standard Model, where they have steered the SM theorists into an uncomfortable situation with respect to quark masses, as we discuss in this section and the next.

Figs. 10 and 30 display the $\alpha$-quantization of the PS mesons pictorially. It is also instructive to consider the accuracy of this $\alpha$-quantization from a numerical viewpoint. The $\pi^{\pm}$, $\pi^o$, $\eta$ and $\eta'$ *lifetimes* are spread out over more than twelve orders of magnitude. By assigning lifetime ratios of $\alpha^4$, $\alpha$ and $\alpha$, sequentially, to these four lifetimes, we fit them into the α-spaced grid of Fig. 10. In order to evaluate the accuracy of this result, we set $1/\alpha = S$, so that the lifetime ratios are now expressed as $S^x$, with $x \cong -4, -1, -1$. We then vary $S$ to give the "best fit" to these four lifetimes, where the best fit is the minimum linear average absolute deviation (AAD) of the exponents $x$ from $x = -4, -1, -1$ integer values. This AAD minimization yields $S = 136.09$ (Fig. 17a). We next make a quadratic $\chi^2$ fit to the data (Eq. 5), using the lifetime values and experimental errors of RPP2004 [8]. This gives $S = 138.80$ (Fig. 17b). Each of these independent determinations of $S$ is within 1.3% of the value $1/\alpha = 137.036$, and their average $S$ value of 137.445 is within 0.30% of the value $1/\alpha$. These results could hardly be more accurate, considering the tremendous range of lifetimes included here. This lifetime α-dependence also extends to the flavored $K^\pm$ and $K^o_S$ mesons, which are related to one another by their $\pi\pi$ decay modes (Fig. 3), and whose experimental lifetime ratio is 138.3.

The α-quantization of the PS meson *masses* in 137 MeV mass units is also impressive. The lowest-energy particle-antiparticle-symmetric production threshold for massive particles is the electron-positron level at 1.022 MeV. The second-lowest symmetric threshold contains the $\pi^o$ and $\pi^\pm$ pion levels at 135.0 and 139.6 MeV. Lacking a formalism for handling the pion and kaon charge splittings of the masses (see Fig. 25), the best guess we can make for the "intrinsic" pion mass $\bar{\pi}$ mass is its average value of 137.27 MeV. The experimental masses of the η and η' mesons are 547.75 and 957.78 MeV, respectively [8]. Thus the $\eta - \bar{\pi}$ and $\eta' - \eta$ mass intervals are 410.48 and 410.03 MeV, respectively, which are each approximately equal to $3 \times 137 = 411$ MeV (Fig. 30). An important aspect of these $\eta - \bar{\pi}$ and $\eta' - \eta$ mass intervals is that they represent *mass linearity* at an accuracy level of 0.1%.

The 0.1% accuracy of the α-quantized $\eta - \bar{\pi}$ and $\eta' - \eta$ mass intervals, when combined with the 0.3% average accuracy of the *lifetime* AAD and $\chi^2$ fits of these mesons to an α-quantized grid, leads to an important phenomenological conclusion. Empirically, the η and η' mesons occur as the *first excited state* and *second excited state*, respectively, of the π meson, and they are on a completely equivalent footing from this standpoint. This conclusion seems borne out by their dominant $\eta' \to \eta + \pi\pi$ and $\eta \to \pi + \pi\pi$ decay channels. However, that is *not* the way the η and η' are fitted into the Standard model, as we now discuss.

The pseudoscalar QCD octet in the Standard Model includes the $\pi^+$, $\pi^o$, $\pi^-$, $K^+$, $K^o$, $\overline{K}^o$, $K^-$ and η mesons. The η' is included with this octet as an associated singlet that has its own adjustable mass parameter. The QCD mass relationships for spin 1/2 fermions involve accurate linear mass intervals. However, linear mass intervals for the PS mesons in the QCD formalism do not work very well, so quadratic mass relationships are invoked for them. The argument advanced for this procedure is that mass is a self-energy phenomenon, and boson self-energies such are those for the PS mesons are calculated by the Klein-Gordon equation, which gives quadratic masses [28]. However, the results of the present paper suggest that elementary particle masses are not a self-energy phenomenon, but instead come from an α-coupling to the mass of the electron, which undercuts the use of Klein-Gordon quadratic masses. Furthermore, the observed $\eta - \overline{\pi}$ and η' – η mass intervals, as described above, are in fact exceedingly linear. A more serious conceptual difference between these two approaches is that the η and η' mesons occur as the first and second "excited states" of the pion in the reciprocal α-quantized mass and lifetime data displayed in Fig. 30, and hence should not be handled differently from one another in the theoretical formalism. However, in the QCD formulation the η and η' occur in different SU(3) groups that are linked by a rather awkward coupling constant [8, 29].

The really crucial difference between the present α-quantized approach and the Standard Model quarks is in the SM use of the same spin 1/2 *u* and *d* quarks to represent the low-mass pseudoscalar mesons as the ones used for the higher-mass proton and neutron. This SM procedure gives the correct isotopic spin states for the pseudoscalar mesons, which is an enormous accomplishment that does not yet emerge in an obvious way from the use of α-quantized masses. However the *u* and *d* quarks can only reproduce both the pion and the proton if their intrinsic masses are small, so that the difference between the proton and pion masses is attributed to their different gluon currents. Historically, the QCD formalism for the PS octet was devised during the 1960's [16], when the quark masses were believed to be so large that nothing much could be said theoretically about their absolute values. The implementation of "asymptotic freedom", where quark masses in their ground states have very small binding energies, changed this situation, but it did not change the formulation of the PS octet. Thus the *u* and *d* SM quarks are now regarded as *current* quarks [8] with small intrinsic masses. By way of contrast, the α-quantized mass states of Tables 1 and 2 in the present paper correspond to *constituent* quarks, with their large intrinsic masses creating the observed masses of the particles. The hadronic substates in Table 2 do have a small hadronic binding energy (Fig. 26), but not one that is sufficient to hold them together. The strong binding energy that gives rise to "asymptotic freedom" arises from interactions between the *charges* rather than interactions between the *masses*. The difference between the current-quark representation in the Standard Model and the constituent-quark representation in the present α-quantized model raises the general question as to whether we can in fact employ different mass representations for essentially the same set of quarks. This is the topic of Sec. C.

## C. Mass freedom within QCD

The beginning of Part I contains a quotation [2] from Feynman in his book *QED* [1], where, after carefully discussing the physical principles involved in leptonic QED calculations and the principles that can be carried over to hadronic QCD calculations, he noted that the physical basis for the mass values used in these theories was still unknown. The astrophysical "dark matter" and "dark energy" masses briefly described in Sec. A above only serve to compound this difficulty This mass problem has its clearest delineation in the quark model theories that have been used to address the spectrum of elementary particles. These theories are presently based on the QCD quarks as used in the Standard Model, whose properties are summarized in Review of Particle Physics [8].

Historically, the quark model *circa* 1969 was first developed under the assumption that quarks are very massive (> 5 GeV) constituent-quark states which are held together by their enormous binding energies [16]. When high-energy accelerators were subsequently developed that pushed the required binding energy up past 5 GeV to much higher values, and free quarks were still not observed, it was realized that conventional binding energies didn't apply to the quark bonding, and the mass problem was then opened up to the twin tracks of current quarks and constituent quarks. This ambiguity is made possible in part because the striking isotopic spin successes of the quark model have to do with the quark charges, which are not directly tied to the quark masses. The Standard Model is presently centered on the use of current quarks [8], although as first the *c* quarks and then the *b* quarks were discovered, their experimental properties were creeping closer and closer to the systematics of constituent quarks. As soon as the concept of very large binding energies was abandoned in favor of asymptotic freedom, then the fact that the Standard Model uses the same *u* and *d* quarks for the low-mass pseudoscalar mesons as it does for the higher-mass baryons and hyperons meant that a current-quark approach was essentially mandated. These quark assignments give the correct isotopic spin rules for the PS mesons, but they have led to problems with the masses. Also, leptons and hadrons belong to completely different mass formalisms in the Standard Model, which has precluded any ability to calculate *e.g.* the proton-to-electron mass ratio.

To recapitulate the present work, we started by examining the spectrum of elementary particle lifetimes. They are simpler to work with than elementary particle masses. This analysis showed that the long-lived threshold-state particle lifetimes occur singly or in groups, with separation intervals in powers of $\alpha = e^2/\hbar c$. These groupings include all of the observed long-lived leptons and hadrons. When we then examined the spectrum of elementary particle masses from this same $\alpha$-quantized viewpoint, we were able to identify an $m_b$ = 70 MeV boson mass and an $m_f$ = 105 MeV fermion mass that serve as basis states for constructing the threshold-state particles, including both leptons and hadrons. Finally, we used the relativistically spinning sphere (RSS) model to tie together the $m_b$ and $m_f$ masses. The mass values that we thereby calculated in Table 2 are constituent-quark masses, so we have gone full circle from the original constituent-quark masses to current-quark masses and back to ($\alpha$-quantized) constituent-quark masses, but now with moderate binding energies. The non-observability of individual quarks arises in the $\alpha$-quantized formulation from the tenacious binding of fractional quark charges, which cannot be separated from their original integer units. Charges can be separated into fractions on a cluster of quarks, but they reunite if the quarks are separated or fragmented. This $\alpha$-quantized lifetime and mass formalism exhibits the QCD flavor structure, as for example in the lifetime groupings of Fig. 11, and it must necessarily incorporate or allow for the plethora of striking results that have been achieved in the Standard Model [8]. But it differs in details of the treatment of masses, and in the handling of leptons and hadrons.

In Feynman's book *QED* cited above [1], he pointed out an important contribution made to the theory of quantum electrodynamics by Bethe and Weisskopf [27], who noted that if cutoffs were systematically applied to divergent integrals, then the results could be carried over from one calculation to another. In their classic two-volume *Concepts of Particle Physics*, which summarized the status of elementary particle physics and quantum chromodynamics, Gottfried and Weisskopf addressed the quark mass problem, commenting as follows [30]:

*"Unfortunately, QCD has nothing whatsoever to say about the quark mass spectrum, nor, for that matter, does any other existing theory."*

In order to move beyond this impasse, we have in the present work turned directly to experiment, where the mass values shown in Table 2 emerge in a manner that is essentially independent of theory.

## D. The interpretation of the fine structure constant $\alpha = e^2/\hbar c \cong 1/137$

The phenomenological linkage between elementary particle lifetimes and masses enables us to use the observed α-quantization of particle *lifetime*s as the motivation to investigate the possibility of finding a reciprocal α-quantization of particle *masses*. This investigation shows that particle masses are in fact α-quantized at the lowest levels—namely, the triad formed by the electron pair, the pion, and the muon pair—and that this α-quantization extends on up to the supersymmetric excitation quantum X and the higher-mass threshold particles. The determination of this mass α-quantization is a worthwhile goal in itself, but it has some broader theoretical implications with respect to the fine structure constant α, which is the coupling constant between the electron pair and the pion. The constant α is itself one of the greatest mysteries in modern physics, as Richard Feynman summarized in the quotation [5] given near the end of Sec. I A. Continuing on here with this same quotation, we have Feynman expressing his frustration at this situation in the following words:

*"You might say the 'hand of God' wrote that number [the constant α], and 'we don't know how He pushed His pencil.' We know what kind of a dance to do experimentally to measure this number very accurately, but we don't know what kind of a dance to do on a computer to make this number come out—without putting it in secretly!*

*A good theory would say that e [Feynman's notation for α] is 1 over 2 pi times the square root of 3, or something. There have been, from time to time, suggestions as to what e is, but none of them has been useful."* [5]

The dimensionless fine structure constant is $\alpha \equiv e^2/\hbar c \cong 1/137$. The numerical value of this constant is 1/137.03599976 [8]. In spite of its extreme accuracy, this numerical value does not convey much physical insight as to its true significance. The constants $e$, $\hbar$ and $c$ appear in many areas of physics, and do not *per se* reveal much information. However, the way they are combined together may furnish some clues. The function of α in the present work is to act as an operator on electrons and positrons and generate a sequence of α-masses of the forms $m_b = m_e/\alpha$ (Eq. 7) and $m_f = 3m_e/2\alpha$ (Eq. 8), where $m_e$ is the electron mass, and where $m_b$ and $m_f$ are boson (spin 0) and fermion (spin 1/2) mass quanta, respectively. This suggests that a mass value $m$ should logically appear as a factor in α. In order to accomplish this, we first write α in the form

$$e^2/r \cong (1/137)(\hbar c/r) . \tag{30}$$

The left side of Eq. (30) now represents an electrostatic energy. Then we use the Compton radius equation

$$r = \hbar/mc \tag{31}$$

to replace the factor of r on the right side of Eq. (30). This gives

$$e^2/r \cong (1/137)(mc^2) . \tag{32}$$

The radius $r$ on the left side of Eq. (32) is the Compton radius that corresponds to the mass value $m$ on the right side of Eq. (32). Putting in dimensions, we have

$$e^2/r = (197.327/137.036)(1/r) = 1.440/r \text{ MeV/fermi} , \tag{33}$$

where the radius $r$ is measured in fermi units of $10^{-13}$ cm. The Compton radius is

$$r = \hbar c/mc^2 = 197.327/mc^2 \text{ fermi} , \tag{34}$$

where the mass energy term $mc^2$ is in units of MeV. If we take this term to be the average pion mass $\bar{\pi}$ = 137.27 Mev, we obtain a pion Compton radius $r$ of 1.438 fermi, which gives an electrostatic energy $e^2/r$ in Eq. (33) of 1.001 MeV. This value is 2% below the energy of an unbound electron-positron pair, and it reflects the ~2% hadronic binding energy in the pion (see Fig. 26). The calculation of the "intrinsic" mass of the pion that should be inserted into Eq. (34) is obscured by both binding energy and isotopic spin mass-splitting effects. The main point we are making in this calculation is not about precise numerical values, but rather about the fact that by combining the fine structure constant $\alpha = e^2/\hbar c$ with the Compton radius $r = \hbar/mc$, we obtain a theoretical expression (Eq. 32) that ties together the mass of the pion with the mass of an electron pair. This point is of interest in connection with the definition of the fundamental mass excitation quantum $m_b = m_e/\alpha = 70.025$ MeV in Eq. (7). But it is also of interest with respect to the definition of the constant $\alpha$. We can rewrite Eq. (32) in the form

$$(e^2/r)/(mc^2) \cong (1/137) \,. \tag{35}$$

We now have the numerical factor ~1/137 expressed as the ratio of an electrostatic energy divided by a mass energy—an electron pair energy over a pion energy, where the electrostatic term consists of two electric charges $e$ separated by a Compton radius $r$ that corresponds to the mass of the pion. This suggests that the factor ~1/137 is in some manner related to the "geometries" of both an electron pair and a pion, where this "geometry" can be a function of charge structures or mass structures or both. Thus the theoretical explanation for the value ~1/137 must come from a clear understanding of particle mass structure and charge structure.

In QED the coupling constant $\alpha$ gives the interaction strength for an electron to produce a photon [6]. In Tables 1 and 2 of the present paper, and in the accompanying figures, we have phenomenologically extended this QED result by having the constant $\alpha$ couple to an electron-positron pair and generate the pion, the muon-antimuon pair, the proton-antiproton pair, and higher excited states. These new results do not constitute a theory, but merely an observation as to what is going on in all of the high energy accelerators. Hopefully the concept of the reciprocal $\alpha$-quantization of elementary particle lifetimes and masses will serve to open new avenues for exploring these mysteries.

ACKNOWLEDGMENTS

I would like to thank Paolo Palazzi and David Akers for supplying preprints and other useful information. Their contributions, together with those of my wife Eleanor, who has supported this effort in a variety of ways for many, many years, were essential in the task of carrying the present work through to completion.ACKNOWLEDGMENTS

I would like to thank Paolo Palazzi and David Akers for supplying preprints and other useful information. Their contributions, together with those of my wife Eleanor, who has supported this effort in a variety of ways for many, many years, were essential in the task of carrying the present work through to completion.

Table 1. The "MX octet" excitation matrix. The eight MX threshold-state particles of Eq. (23) are shown separated into hadronically-bound and unbound excitations, and into unflavored and flavored excitations, using the MX notation of Eqs. (24) and (25). (Two related excitations are shown in parentheses.) These MX states accurately map the low-mass threshold particles of Table 2. Also, they blend leptons and hadrons into one comprehensive α-quantized pattern. Without combining these two particle types, we would not obtain this MX structure. Also, we would not obtain an accurate parameter-free value for the proton-to-electron mass ratio.

|  | Hadronically-bound states | Unbound states |
|---|---|---|
| Unflavored states | $M_\pi^{0,1,2} = \pi, \eta, \eta'$ | $M_\mu^{0,2,4} = \mu, p, \tau$ |
| Flavored states | $M_\phi^1 = \phi = s\bar{s}$ | $M_K^1 = K$ |
|  | $(\eta' = K\bar{K})$ | $(M_\mu^1 = s)$ |

Table 2. Mass calculations for the MX octet particles of Table 1. The experimental masses are averaged over isotopic spin states. The theoretical masses are calculated from Eqs. (24) and (25). The hadonic binding energy (HBE) of 2.6% (Fig. 26) is the only adjustable parameter in these calculations. The α-leap platforms $M_\pi, M_\varphi, M_K, M_\mu$ are defined in Eqs. (14 - 17). Two particles (π and μ) appear directly on these platforms, and the other six (η, η', K, p, τ and ϕ) appear at accurate X-quantized mass intervals above these platforms (Fig. 24), where the supersymmetric excitation quantum X = 420 MeV is defined in Eq. (18). The average *absolute mass deviation* (AMD) for these eight basic particle masses is 3.57 MeV, and the average *absolute percent deviation* (APD) is 0.40%.

| Particle | spin | Platform state | HBE | Exp. mass (MeV) | Calc. mass (MeV) | AMD (MeV) | APD |
|---|---|---|---|---|---|---|---|
| $e$ | 1/2 | (input) | | | | | |
| π | 0 | $M_\pi^0$ | 2.6% | 137.3 | 137.4 | 0.13 | 0.10% |
| η | 0 | $M_\pi^1$ | 2.6% | 547.3 | 546.6 | 0.67 | 0.12% |
| η' | 0 | $M_\pi^2$ | 2.6% | 957.8 | 955.9 | 1.92 | 0.20% |
| K | 0 | $M_K^1$ | 0 | 495.7 | 490.7 | 4.99 | 1.01% |
| μ | 1/2 | $M_\mu^0$ | 0 | 105.7 | 105.6 | 0.11 | 0.10% |
| p, n | 1/2 | $M_\mu^2$ | 0 | 938.9 | 945.9 | 6.93 | 0.74% |
| τ | 1/2 | $M_\mu^4$ | 0 | 1777.0 | 1786.2 | 9.16 | 0.52% |
| ϕ | 1 | $M_\phi^1$ | 2.6% | 1019.5 | 1024.1 | 4.61 | 0.45% |

Appendix A. The elementary particle lifetime data base, taken from RPP2004 [8]. The mean life τ is listed for each particle. The measured full width $\Gamma = \hbar/\tau$ is given for the short-lived particles. The 36 long-lived threshold particles are listed first, in order of decreasing lifetimes. The 120 long-lived particles are arranged in order of type and increasing mass.

**Table A. Lifetime data base.**

| | Mean life τ (sec) | Full width Γ (MeV) | Particle |
|---|---|---|---|
| 1 | 8.8570E+02 | | neutron |
| 2 | 2.1970E-06 | | muon |
| 3 | 5.1800E-08 | | $K^°_L$ |
| 4 | 2.6033E-08 | | $\pi^\pm$ |
| 5 | 1.2384E-08 | | $K^\pm$ |
| 6 | 2.9000E-10 | | $\Xi^°$ |
| 7 | 2.6320E-10 | | $\Lambda$ |
| 8 | 1.6390E-10 | | $\Xi^-$ |
| 9 | 1.4790E-10 | | $\Sigma^-$ |
| 10 | 8.9530E-11 | | $K^°_S$ |
| 11 | 8.2100E-11 | | $\Omega^-$ |
| 12 | 8.0180E-11 | | $\Sigma^+$ |
| 13 | 1.6710E-12 | | $B^\pm$ |
| 14 | 1.5360E-12 | | $B^°$ |
| 15 | 1.4610E-12 | | $B^°_s$ |
| 16 | 1.3900E-12 | | $\Xi_b$ |
| 17 | 1.2290E-12 | | $\Lambda^°_b$ |
| 18 | 1.0400E-12 | | $D^\pm$ |
| 19 | 4.9000E-13 | | $D_s$ |
| 20 | 4.6000E-13 | | $B^\pm_c$ |
| 21 | 4.4200E-13 | | $\Xi^+_c$ |
| 22 | 4.1030E-13 | | $D^°$ |
| 23 | 2.9060E-13 | | $\tau$ |
| 24 | 2.0000E-13 | | $\Lambda^+_c$ |
| 25 | 1.1200E-13 | | $\Xi^°_c$ |
| 26 | 6.9000E-14 | | $\Omega^°_c$ |
| 27 | 8.4000E-17 | | $\pi^°$ |
| 28 | 5.1024E-19 | 1.2900E-03 | $\eta$ |
| 29 | 7.4000E-20 | | $\Sigma^°$ |
| 30 | 2.5027E-20 | 2.6300E-02 | $\Upsilon_{3s}$ |
| 31 | 1.5307E-20 | 4.3000E-02 | $\Upsilon_{2s}$ |
| 32 | 1.2419E-20 | 5.3000E-02 | $\Upsilon_{1s}$ |
| 33 | 7.2331E-21 | 9.1000E-02 | $J/\psi_{1s}$ |
| 34 | 6.8564E-21 | 9.6000E-02 | $D^{*\pm}(2010)$ |
| 35 | 3.2585E-21 | 2.0200E-01 | $\eta'$ |
| 36 | 2.3424E-21 | 2.8100E-01 | $J/\psi_{2s}$ |
| 37 | 1.5451E-22 | 4.2600E+00 | $\phi(1020)$ |
| 38 | 7.7528E-23 | 8.4900E+00 | $\omega(782)$ |
| 39 | 3.8047E-23 | 1.7300E+01 | $\eta_{c1s}(2980)$ |
| 40 | 3.0989E-25 | 2.1240E+03 | $W^{\pm}$ |
| 41 | 2.6379E-25 | 2.4952E+03 | $Z^{\circ}$ |
| 42 | 4.3793E-24 | 1.5030E+02 | $\rho(770)$ |
| 43 | 9.4030E-24 | 7.0000E+01 | $f_o(980)$ |
| 44 | 8.7762E-24 | 7.5000E+01 | $a_o(980)$ |
| 45 | 1.8284E-24 | 3.6000E+02 | $h_1(1170)$ |
| 46 | 4.6353E-24 | 1.4200E+02 | $b_1(1235)$ |
| 47 | 3.5560E-24 | 1.8510E+02 | $f_2(1270)$ |
| 48 | 2.7312E-23 | 2.4100E+01 | $f_1(1285)$ |
| 49 | 1.1967E-23 | 5.5000E+01 | $\eta(1295)$ |
| 50 | 6.1515E-24 | 1.0700E+02 | $a_2(1320)$ |
| 51 | 1.1989E-23 | 5.4900E+01 | $f_1(1420)$ |
| 52 | 3.0615E-24 | 2.1500E+02 | $\omega(1420)$ |
| 53 | 2.4838E-24 | 2.6500E+02 | $a_o(1450)$ |
| 54 | 1.6455E-24 | 4.0000E+02 | $\rho(1450)$ |
| 55 | 6.0386E-24 | 1.0900E+02 | $f_o(1500)$ |
| 56 | 9.0166E-24 | 7.3000E+01 | $f_2'(1525)$ |
| 57 | 2.0896E-24 | 3.1500E+02 | $\omega(1650)$ |
| 58 | 3.9179E-24 | 1.6800E+02 | $\omega_3(1670)$ |
| 59 | 2.5414E-24 | 2.5900E+02 | $\pi_2(1670)$ |
| 60 | 4.3881E-24 | 1.5000E+02 | $\phi(1680)$ |
| 61 | 4.0883E-24 | 1.6100E+02 | $\rho_3(1690)$ |
| 62 | 2.6328E-24 | 2.5000E+02 | $\rho(1700)$ |
| 63 | 4.7015E-24 | 1.4000E+02 | $f_o(1710)$ |
| 64 | 3.1798E-24 | 2.0700E+02 | $\pi(1800)$ |
| 65 | 7.5657E-24 | 8.7000E+01 | $\phi_3(1850)$ |
| 66 | 3.2585E-24 | 2.0200E+02 | $f_2(2010)$ |
| 67 | 2.0378E-24 | 3.2300E+02 | $a_4(2040)$ |
| 68 | 2.9649E-24 | 2.2200E+02 | $f_4(2050)$ |
| 69 | 4.4175E-24 | 1.4900E+02 | $f_2(2300)$ |
| 70 | 2.0634E-24 | 3.1900E+02 | $f_2(2340)$ |
| 71 | 1.2957E-23 | 5.0800E+01 | $K^*(892)$ |
| 72 | 7.3135E-24 | 9.0000E+01 | $K_1(1270)$ |
| 73 | 3.7828E-24 | 1.7400E+02 | $K_1(1400)$ |
| 74 | 2.8371E-24 | 2.3200E+02 | $K^*(1410)$ |
| 75 | 2.2388E-24 | 2.9400E+02 | $K^*_o(1430)$ |
| 76 | 6.6824E-24 | 9.8500E+01 | $K^*_2(1430)^{\pm}$ |
| 77 | 6.0386E-24 | 1.0900E+02 | $K^*_2(1430)^{\circ}$ |
| 78 | 2.0441E-24 | 3.2200E+02 | $K^*(1680)$ |
| 79 | 3.5388E-24 | 1.8600E+02 | $K_2(1770)$ |
| 80 | 4.1397E-24 | 1.5900E+02 | $K^*_2(1780)$ |
| 81 | 2.3848E-24 | 2.7600E+02 | $K_2(1820)$ |
| 82 | 3.3243E-24 | 1.9800E+02 | $K^*_4(2045)$ |
| 83 | 3.4826E-23 | 1.8900E+01 | $D_1(2420)^\circ$ |
| 84 | 2.8618E-23 | 2.3000E+01 | $D^*_2(2460)^\circ$ |
| 85 | 2.6328E-23 | 2.5000E+01 | $D^*_2(2460)^\pm$ |
| 86 | 4.3881E-23 | 1.5000E+01 | $D_{sJ}(2573)^\pm$ |
| 87 | 1.4378E-23 | 4.5780E+01 | $B^*(5325)$ |
| 88 | 5.1423E-24 | 1.2800E+02 | $B^*_J(5732)$ |
| 89 | 6.5169E-23 | 1.0100E+01 | $\chi_{c0}(1p)$ |
| 90 | 7.2331E-22 | 9.1000E-01 | $\chi_{c1}(1p)$ |
| 91 | 3.1195E-22 | 2.1100E+00 | $\chi_{c2}(1p)$ |
| 92 | 2.7890E-23 | 2.3600E+01 | $\psi(3770)$ |
| 93 | 1.0616E-23 | 6.2000E+01 | $\psi(4040)$ |
| 94 | 8.4386E-24 | 7.8000E+01 | $\psi(4160)$ |
| 95 | 1.5307E-23 | 4.3000E+01 | $\psi(4415)$ |
| 96 | 3.2911E-23 | 2.0000E+01 | $\Upsilon_{4s}$ |
| 97 | 5.9837E-24 | 1.1000E+02 | $\Upsilon(10860)$ |
| 98 | 8.3318E-24 | 7.9000E+01 | $\Upsilon(11020)$ |
| 99 | 1.8806E-24 | 3.5000E+02 | $N(1440)P_{11}$ |
| 100 | 5.4851E-24 | 1.2000E+02 | $N(1520)D_{13}$ |
| 101 | 4.3881E-24 | 1.5000E+02 | $N(1535)S_{11}$ |
| 102 | 4.3881E-24 | 1.5000E+02 | $N(1650)S_{11}$ |
| 103 | 4.3881E-24 | 1.5000E+02 | $N(1675)D_{15}$ |
| 104 | 5.0632E-24 | 1.3000E+02 | $N(1680)F_{15}$ |
| 105 | 6.5821E-24 | 1.0000E+02 | $N(1700)D_{13}$ |
| 106 | 6.5821E-24 | 1.0000E+02 | $N(1710)P_{11}$ |
| 107 | 4.3881E-24 | 1.5000E+02 | $N(1720)P_{13}$ |
| 108 | 1.3217E-24 | 4.9800E+02 | $N(1900)P_{13}$ |
| 109 | 1.4627E-24 | 4.5000E+02 | $N(2190)G_{17}$ |
| 110 | 1.6455E-24 | 4.0000E+02 | $N(2220)H_{19}$ |
| 111 | 1.6455E-24 | 4.0000E+02 | $N(2250)G_{17}$ |
| 112 | 1.0126E-24 | 6.5000E+02 | $N(2600)I_{1,11}$ |
| 113 | 5.4851E-24 | 1.2000E+02 | $\Delta(1232)P_{33}$ |
| 114 | 1.8806E-24 | 3.5000E+02 | $\Delta(1600)P_{33}$ |
| 115 | 4.3881E-24 | 1.5000E+02 | $\Delta(1620)S_{31}$ |
| 116 | 2.1940E-24 | 3.0000E+02 | $\Delta(1700)D_{33}$ |
| 117 | 3.2911E-24 | 2.0000E+02 | $\Delta(1900)S_{31}$ |
| 118 | 1.8806E-24 | 3.5000E+02 | $\Delta(1905)F_{35}$ |
| 119 | 2.6328E-24 | 2.5000E+02 | $\Delta(1910)P_{31}$ |
| 120 | 3.2911E-24 | 2.0000E+02 | $\Delta(1920)P_{33}$ |
| 121 | 1.8806E-24 | 3.5000E+02 | $\Delta(1930)D_{35}$ |
| 122 | 2.1940E-24 | 3.0000E+02 | $\Delta(1950)P_{37}$ |
| 123 | 1.6455E-24 | 4.0000E+02 | $\Delta(2420)H_{3,11}$ |
| 124 | 1.3164E-23 | 5.0000E+01 | $\Lambda(1405)S_{01}$ |
| 125 | 4.2193E-23 | 1.5600E+01 | $\Lambda(1520)D_{03}$ |

```
126    4.3881E-24    1.5000E+02    Λ(1600)P_{01}
127    1.8806E-23    3.5000E+01    Λ(1670)S_{01}
128    1.0970E-23    6.0000E+01    Λ(1690)D_{03}
129    2.1940E-24    3.0000E+02    Λ(1800)S_{01}
130    4.3881E-24    1.5000E+02    Λ(1810)P_{01}
131    8.2276E-24    8.0000E+01    Λ(1820)F_{05}
132    6.9285E-24    9.5000E+01    Λ(1830)D_{05}
133    6.5821E-24    1.0000E+02    Λ(1890)P_{03}
134    3.2911E-24    2.0000E+02    Λ(2100)G_{07}
135    3.2911E-24    2.0000E+02    Λ(2110)F_{05}
136    4.3881E-24    1.5000E+02    Λ(2350)H_{09}
137    1.8284E-23    3.6000E+01    Σ(1385)P_{13}
138    6.5821E-24    1.0000E+02    Σ(1660)P_{11}
139    1.0970E-23    6.0000E+01    Σ(1670)D_{13}
140    7.3135E-24    9.0000E+01    Σ(1750)S_{11}
141    5.4851E-24    1.2000E+02    Σ(1775)D_{15}
142    5.4851E-24    1.2000E+02    Σ(1915)F_{15}
143    2.9919E-24    2.2000E+02    Σ(1940)D_{13}
144    3.6567E-24    1.8000E+02    Σ(2030)F_{17}
145    6.5821E-24    1.0000E+02    Σ(2250)
146    7.2331E-23    9.1000E+00    Ξ(1530)°P_{13}
147    6.6486E-23    9.9000E+00    Ξ(1530)⁻P_{13}
148    2.7425E-23    2.4000E+01    Ξ(1820)D_{13}
149    1.0970E-23    6.0000E+01    Ξ(1900)
150    3.2911E-23    2.0000E+01    Ξ(2030)
151    1.1967E-23    5.5000E+01    Ω(2250)
152    1.8284E-22    3.6000E+00    Λ_{c}(2593)⁺
153    3.2911E-22    2.0000E+00    Σ_{c}(2455)⁺⁺
154    4.1138E-22    1.6000E+00    Σ_{c}(2455)°
155    3.6567E-23    1.8000E+01    Σ_{c}(2520)⁺⁺
156    5.0632E-23    1.3000E+01    Σ_{c}(2520)°

Not included in analysis

157    1.5487E-24    4.2500E+02    a_{1}(1230)
158    1.6455E-24    4.0000E+02    π(1300)
159    1.8806E-24    3.5000E+02    f_{o}(1370)
160    2.1940E-24    3.0000E+02    π_{1}(1400)
161    1.2906E-23    5.1000E+01    η(1405)
162    7.5657E-24    8.7000E+01    η(1475)
163    2.1097E-24    3.1200E+02    π_{1}(1600)
164    3.6365E-24    1.8100E+02    η_{2}(1645)
165    1.3857e-24    4.7500E+02    f_{2}(1950)
```

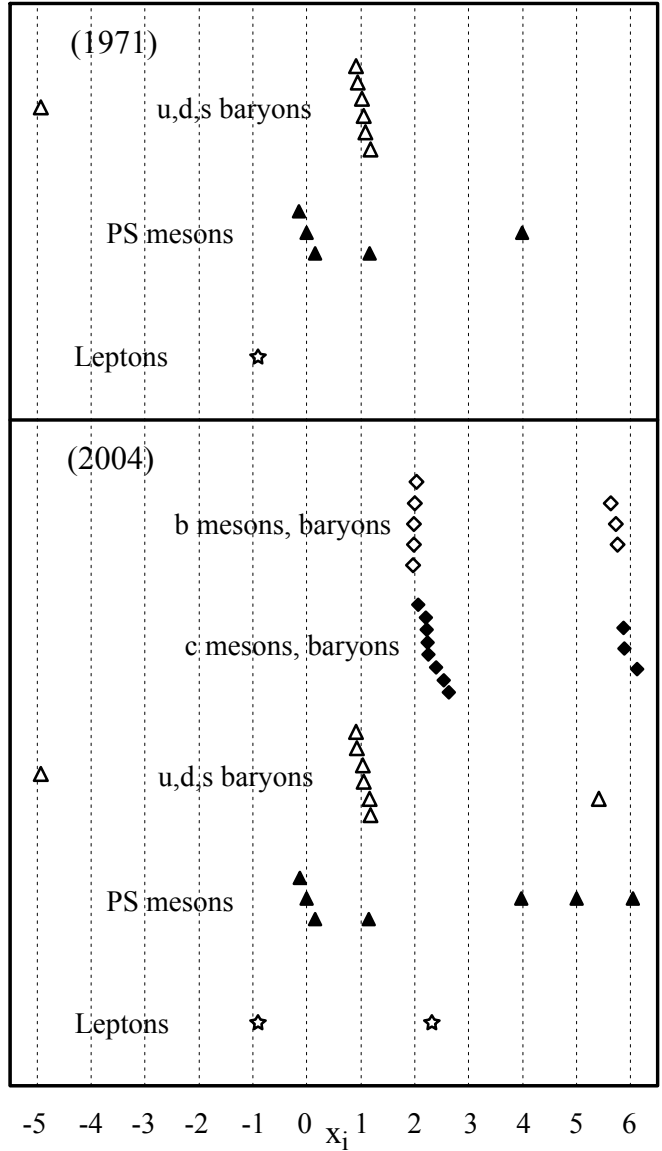

Figure 1. The threshold elementary particle lifetimes, plotted as logarithms $x_i$ to the base $\alpha$, using $\pi^{\pm}$ as the reference lifetime (Eq. 4). The top figure shows the lifetimes as they were reported in 1971 [11], and the bottom figure shows them in 2004 [8]. The lifetime $\alpha$-spacing was already apparent in 1971, and it was progressively expanded [13, 14] as higher-mass particle states were discovered with the aid of higher-energy accelerators.

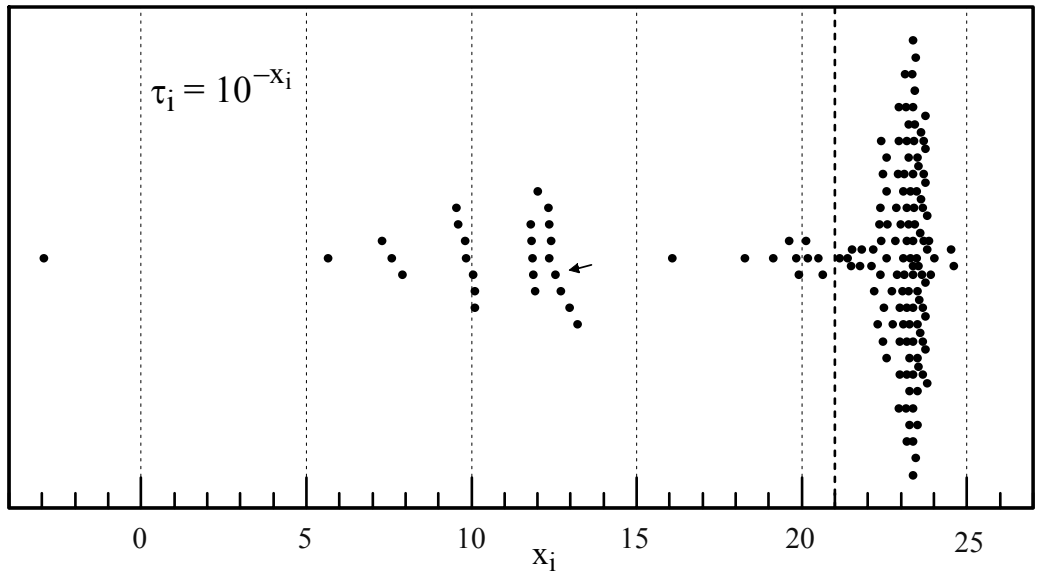

Figure 2. The 156 well-measured elementary particle lifetimes $\tau_i$ listed in RPP2004 [8], plotted as logarithms $-x_i$ to the base 10 along the abscissa. The dashed line at $\tau = 10^{-21}$ sec (1 zeptosec) divides these lifetimes into 36 long-lived threshold states and 120 short-lived excited states. The long-lived states occur in discrete clusters, whereas the short-lived states occur with essentially a continuum of values. The arrow points to the tau lepton.

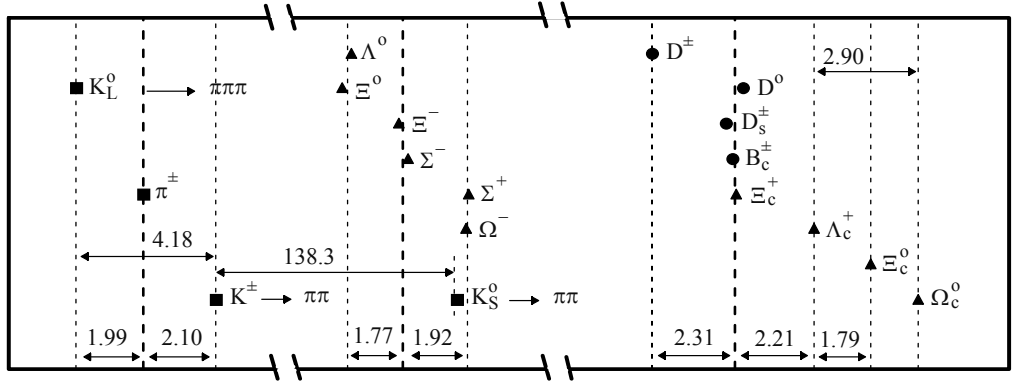

Figure 3. Lifetime groups from Fig. 2 that exhibit similar factor-of-two "hyperfine" (HF) lifetime ratios. The plot here is an expanded portion of the plot in Fig. 2, but with the distances between the groups shortened. The heavy vertical lines represent the "central" lifetimes in these groups, and the light lines indicate the experimental HF deviations (averaged over subgroups). Triangles represent fermions, and squares and circles represent bosons. The *left-group* ($\pi$, K) particles are pseudoscalar mesons, the *central-group* fermions (and the $K_S^o$) are *s*-quark excitations, and the *right-group* bosons and fermions are all *c*-quark excitations. Thus the approximately factor-of-two HF deviations are between closely-related particles. The HF "corrections" to these data are applied (Fig. 4) by simply multiplying or dividing by two so as to bring them close to the central value. The $K^\pm$ and $K_S^o$ mesons are related by their common $\pi\pi$ decay modes, and their measured lifetime ratio is $K_S^o / K^\pm$ = 1/138.3, which is very close to the value $\alpha \cong 1/137.0$, where $\alpha$ is the fine structure constant $\alpha = e^2/\hbar c$. A weighted average of the seven approximate factor-of-two HF ratios shown at the bottom of Fig. 3 gives the value 2.01. Thus the small deviations of these HF ratios from being exactly two seem random in nature, and probably arise from a variety of small corrections. The $\Lambda_c$ to $\Omega_c$ lifetime ratio of about three reflects the ratios of their flavored quarks, and this pair is assigned an HF correction of three.

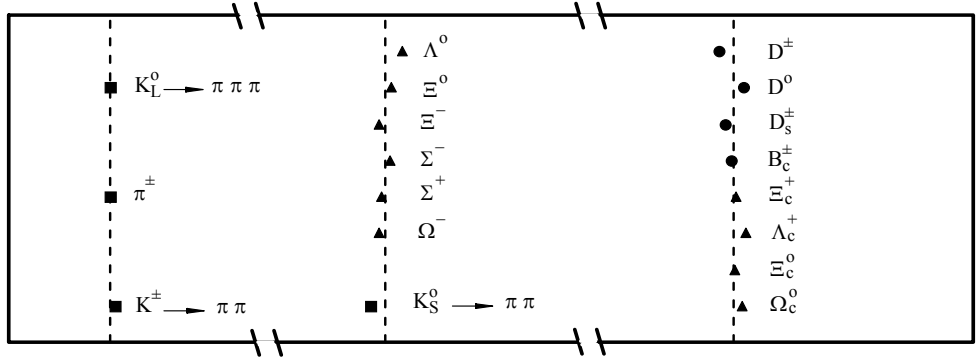

Figure 4. The lifetime groups of Fig. 3 after factor-of-two (factor-of-three for $\Lambda_c^+ - \Omega_c^o$) HF corrections have been applied. As can be seen, all of the "corrected" lifetimes in each group are in fairly close correspondence with the "central" lifetime value of the group, which is defined in Fig. 3. These HF corrections permit a more accurate analysis of the overall lifetime group spacings (see Fig. 5). The lifetime group spacing analysis is described in Sec. II E and displayed in Figs. 6 - 8.

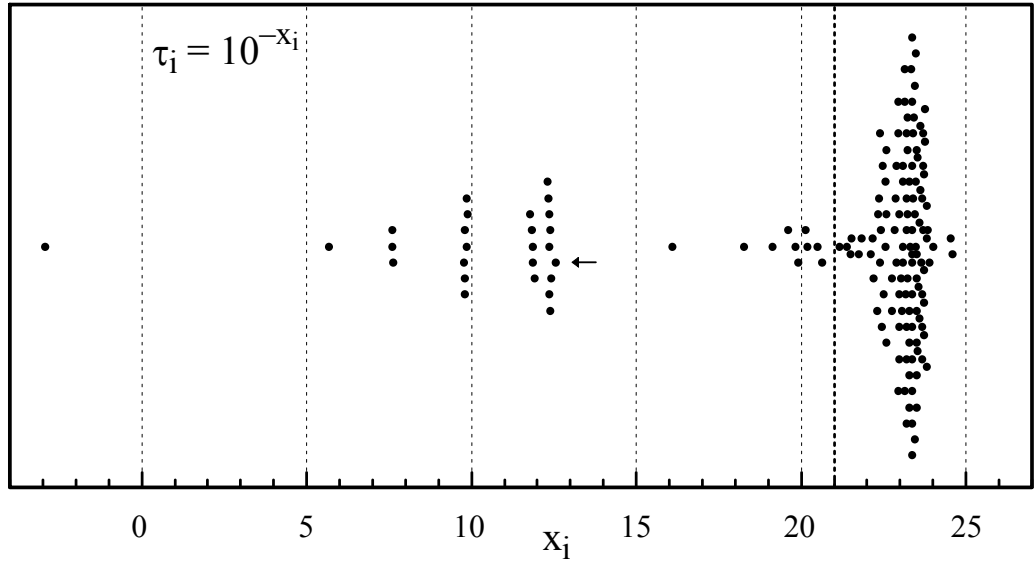

Figure 5. This is the same plot as Fig. 2, but with factor-of-two HF corrections applied, as indicated in Fig. 3 and shown in Fig. 4. This "corrected" lifetime plot clarifies the overall structure of the lifetime groups. The arrow points to the tau lepton, which was not included in Figs. 3 and 4. Numerical analyses of the interval spacings between the long-lived lifetime groups to the left of the $\tau = 10^{-21}$ sec (1 zsec) vertical dashed line are described in Sec. II E.

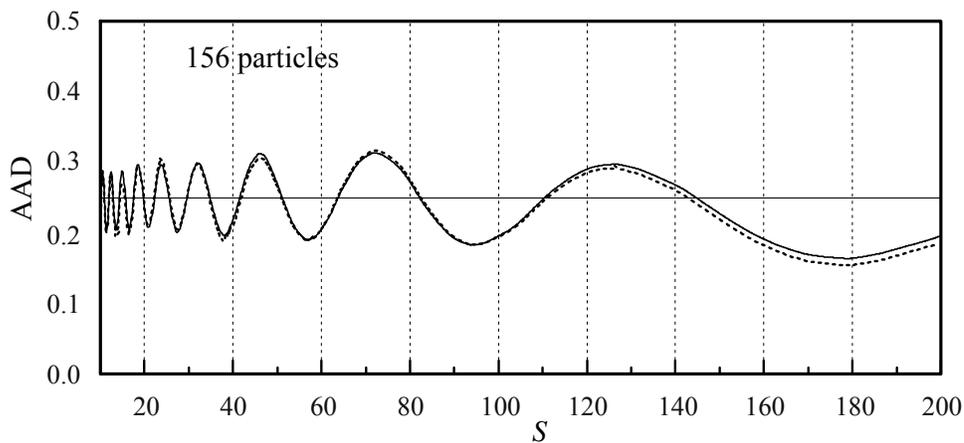

Figure 6. Plots of the *average absolute deviation* AAD vs $S$ (see Eq. 3) for the uncorrected lifetimes of Fig. 2 (solid line) and the HF-corrected lifetimes of Fig. 5 (dotted line). The horizontal line at AAD = 0.25 represents the AAD value expected for a random distribution of lifetimes. Both of these curves indicate random distributions, showing that the AAD in each case is dominated by the 120 short-lived and randomly-spaced excited states. The close agreement between the two curves suggests that the small HF corrections do not strongly affect the much larger assumed $S$ spacings.

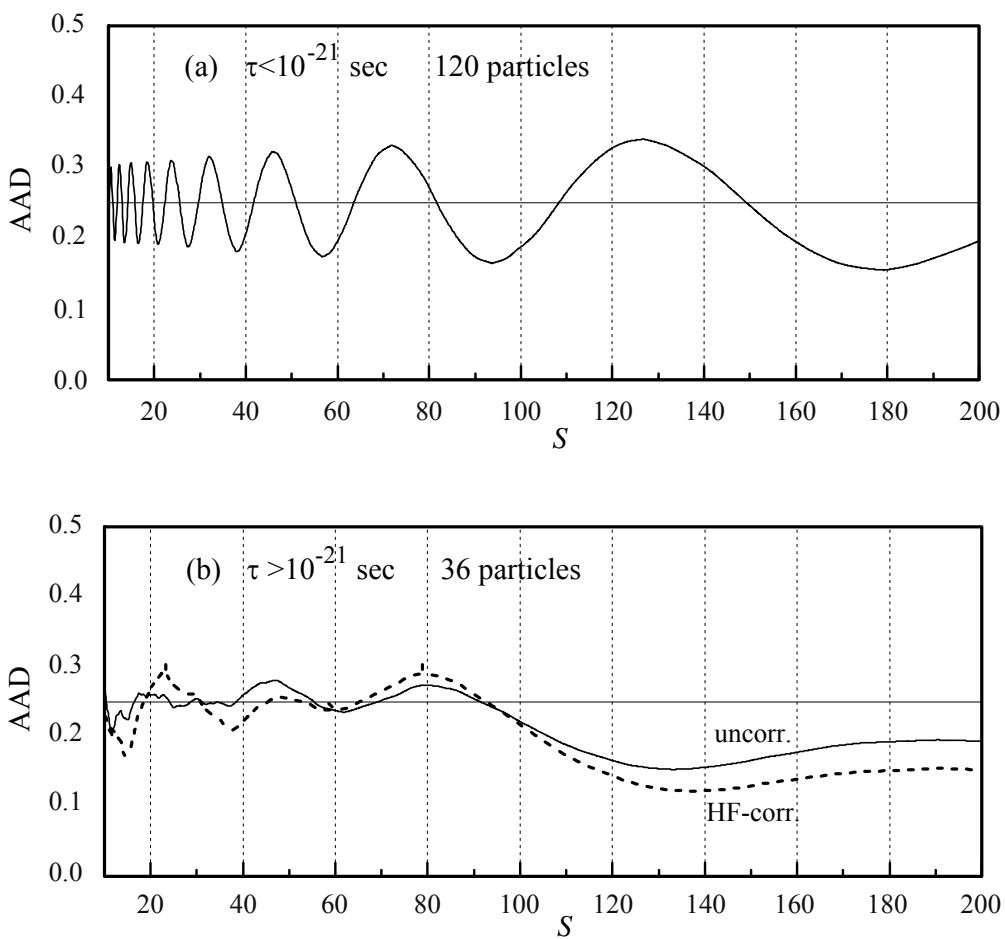

Figure 7. Plots of the AAD vs $S$ distributions for the cases where the 156 lifetimes of Fig. 6 are separated into (a) 120 short excitation-state lifetimes, and (b) 36 long threshold-state lifetimes. The short lifetimes exhibit a random distribution of $x_i$ values in Eq. (2). The long lifetimes, however, each have an AAD minimum at $S \cong 135$, where the HF-corrected lifetimes of Fig. 5 are at lower AAD values than the uncorrected lifetimes of Fig. 2. Expanded plots of these AAD minima are shown in Fig. 8.

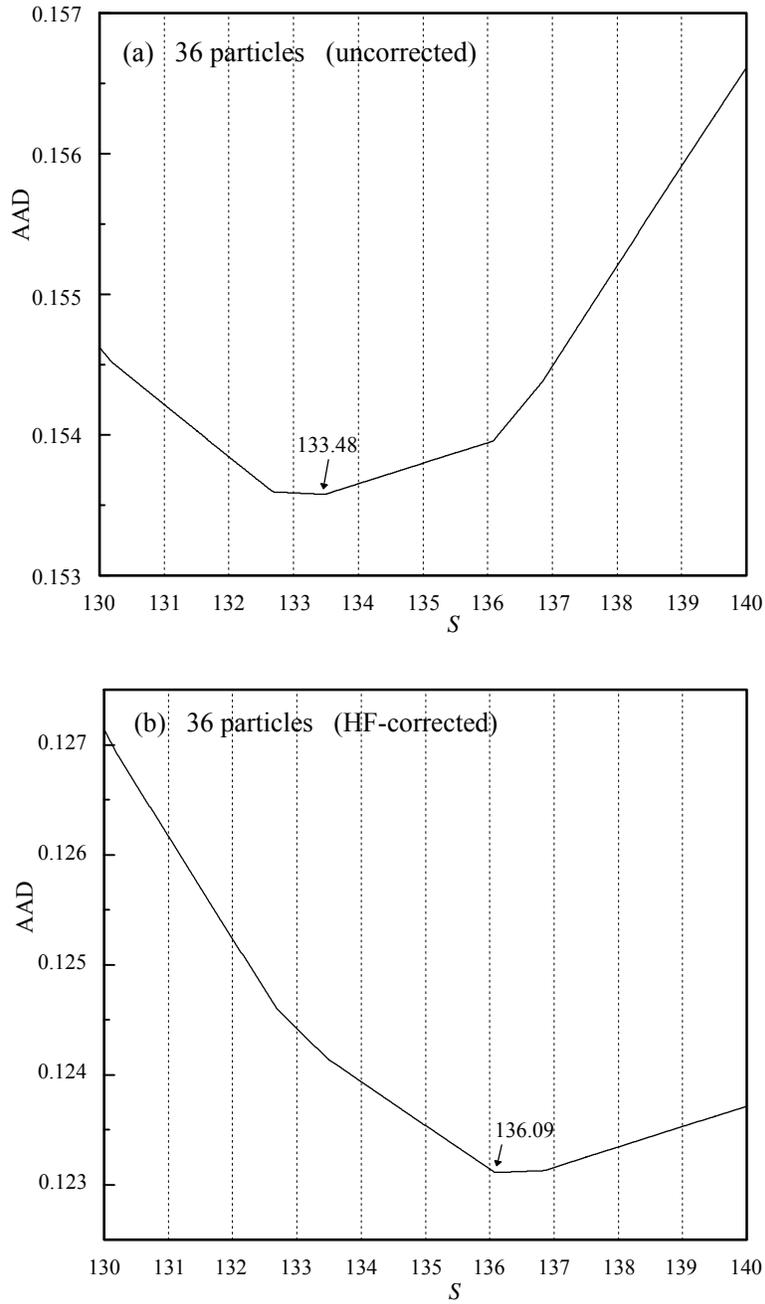

Figure 8. Detailed plots of the AAD vs $S$ minima for the 36-particle long-lifetime curves shown in Fig. 7b. In Fig. 8a (uncorrected lifetimes), the AAD minimum is 0.1536 at $S = 133.48$. In Fig. 8b (HF-corrected lifetimes), the AAD minimum is 0.1231 at $S = 136.09$. The agreement between the location of the Fig. 8b dip and the reciprocal of the fine structure constant, $\alpha^{-1}$, is accurate to 99.3%. These results demonstrate that $\alpha$ is in fact the proper scaling factor for these particle lifetimes.

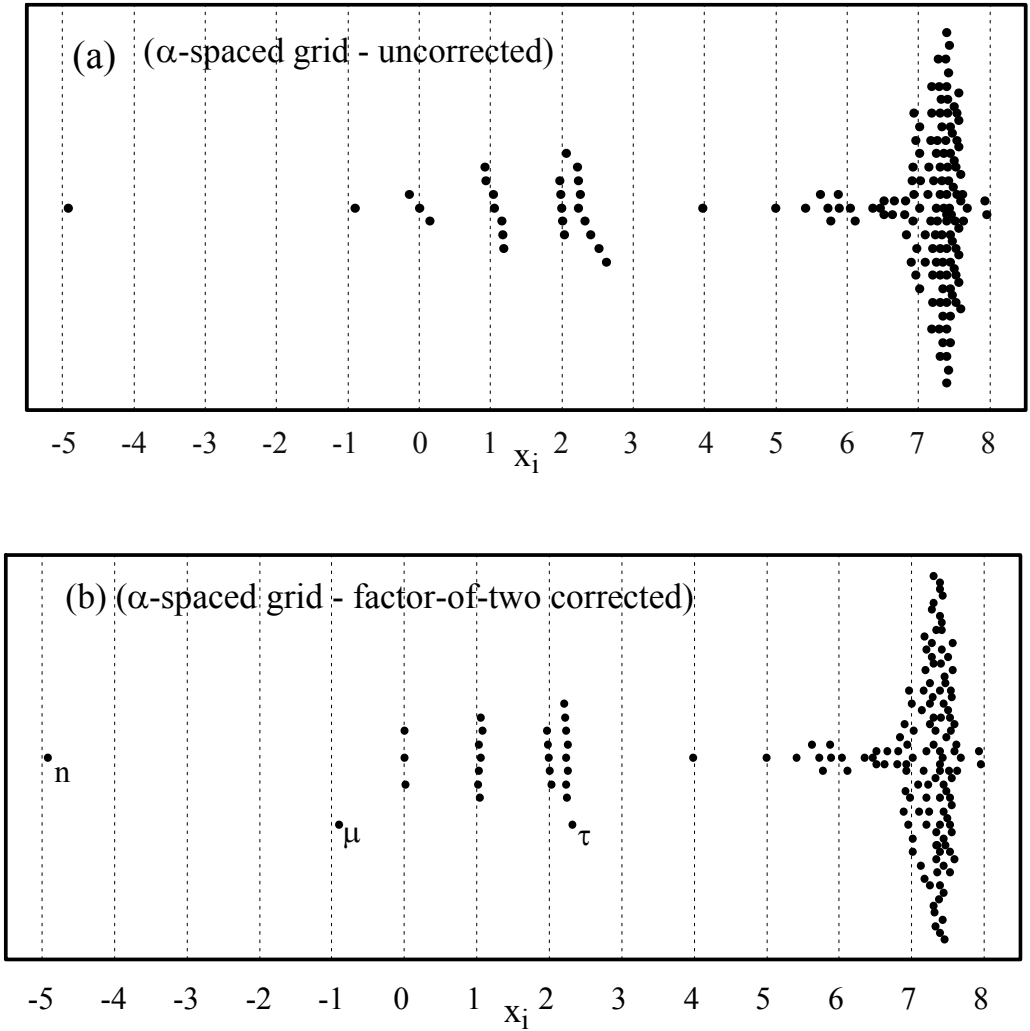

Figure 9. The elementary particle lifetime groups for (a) uncorrected and (b) factor-of-two HF-corrected data, plotted on the α-spaced grid of Eq. (4). In Fig. 9b, the leptons μ and τ are displayed downward to clarify the groupings of the hadronic states. Also, the distribution of the continuum of short-lived states along the ordinate is somewhat different in the two figures. Note the accuracy of the α-spacing over many orders of magnitude, and also its comprehensiveness: all of the well-measured lifetimes are included here.

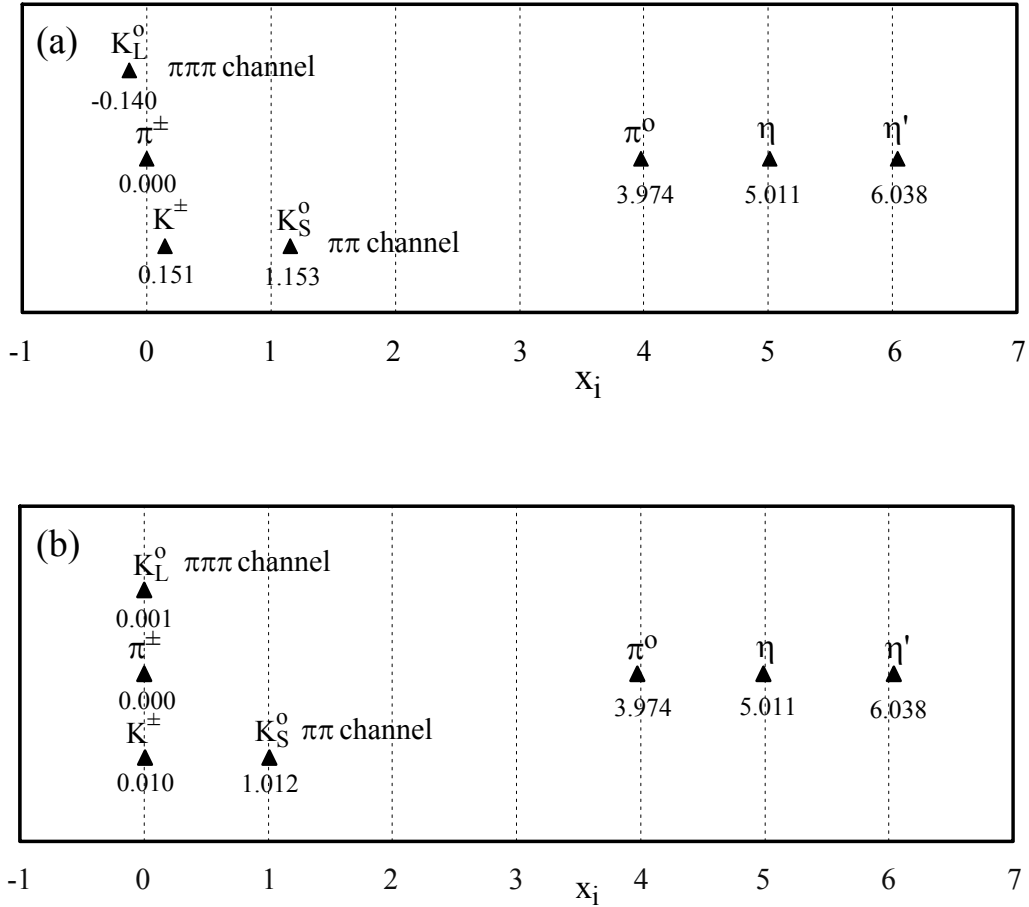

Figure 10. The pseudoscalar meson octet, shown in uncorrected (10a) and factor-of-two HF-corrected (10b) forms. The numerical values of the lifetime logarithms $x_i$ to the base $\alpha$ (Eq. 4) are shown under the data points. $\pi^{\pm}$ is the reference lifetime. The $\pi^o$ lifetime is shorter by a factor of $\alpha^4$, and the $\eta$ and $\eta'$ lifetimes are successively shorter by additional factors of $\alpha$. The lifetime of the $K_L^o$ meson, which has a $\pi\pi\pi$ decay channel, is a factor of four longer than that of the $K^{\pm}$ meson, which has a $\pi\pi$ decay channel. The closely-related $K_S^o$ meson, which also has a $\pi\pi$ decay channel, is a factor of almost exactly $\alpha$ shorter in lifetime than the $K^{\pm}$.

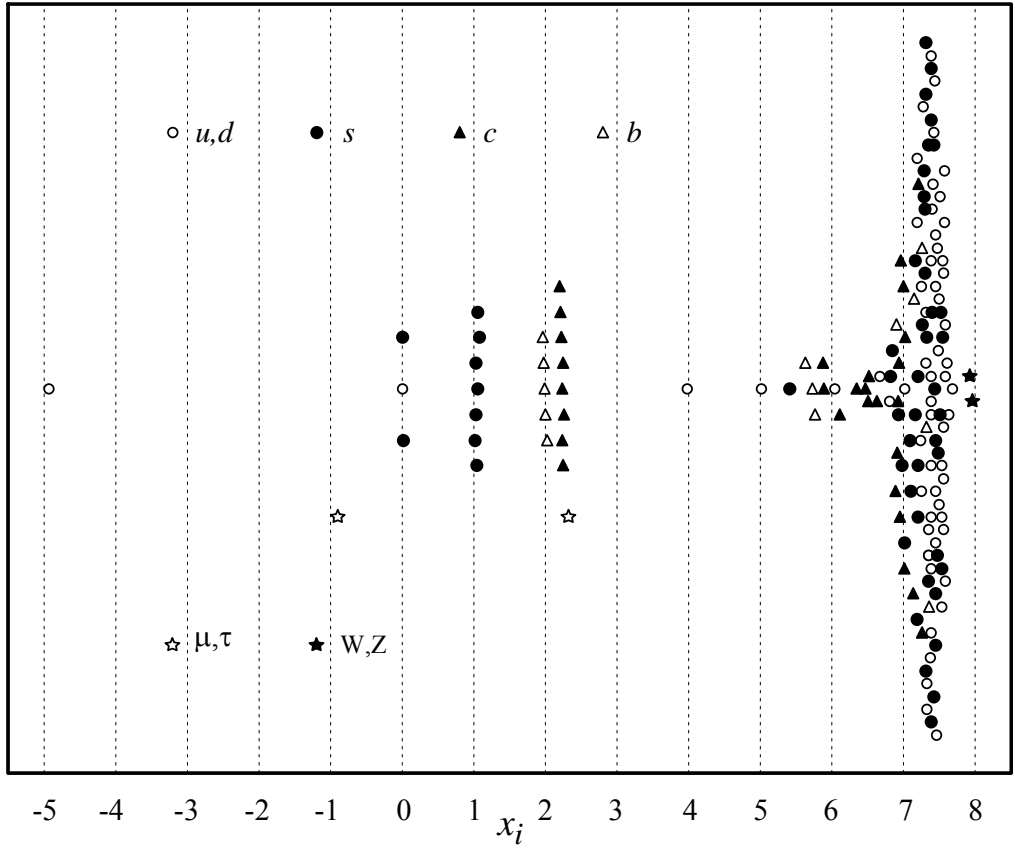

Figure 11. The HF-corrected lifetimes of Fig. 9b, plotted in powers of $\alpha$ with $\pi^{\pm}$ as the reference lifetime, and shown sorted into their dominant-quark content. Each $\alpha$-spaced group corresponds to a single type of quark. This lifetime display establishes the physical significance of the Standard Model quarks, independently of their isotopic spin systematics. The factor-of-three splitting between the unpaired *b*-quark and *c*-quark excitations near $x_i = 2$ (see Fig. 14) is reflected in their factor-of-three mass ratio (Fig. 20).

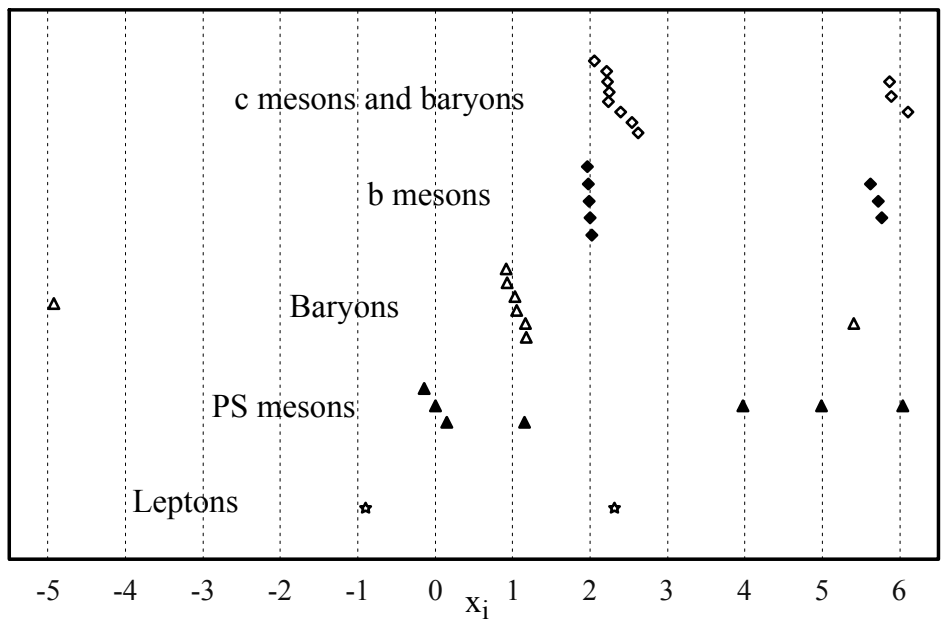

Figure 12. The 36 uncorrected long-lifetime particles of Fig. 9a, shown sorted out into family types. Each very-long-lifetime meson and baryon lifetime group has a corresponding medium-lifetime group that is displaced by about a factor of $\alpha^4$ in lifetime. The very-long-lifetime particles ($x_i < 3$) have decays triggered by single quarks. The medium-lifetime particles ($x_i > 3$) have decays involving pairs of quarks, plus some radiative decays for neutral decay modes. These quark families are described in more detail in Fig. 13.

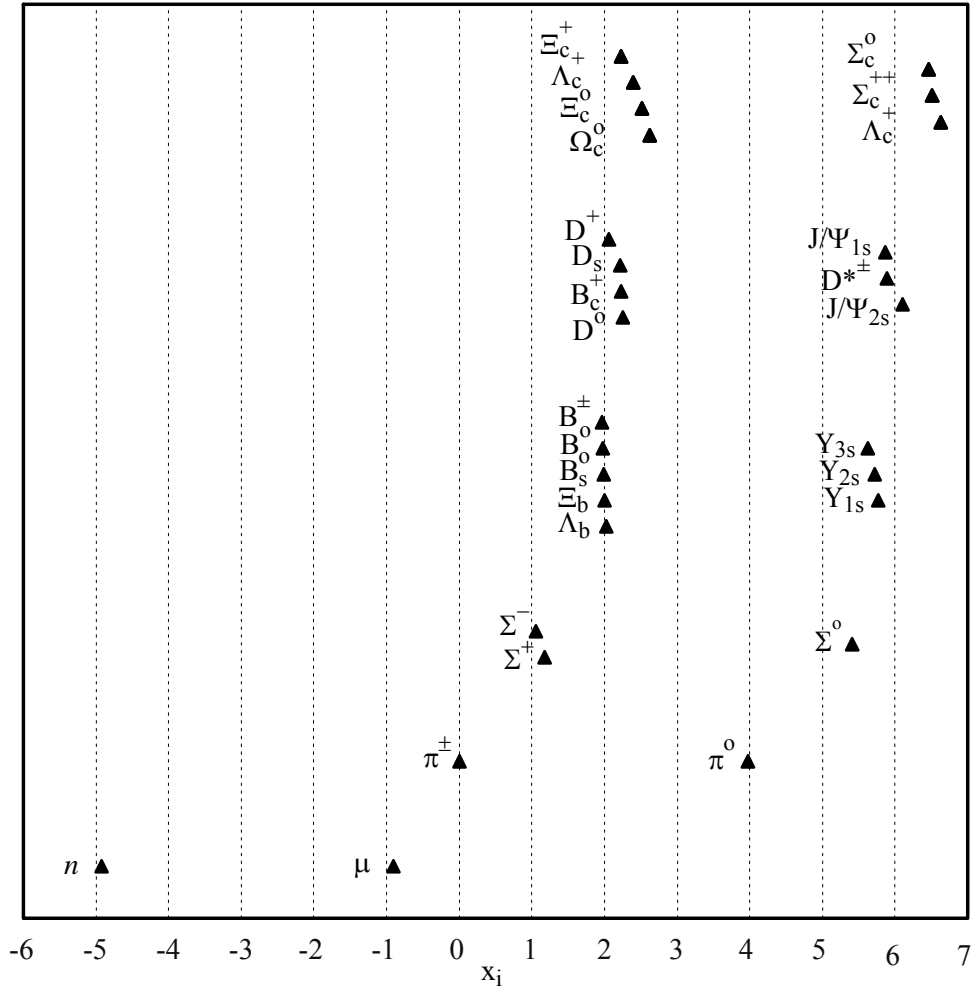

Figure 13. Characteristic $\alpha^4$ lifetime separation intervals among related long-lived ($\tau > 10^{-21}$ sec) elementary particles. The particles with lifetime logarithms $x_i < 3$ have single-quark decays. The particles with lifetime logarithms $x_i > 3$ have either matching quark-antiquark decays or radiative decays, and their lifetimes are each approximately a factor of $\alpha^4$ shorter than the corresponding unpaired-quark decays. Thus the $\alpha^4$ scaling factor is an essential ingredient in theories of these decays. Also, the overall decay systematics of these particle families systematically shifts by one power of $\alpha$ from one group to another with increasing mass, up to the point where the $c$ and $b$ quark excitations are reached. The $c$ and $b$ quark lifetime groups are discussed in detail in Sec. II G.

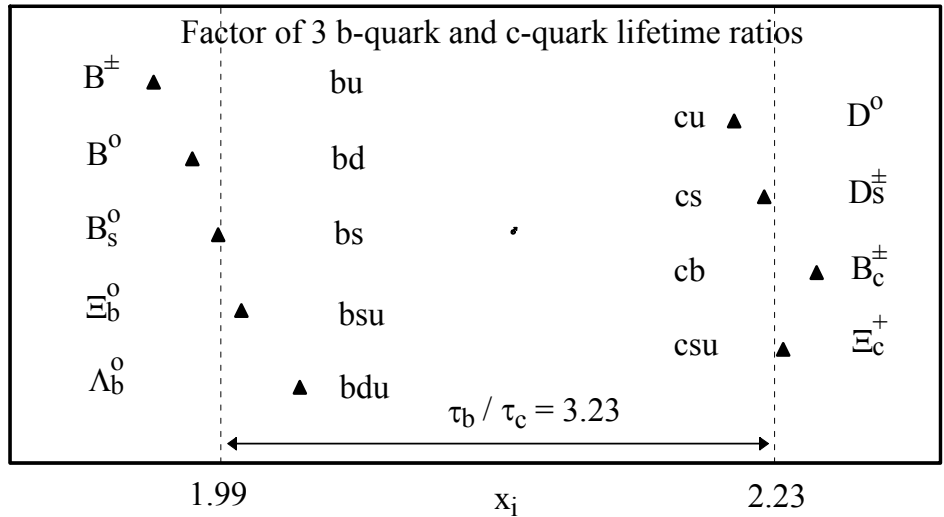

Figure 14. An expanded plot of the "central lifetime" $b$ and $c$ mesons and baryons that do not require factor-of-two HF corrections. The vertical dotted lines show the average $x_i$ value for each group (Eq. 4). As can be seen, all particles that contain a $c$ quark fall into the $x_i \cong 2.23$ group, and all particles that contain a $b$ quark but no $c$ quark fall into the $x_i \cong 1.99$ group. Thus the lifetime dominance rule for flavored quarks is $c > b > s$. The lifetime ratio of the $b$-quark and $c$-quark dotted lines is 3.225. The mass ratio of the $\Upsilon_1 \equiv b\bar{b}$ and $J/\Psi_{1S} \equiv c\bar{c}$ resonances is 3.055 (Fig. 20).

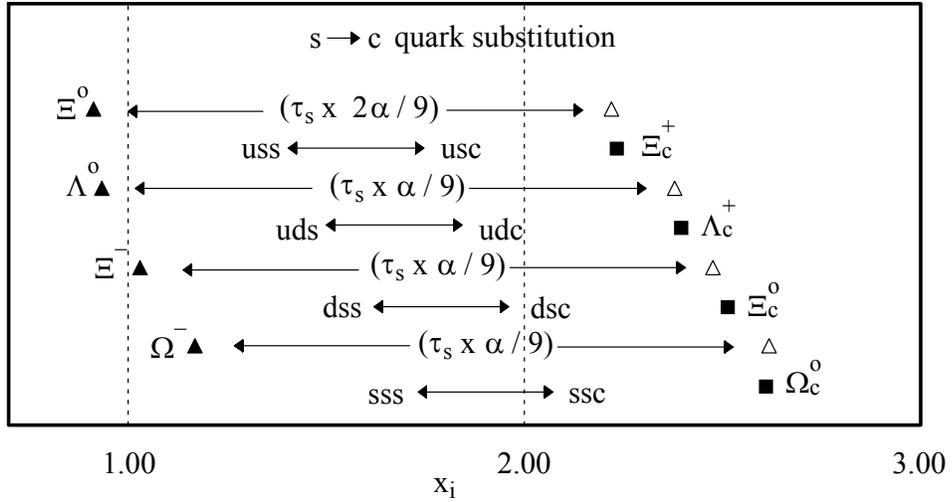

Figure 15. Comparison of the baryon lifetimes when an *s* quark is replaced by a *c* quark. The lifetimes are each shortened by a factor of $\alpha/9$, with an additional factor-of-two HF correction applied in the case of the $\Xi^o \rightarrow \Xi_c^+$ transition. The solid symbols show the experimental lifetimes, and the open triangles show the calculated lifetimes using the $\alpha/9$ scaling factor for the *s* to *c* quark substitutions. Since *b*-quark and *c*-quark lifetimes characteristically differ by a factor of 3 (Fig. 14), and since the $s \rightarrow b$ scaling is $2\alpha/3$ (Fig. 16), the "fundamental" $s \rightarrow c$ scaling here is logically $2\alpha/9$.

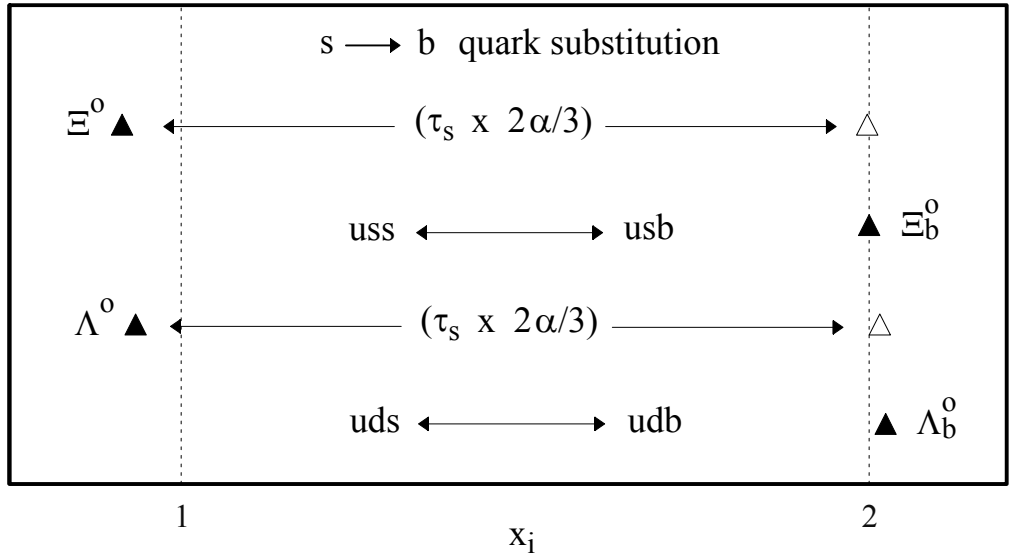

Figure 16. Comparisons of the baryon lifetimes when an *s* quark is replaced by a *b* quark. The lifetimes are each shortened by a factor of $2\alpha/3$. The solid triangles show the experimental lifetimes, and the open triangles show the calculated lifetimes using the $2\alpha/3$ scaling factor for the *s* to *b* quark substitutions.

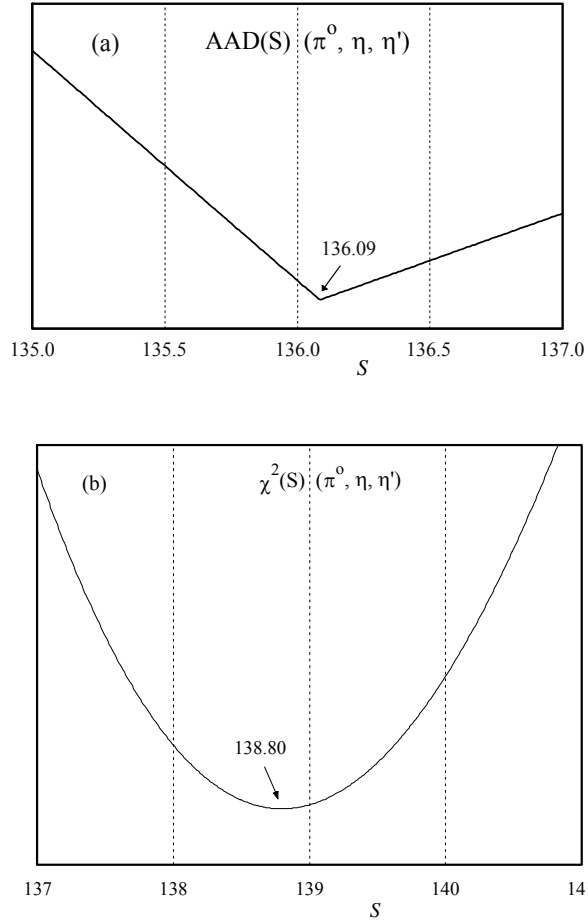

Figure 17. Determinations of $S$ by fits to the $\pi^o$, $\eta$, and $\eta'$ pseudoscalar meson lifetimes of Set A in Sec. II H. The AAD fit (Eq. 3) in Fig. 17a shows an AAD minimum at $S = 136.09$ (the same as in Fig. 8b for a different data set). The $\chi^2$ fit (Eq. 5) in Fig. 17b shows a $\chi^2$ minimum at $S = 138.80$. These two determinations of $S$ closely bracket the value $\alpha^{-1} \cong 137.04$, and they indicate that the fine structure constant $\alpha$ is the relevant scaling factor for these lifetimes. The $\pi^o$, $\eta$, and $\eta'$ lifetimes and the $\pi^\pm$ reference lifetime altogether span more than 12 orders of magnitude, and their $\alpha$-quantization is accurately maintained over this entire range of values (see Fig. 10).

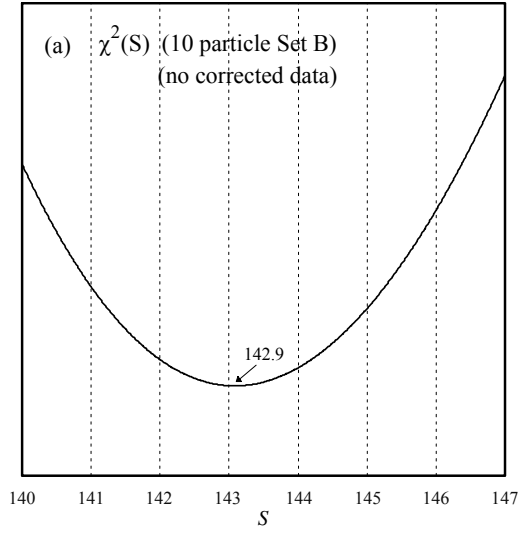

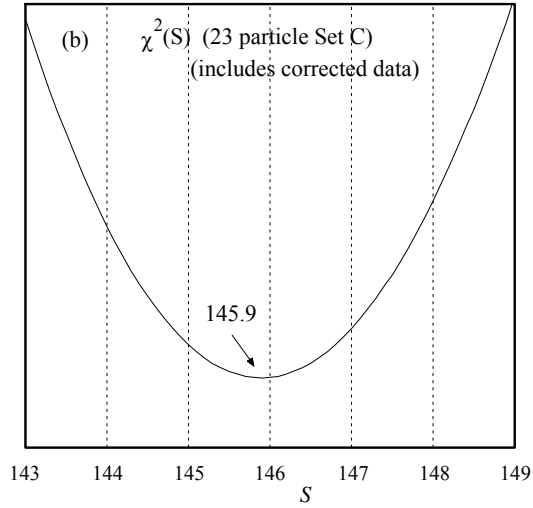

Figure 18. Determinations of S by $\chi^2$ fits to the 10-particle Set B and 23-particle Set C lifetimes, as defined in the text. The Set B particles of Fig. 18a do not involve any microscopic HF hyperfine corrections. The Set C particles of Fig. 18b require both factor-of-2 HF (Fig. 3) and factor-of-3 flavor (Fig. 14) corrections. The Set B minimum is at $S = 142.9$, and the Set C minimum is at $S = 145.9$. The upward shift of the S curve minimum for the Set C data reflects the fact that the actual "corrections" for the $b$–$c$ lifetime flavor splitting are slightly larger (see Fig. 14) than the value 3 used here. These HF and flavor "corrections" are empirical and only approximately accurate, but they serve to clarify the overall assessment of the lifetime systematics. The Set B and Set C expanded data results suggest that the global $\alpha$-quantization of these lifetimes is more-or-less independent of the much smaller HF and flavor corrections.

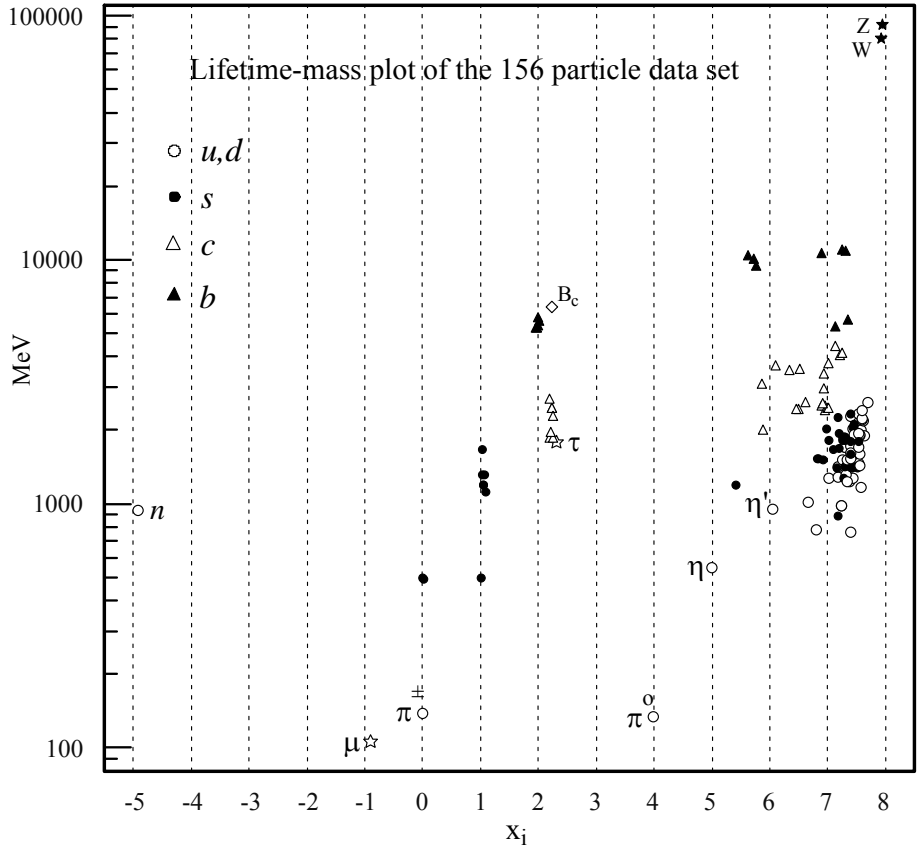

Figure 19. A plot of lifetimes versus masses for the 156 particle data set. The abscissa shows the same factor-of-two corrected lifetimes as displayed in Fig. 11. The ordinate gives the mass values of these particles. The effect of the successively increasing masses of the *s*, *c*, and *b* flavored quarks is clearly in evidence. The $B_c$ meson has the mass of the *b* quark and the lifetime of the *c* quark. The mass and lifetime of the τ lepton are comparable to the masses and lifetimes of the *c* quarks. The $\pi^{\pm}$, $\pi^o$, η and η' spin 0 bosons and the *n*, μ and τ spin 1/2 fermions are the only non-flavored long-lived ($\tau > 10^{-21}$ sec) particles in this data set. Together with the stable electron and proton, they play an important role in the search for a reciprocal mass α-quantization, as is discussed in detail in Sec IV.

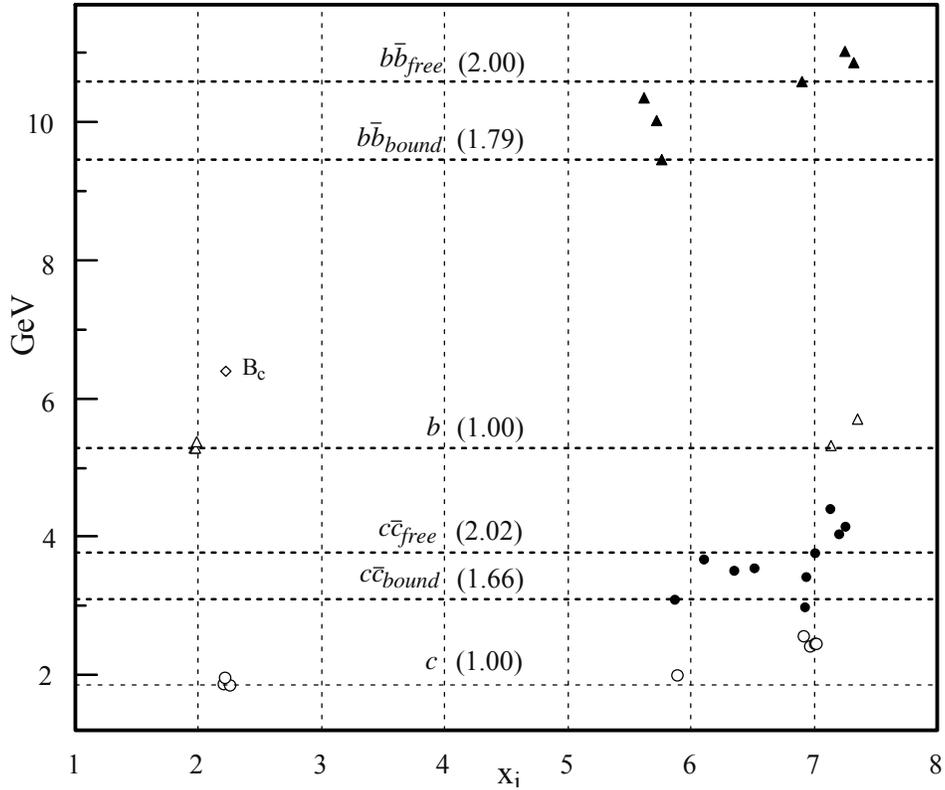

Figure 20. A lifetime versus mass plot of the *c*-quark (circles) and *b*-quark (triangles) meson resonances. The $B_c$ meson, which contains both a *b* and a *c* quark, has the large mass of the *b* quark, but the factor-of-three shorter lifetime of the *c*-quark (see Fig. 14). The unpaired-quark mesons have lifetimes which are a factor of $\alpha^{-4}$ longer than those of their paired-quark counterparts. They also have mass values which are about half of the paired-quark values, which suggests a constituent-quark mass structure. The dotted lines indicate the threshold mass values at which single quark, bound paired-quark, and unbound paired-quark states first appear. The numerical values in parentheses above the dotted lines indicate the relative mass ratios of these thresholds. The mass values of the *b*-quark states are systematically a factor of three larger than those of the *c*-quark states. Specifically, the mass ratio of the thresholds for producing single *b* and *c* quarks is 2.83, the $b\bar{b}$ to $c\bar{c}$ threshold mass ratio is 3.05, and the unbound $b\bar{b}$ to $c\bar{c}$ threshold mass ratio is 2.81.

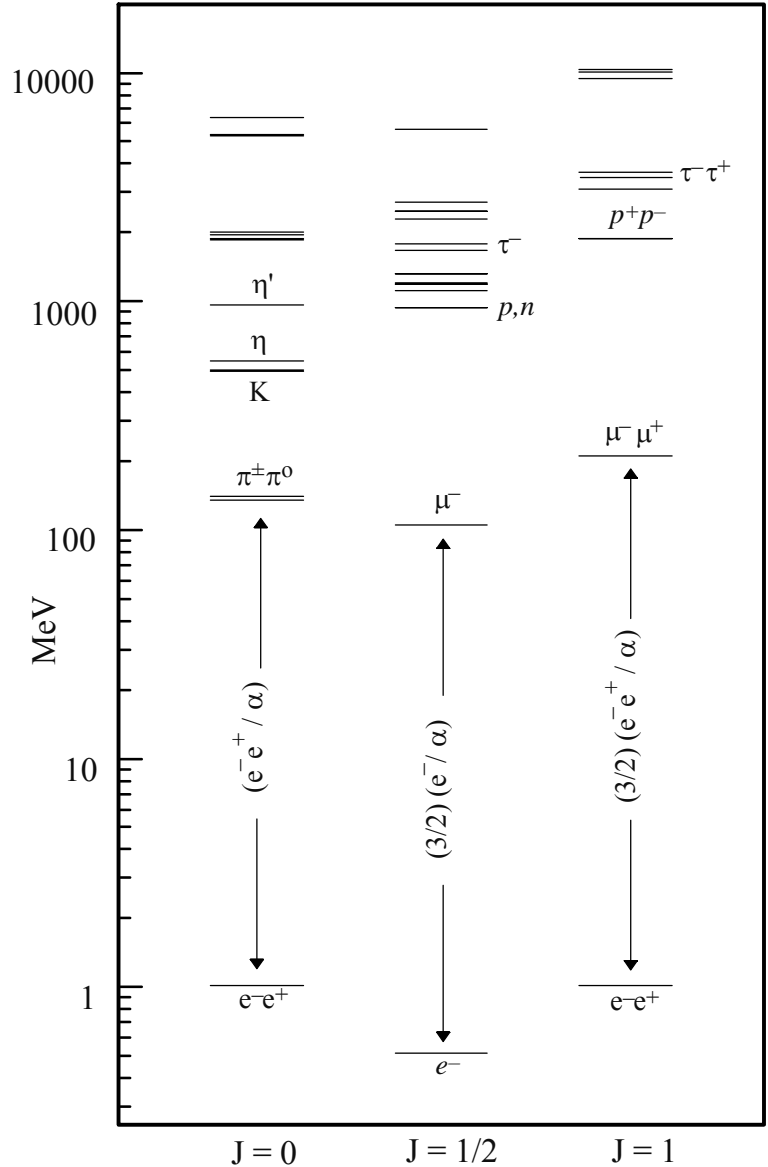

Figure 21. The mass values of all massive elementary particles with half-lives longer than $10^{-21}$ sec, shown arranged in spin-channel excitation towers. Also indicated are the quantized "α-leap" mass intervals above the electron "ground states", where α serves as the coupling constant between leptonic and hadronic masses. The spin 0 bosonic α-leap in mass is $e\bar{e}/\alpha$, and the spin 1 fermionic α-leap is $(3/2)(e\bar{e}/\alpha)$, a factor of 3/2 larger. The *unflavored* particles included here are $e^{\pm}$, $\pi^{\pm}$, $\pi^{0}$, $\eta$, $\eta'$, $\mu^{\pm}$, $\tau^{\pm}$, $p^{+}$, $n$, as labeled in the figure. These particles are sufficient to delineate a dominant excitation quantum X, as we demonstrate in Fig. 22.

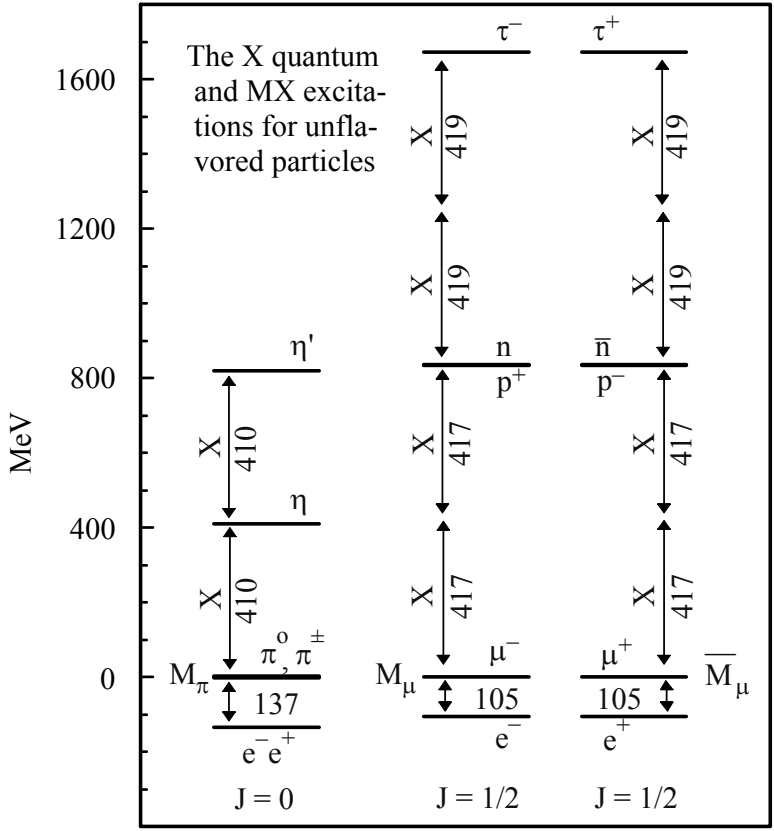

Figure 22. Mass levels for the *unflavored* hadron and lepton states of Fig. 21, plotted as excitation towers above the platforms $M_\pi$ (J = 0), $M_\mu$ (J = 1/2) and $\bar{M}_\mu$ (J = 1/2) (Eqs. 14 and 17), where the J = 1 platform $M_{\mu\mu}$ (Eq. 15a) is shown here divided into its particle and antiparticle channels. The excitation process travels in tandem up the $M_\mu$ and $\bar{M}_\mu$ channels. This mass diagram illustrates the universal nature of the excitation quantum X (Eq. 18) for these threshold states. The experimental values for X are slightly smaller in the $M_\pi$ excitation tower than in the $M_\mu$ towers because the π, η, η' bosons have small hadronic binding energies (Fig. 26) that do not occur in the separated μ, p, τ fermion channels. In this two-stage $e^-e^+ \to M \to M+nX$ excitation process, an electron α-leap to the platform M is followed by several X excitations This MX systematics gives a parameter-free calculation of the proton-to-electron mass ratio. Without considering both leptons and hadrons together, this MX excitation mechanism would be difficult to unravel.

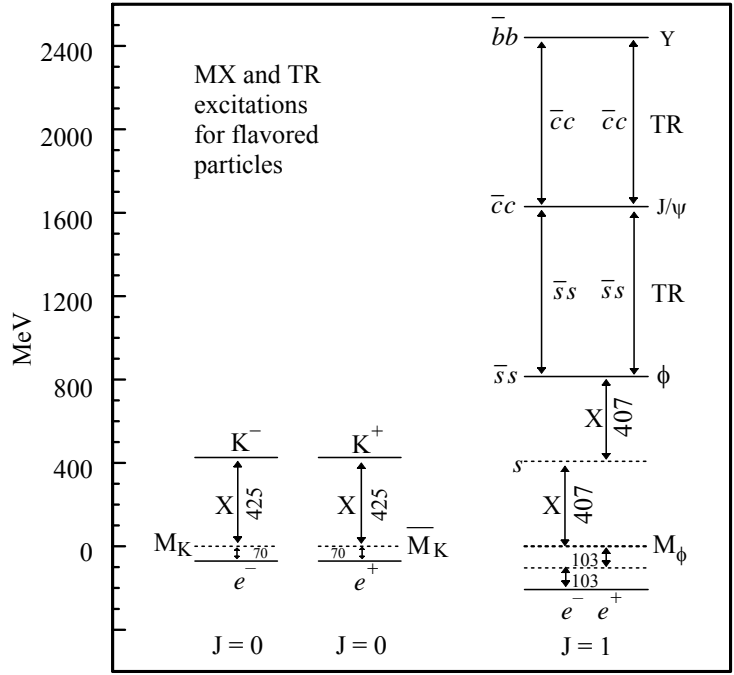

Figure 23. Mass values for the *flavored* hadron threshold states of Fig. 21, plotted as excitation towers above the platforms $M_K$ (J = 0), $\overline{M}_K$ (J = 0) and $M_\phi$ (J = 1) (Eqs. 16 and 15), where the J = 0 $M_\pi$ platform has been divided here into J = 0 particle and antiparticle channels. The kaon, a single-X excitation that must be produced in pairs, is the only member of the $M_K$ excitation tower. It occurs in different forms as $K^o_L$, $K^\pm$ and $K^o_S$. The *s* quark is also a single-X excitation (produced in pairs), and the $s\bar{s} = \phi(1020)$ vector meson is the only hadronically-bound X level of the $M_\phi$ tower. However, successive mass triplings (TR) of the $s\bar{s}$ level generate the $c\bar{c} = J/\psi$ and $b\bar{b} = \Upsilon$ threshold states above the $s\bar{s}$ (see Fig. 28). The mass-tripled levels are not plotted to scale.

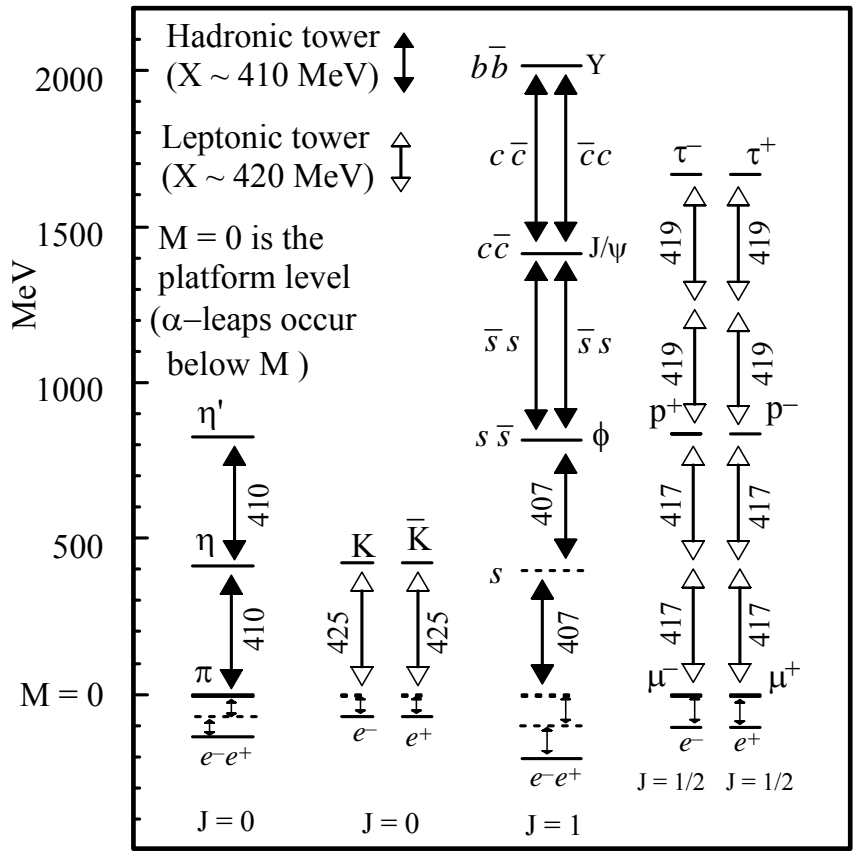

Figure 24. The α-quantized MX and TR excitation towers for threshold-state particle production. The initial $m_b$ and $m_f$ α-leaps (below M = 0) of Eqs. (10 - 13) to the platforms M are followed by unbound (open arrows) and hadronically bound (solid arrows) X excitations. Also shown are the hadronically bound vector meson TR mass triplings (solid arrows), which are not plotted to scale. The MX excitations shown here, which combine Figs. 22 and 23, generate the MX octet particles of Tables 1 and 2.

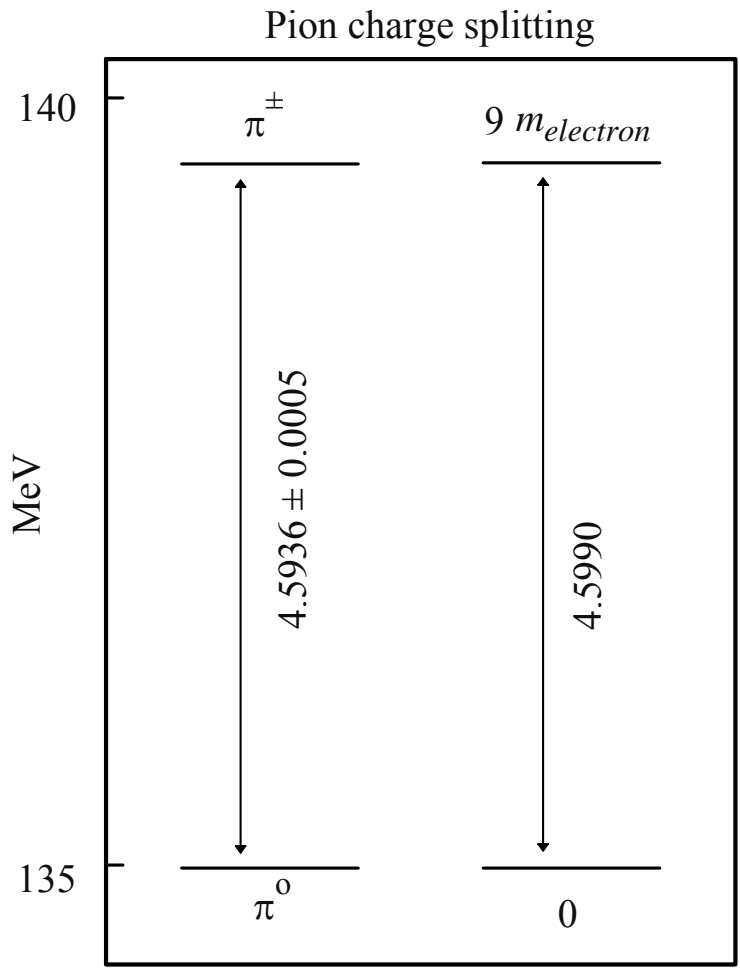

Figure 25. Charge splitting in the pi mesons. The $\pi^{\pm} - \pi^{o}$ mass difference of 3.3%, the largest of any isotopic spin multiplet except the $Z^o$ and $W^{\pm}$, is accurately equal to nine electron masses. This indicates a charge-splitting in the $m_b$ basis states of Eq. (7) that make up the pion. But this charge-splitting must also reproduce the $K^o$ - $K^{\pm}$ mass difference, which has a comparable magnitude and the opposite sign. Charge-splitting is not included in the mass formulae of the present paper.

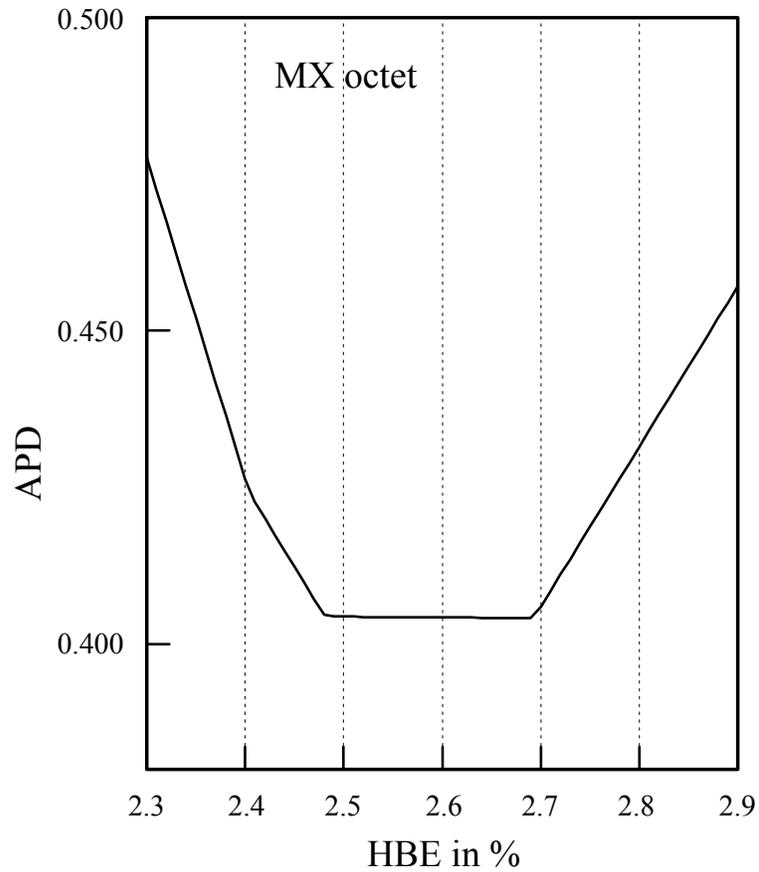

Figure 26. The average *absolute percent deviation* (APD) between the calculated and experimental masses of the MX octet of Table 2, plotted against the *hadronic binding energy* (HBE) used in the calculations. The broad minimum at HBE ≅ 2.5% - 2.7% is in general agreement with other determinations of HBE discussed in Sec. IV J. The discontinuous slope of the APD curve is due to the small number of HBE-dependent particles in the calculation. The average APD value of 0.4% for the eight threshold particles of the MX octet (only four of which depend on the HBE) constitutes an accurate fit to these mass values, especially since this fit involves no adjustable parameters except the HBE.

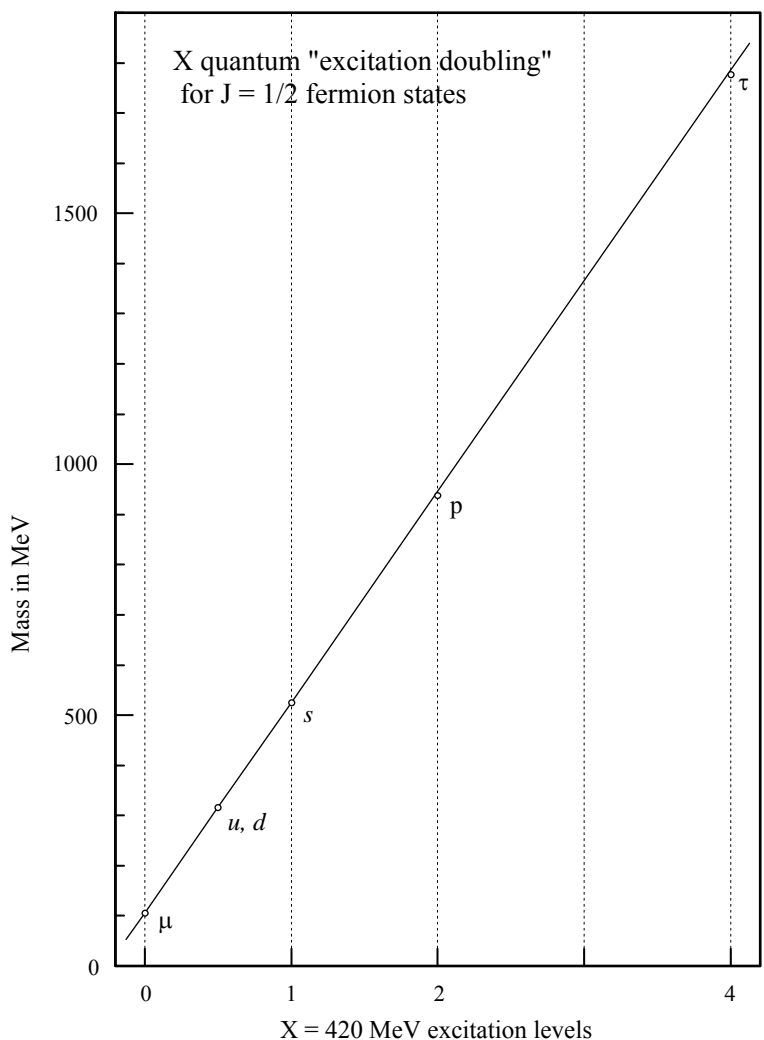

Figure 27. X-quanta "excitation doubling" for the spin 1/2 particles on the $M_\mu$ excitation platform. The straight line gives the theoretical X-quantum excitation mass values for the $M_\mu$ platform, and the circles give the experimental or theoretical masses for the occupied levels. These occupied levels are: (0) μ lepton; (1/2 X) $u,d$ quarks (not directly produced); (1X) $s$ quark; (2X) $p,n$ nucleons; (4X) τ lepton. The 3X level is not filled. The masses of these fermion states are multiples of the $M_\mu$ = μ platform mass. The $u$ and $d$ quarks are not integral-X excitations, and thus are not directly created in pairs as observed resonances. An "excitation doubling" diagram for spin 1 vector mesons is shown in Fig. 28.

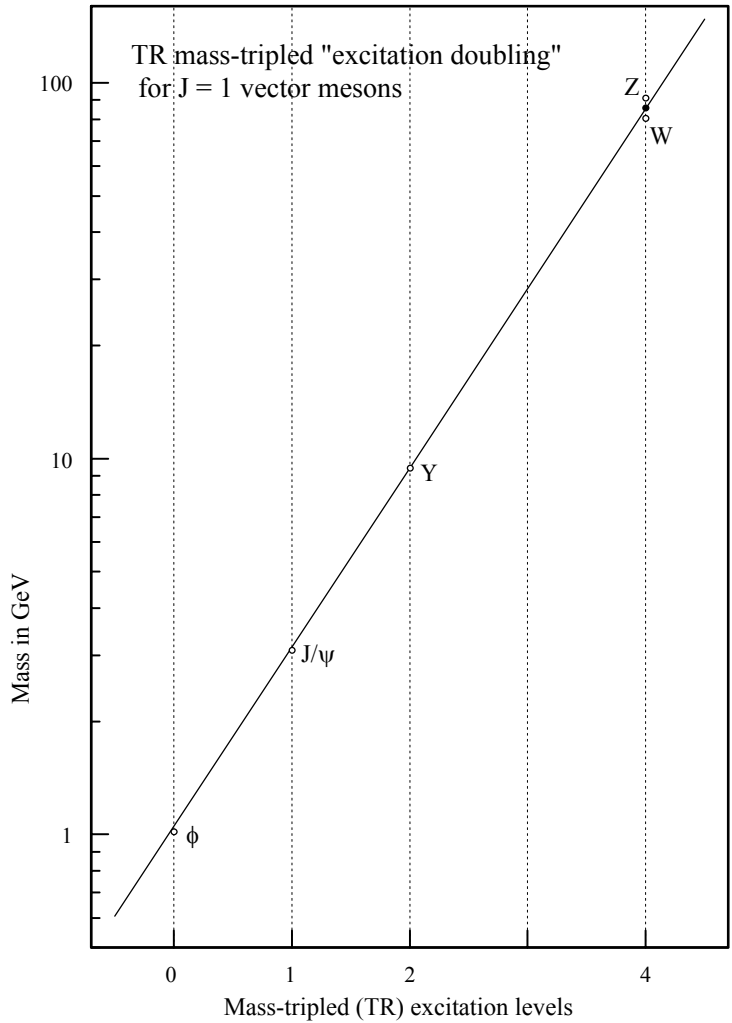

Figure 28. Mass-tripled (TR) "excitation doubling" for the spin 1 vector mesons on the $M_\phi$ excitation platform. The straight line gives the theoretical TR excitation mass values for the $M_\phi$ platform, and the circles give the experimental masses for the occupied levels. These levels, all filled with vector mesons, are: (0) $\phi = s\bar{s}$; (1TR) $J/\psi = c\bar{c}$; (2TR) $\Upsilon = b\bar{b}$; (4TR) W,Z . The open circles for the W and Z are the experimental mass values, and the filled circle is for their average mass value. The masses of these vector mesons are successive mass-triplings of the experimental $\phi$ meson mass. The 3TR level, just like the 3X level in Fig. 27, is not filled, which may be due to the operation of the "excitation-doubling" mechanism.

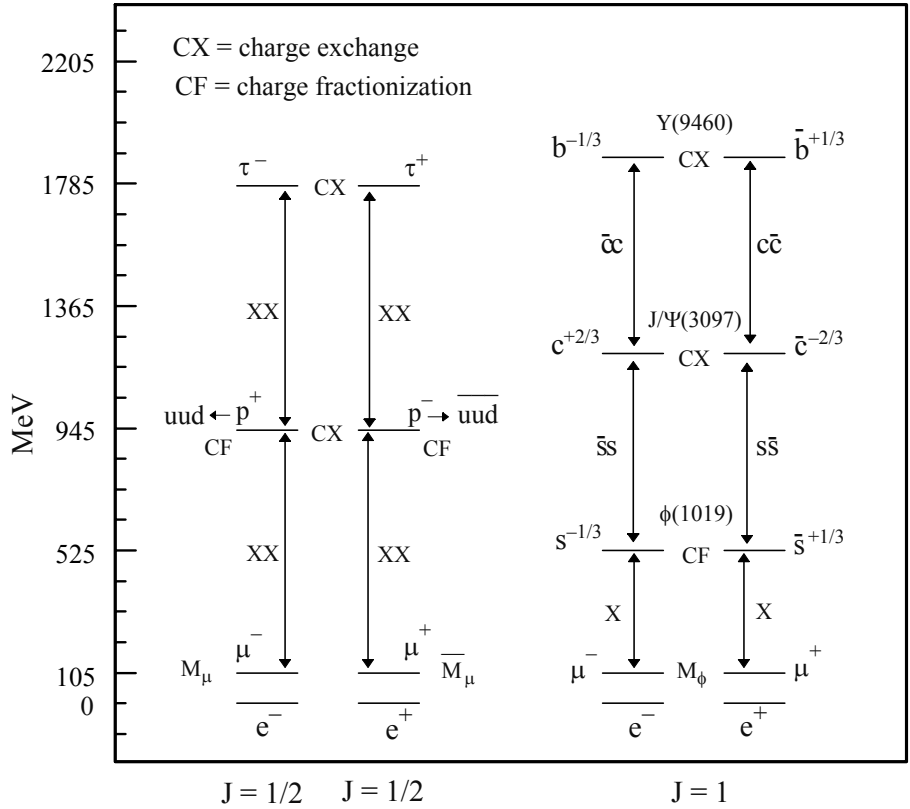

Fig. 29 Charge-exchange (CX) and charge-fractionization (CF) processes in the $M_{\mu\mu}$ and $M_\phi$ excitation towers. The first excitation level in the $M_\phi$ tower is the $\phi = s\bar{s}$ level, produced by an X excitation in each channel and a CF charge redistribution. The second and third levels (not plotted to scale) are the $J/\psi = c\bar{c}$ and $\Upsilon = b\bar{b}$ levels, produced by successive TR mass triplings. The charge on the quark changes from $-1/3\ e$ to $+2/3\ e$ and back to $-1/3\ e$ in this process, signaling a $1e$ CX charge transfer at each level. A somewhat similar excitation chain occurs in the $M_{uu}$ excitation tower, which is shown here separated into its $M_\mu$ and $\bar{M}_\mu$ particle and antiparticle channels. After the initial α-leap to the $M_\mu$ platform, which produces the muon, the first excitation level is reached by an XX channel excitation, which produces the proton. A second XX excitation produces the tau. The fact that the proton, which is in the negatively-charged particle channel, carries a positive charge, shows that a $2e$ CX charge transfer has occurred at the first level. A second $2e$ CX charge transfer gives the tau with its expected negative charge. The charge exchange in the proton is crucial, because the proton, which contains nine $m_f$ mass quanta, decomposes into three $u$ and $d$ quarks, each containing three $m_f$ mass quanta, and the $p^+$ positive electric charge has a CF charge fractionization. The positive proton is trapped in the negative-charge channel and has no decay mode back down to the negative $e^-$ ground state.

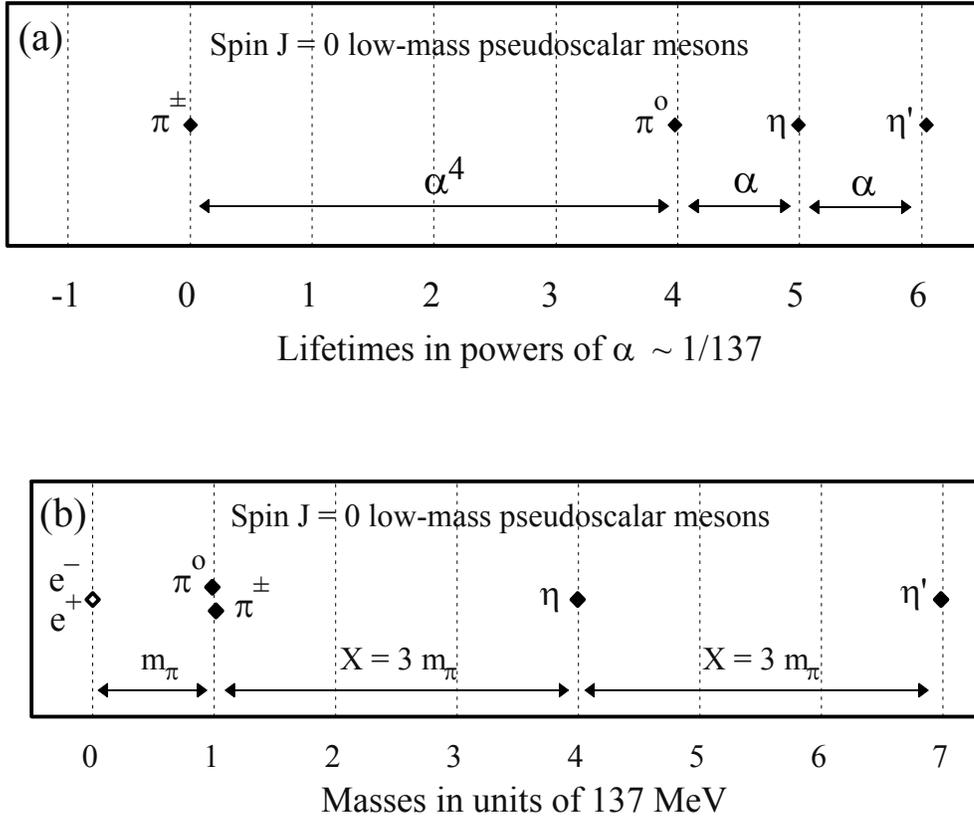

Figure 30. The reciprocal α-quantization of the unflavored pseudoscalar meson lifetimes and masses. These low-mass, spin 0, particle-antiparticle-symmetric hadrons are the simplest and thus probably the most important elementary particles to examine from the viewpoint of their compositional structure. The results displayed here could hardly be more clear-cut. Since lifetimes and mass widths are quantum mechanical observables which are tied together in Eq. (6) by Planck's constant, a *lifetime* α-dependence that is in factors of 1/137 relates to a corresponding reciprocal *mass width* α-dependence that is in factors of 137, and plausibly to a reciprocal *mass* α-dependence in factors of 137. Fig. 30a shows the very accurate 1/137 lifetime scaling with increasing mass (see Figs. 10 and 17), and Fig. 30b shows an equally accurate 137 MeV mass scaling with increasing mass. The fact that these particular mass units fit the mass α-quantization so accurately is not serendipitous, but follows from the fact that the spin 0 $e^-e^+$ electron pair, which serves as the generator for the α-quantization, has a mass of almost exactly 1 MeV. The completeness and accuracy of these results stands as a testament to the validity of the concept of reciprocal lifetime and mass α-quantization.

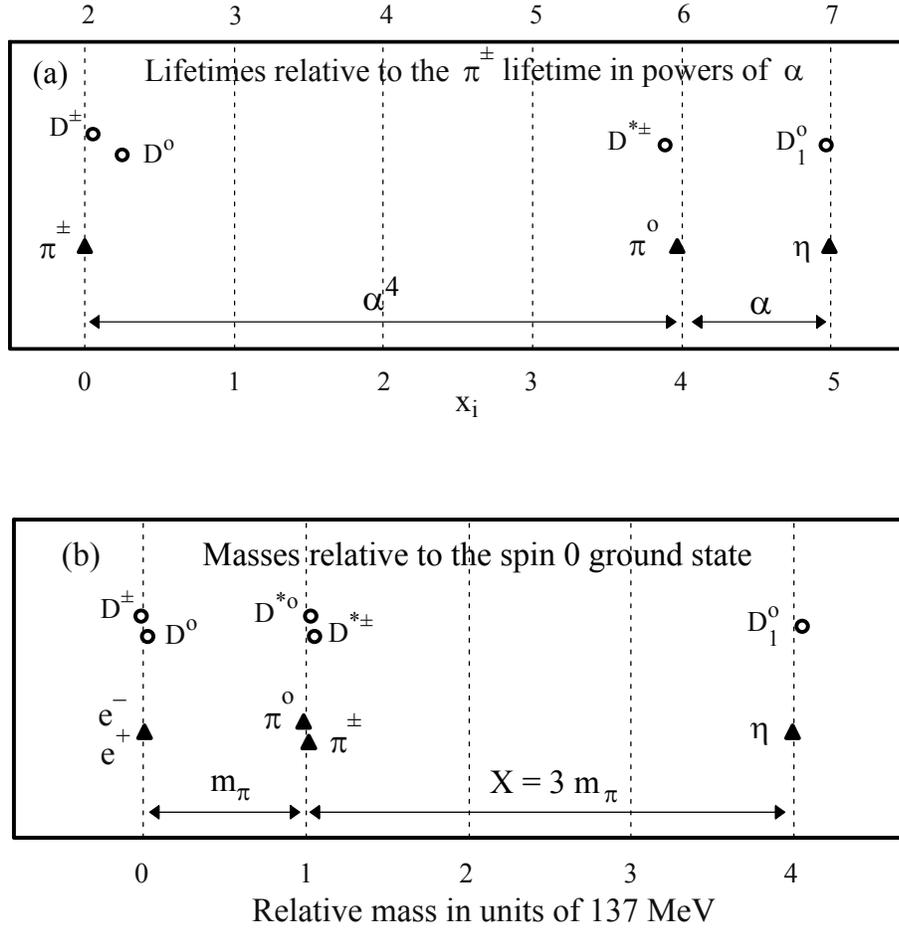

Figure 31. The reciprocal α-quantization of the flavored D meson lifetimes and masses, shown here together with the similar reciprocal α-quantization of the pseudoscalar mesons that was portrayed in Fig. 30. The D meson lifetimes use the α-spaced grid at the top of Fig. 31a, which is shifted by two powers of α with respect to the grid shown at the bottom of Fig. 31a, where $\pi^\pm$ is the reference lifetime for all of these states. The masses in Fig. 31b are plotted relative to their spin 0 "ground states". The similarity of these lifetime and mass α-quantizations illustrates the fact that reciprocal α-quantization is a phenomenon which occurs with both unflavored and flavored ground states. Fig. 32 further extends the mass α-quantization displayed in Fig. 31b.

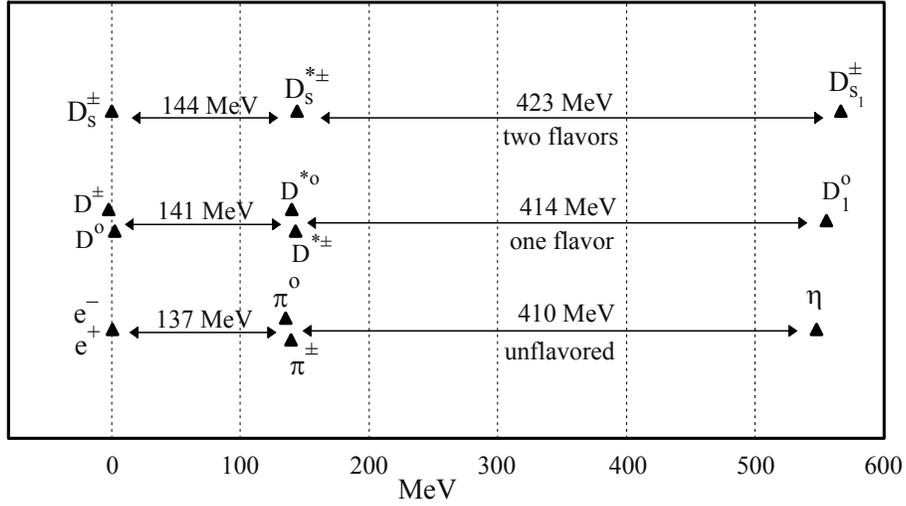

Figure 32. Excitation sequences for spin 0 unflavored, single-flavored, and double-flavored meson states. In each case, a spin 0 "ground state" is excited by an energy interval $m_\pi \sim 140$ MeV and then by an energy interval $X \sim 420$ MeV. In the unflavored channel, the first excitation is the $\alpha$-quantization transition from leptonic to hadronic states. In the flavored channels, the first excitation is an $\alpha$-quantized $m_\pi$ hadronic mass quantum. In all three channels, the second excitation is the dominant excitation quantum $X = 3\,m_\pi$.

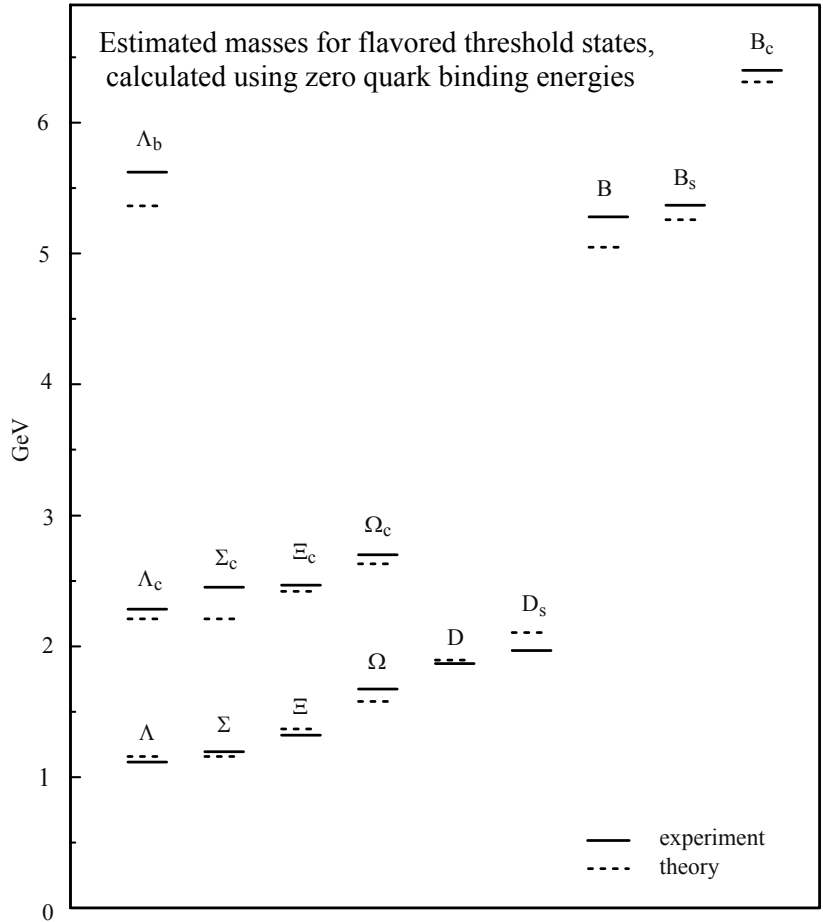

Figure 33. Mass estimates for the *s*, *c*, and *b* flavored hyperons and mesons. The constituent-quark masses of Eq. (26) are used in the zero-binding-energy approximation. The magnitudes of these flavored masses are given correctly, but not with the better than 1% precision of the MX octet states of Table 2. The fact that many theoretical masses are lower than the corresponding experimental masses indicates that their quark structures are more complex than simple quark model assignments suggest.